\documentclass[amsmath,
amssymb,
a4paper,
aps,prx,
twocolumn, reprint, 
fleqn,
showpacs,
floatfix, longbibliography]{revtex4-2}

\usepackage{amsmath}
\usepackage[table]{xcolor}
\usepackage[T1]{fontenc}
\usepackage{lmodern}
\usepackage{bbm}
\usepackage{microtype}
\usepackage{mathtools}
\usepackage{mathcomp}
\usepackage{color}
\usepackage{tabularx}
\usepackage{stmaryrd}
\usepackage{booktabs}
\usepackage{multirow}

\usepackage{xcolor}
\usepackage[
format=hang,
indention=-1.0cm,
justification=RaggedRight,
font=scriptsize
]{caption}

\usepackage{graphicx} 
\usepackage{dcolumn} 
\usepackage{bm} 
\usepackage[version=4]{mhchem}
\usepackage{stackrel}
\usepackage{verbatim}   

\usepackage{hyperref}   
\usepackage{float}

\usepackage[geometry]{ifsym}
\usepackage[utf8]{inputenc}
\usepackage[english]{babel}
\usepackage{calc}
\usepackage{graphicx}
\usepackage{textgreek}

\graphicspath{{Figures/}} 
\newlength\myheight
\newlength\mydepth
\settototalheight\myheight{Xygp}
\newcommand{\be}{\begin{equation}}
\newcommand{\ee}{\end{equation}}
\newcommand{\bea}{\begin{eqnarray}}
\newcommand{\eea}{\end{eqnarray}}

\newcommand{\blackbowtie}{\mathrel \blacktriangleright \joinrel \mathrel \blacktriangleleft}
\newcommand{\halfblackbowtie}{\mathrel \triangleright \joinrel \mathrel \blacktriangleleft}

\newcommand{\ffor}{\mathrel \triangleright \joinrel \mathrel { \triangleright}}

\newcommand{\blackffor}{\mathrel \blacktriangleright \joinrel \mathrel {\!\blacktriangleright}}

\newcommand{\blackwhiteffor}{\mathrel \blacktriangleright \joinrel \mathrel {\!\triangleright}}

\newcommand{\Lmir}{{\text{\reflectbox{$\mathbb{L}$}}}}

\newcommand{\copro}{\text{\footnotesize{$\Upsilon$}}}

\newcommand{\mathleft}{\@fleqntrue\@mathmargin0pt}
\newcommand{\mathcenter}{\@fleqnfalse}

\usepackage{academicons}
\definecolor{orcidlogocol}{HTML}{A6CE39}

\renewcommand{\selectlanguage}[1]{}
\usepackage{orcidlink}

\begin{document}

\title{On data and dimension in chemistry - irreversibility, concealment and emergent conservation laws} 
\author{Alex Blokhuis\orcidlink{0000-0002-4594-596X}}
\email{alex_blokhuis@hotmail.com}
\affiliation{Stratingh Institute for Chemistry, University of Groningen, Nijenborgh 4, 9747 AG Groningen, the Netherlands}
\affiliation{Groningen Institute for Evolutionary Life Sciences, University of Groningen, Nijenborgh 4, 9747 AG Groningen, the Netherlands}
\affiliation{University of Strasbourg $\&$ CNRS, UMR7140, 67083 Strasbourg, France} 
\affiliation{Instituto IMDEA Nanociencia, Calle Faraday 9, 28049 Madrid, Spain} 

\author{Martijn van Kuppeveld\orcidlink{0000-0001-9045-9322}}\affiliation{Radboud Center for Natural Philosophy, Radboud University Nijmegen, Heyendaalseweg
135, 6525 AJ Nijmegen, The Netherlands}
\affiliation{$\text{Stratingh}$ Institute for Chemistry, University of Groningen, Nijenborgh 4, 9747 AG Groningen, the Netherlands}

\author{Daan van de Weem}\affiliation{Scuola Internazionale Superiore di Studi Avanzati (SISSA), Via Bonema 265, 34136 Trieste, Italy}

\author{Robert Pollice\orcidlink{0000-0001-8836-6266}}
\email{r.pollice@rug.nl}
\affiliation{Stratingh Institute for Chemistry, University of Groningen, Nijenborgh 4, 9747 AG Groningen, the Netherlands}

\date{June 2023}
\date{\today} 
\begin{abstract} 
Chemical systems are interpreted through the species they contain and the reactions they may undergo, i.e., their chemical reaction network (CRN). In spite of their central importance to chemistry, the structure of CRNs continues to be challenging to deduce from data. Although there exist structural laws relating species, reactions, conserved quantities and cycles, there has been limited attention to their measurable consequences. One such is the \emph{dimension} of the chemical data: the number of independent reactions or equivalently independent species, which corresponds to the number of measured variables minus the number of constraints. In this paper we attempt to relate the experimentally observed dimensional features to conservation laws and underlying CRN structure. Our approach extends to any Markov model as well as many nonlinear models in statistical physics and furnishes new analytical tools to find exact solutions.

In particular, we investigate the effects of species that are concealed and reactions that are irreversible. For instance, irreversible reactions can have proportional rates. The resulting reduction in degrees of freedom can be captured by the \emph{co-production law} relating co-production relationships to emergent non-integer conservation laws and broken cycles. This law resolves a recent conundrum posed by a machine-discovered candidate for a non-integer conservation law, and characterizes certain types of CRN behavior.
We also obtain laws that allow us to relate data dimension to network structure in cases where some species cannot be discerned or distinguished by a given analytical technique, allowing to narrow down candidate CRNs from experimental data more effectively.

\end{abstract}

\maketitle
\section{Introduction}

Our capacity to build \cite{kriukov_history_2022,ivanov_computing_2023,zhang_complex_2018,sharko_redox-controlled_2023,singh_devising_2020,semenov_rational_2015,meijer_hierarchical_2017,Chen2023,ikeda_installing_2014,donau_phase_2022} and understand\cite{aris_prolegomena_1965,polettini_irreversible_2014,feinberg_foundations_2019,sughiyama_chemical_2022,dal_cengio_geometry_2023,avanzini_circuit_2023,Schnakenberg1976,hill_free_1989,baez2018biochemical,baez_compositional_2017,blokhuis_universal_2020,Vassena2024,deshpande_autocatalysis_2014,andersen_defining_2021,avanzini2023methods,Zhang2023_FT} chemical systems of increasing complexity is intimately tied to our understanding of their reaction networks and our ability to elucidate them. Whereas sizeable molecular structures can be elucidated today, reconstructing the structure of even small reaction networks remains challenging\cite{unsleber_exploration_2020}. Our understanding of theoretical CRN dynamics does not yet fully address the practical matter of relating their structure to experimental observation.



From a data perspective, prospects are ever brighter: There have been leaps in standardization and reproducibility due to increased automation in, e.g., synthesis\cite{manzano_autonomous_2022,zhang_advanced_2018,li_synthesis_2015,kitson_digitization_2018} and reaction monitoring\cite{malig_real-time_2017,malig_development_2020,rougeot_automated_2017,bentrup_combining_2010,theron_simultaneous_2016}, which has enabled the collection of comparably large high quality datasets. Simultaneously, new analytical methods continue to be developed to resolve what was previously invisible or indistinguishable. A few such methods are delayed reactant labeling \cite{jasikova_reaction_2015,hilgers_monitoring_2022}, nonlinear effects for enantiomers and excess\cite{geiger_hyperpositive_2020,blackmond_kinetic_2000,puchot_nonlinear_1986}, its extension to derivatives and rates\cite{pollice_expansion_2016}, time-dependent illumination protocols\cite {chouket_extra_2022}, oscillating temperature\cite{closa_identification_2015} or concentration protocols. As chemical data becomes more precise and captures additional details, deeper signatures of the underlying systems are revealed\cite{blackmond_kinetic_2015}.

The ever increasing availability of larger datasets characterizing CRNs in turn increases the utility of automating analysis steps. The interpretability of chemical data - for instance from spectroscopy - can be improved dramatically through multivariate curve resolution (MCR) techniques that seek to decompose a series of spectra of a mixture into a factorization of spectra of pure components and their concentrations\cite{kalka_spectra_2021,garrido_multivariate_2008,frans_reiterative_1984,ruckebusch_resolving_2016,bijlsma_determination_2001,voronov_multivariate_2014,monakhova_independent_2010,monakhova_chemometrics-assisted_2010}. An intrinsic obstacle for MCR approaches - known as rotational ambiguity - is that this factorization is not unique. 


Usually, chemical data does not capture all species present in any given system, and many CRN hypotheses could be proposed to explain said data. Naturally, one can directly compare the measurement data to the output of a candidate CRN. However, limited conclusions can be drawn from this fit alone\cite{vogt_conformational_2012,gianni_distinguishing_2014,morton_interpreting_1995}, as one can not exhaust all other possible CRNs this way. Recent approaches include alternative means of assessing more CRNs or assessing them more rigorously, e.g., through mechanism test functions\cite{closa_identification_2015}, Bayesian ERN analysis\cite{baltussen_bayesian_2022}, first-passage time distributions \cite{li_mechanisms_2013,thorneywork_direct_2020}, and increasingly generalized assessments of reaction orders\cite{blackmond_reaction_2005,bures_simple_2016,bures_variable_2016,pollice_2019}. However, exhaustive CRN assessment is as of yet not within reach. 




In 1963, R. Aris reported on the experimental possibility of establishing the number of independent reactions from concentration-time data in a stirred tank reactor\cite{aris_independence_1963} through linear algebra. Motivated by such insight and examples set by other fields, CRN theory started to develop shortly after\cite{aris_prolegomena_1965}. One important pillar of CRN theory is the establishment of structural criteria\cite{feinberg_complex_1972,horn_necessary_1972,feinberg_foundations_2019,hirono_structural_2021,Vassena2024} for the onset or absence of complex behavior (e.g., oscillations, multistability, chaos). Another pillar is nonequilibrium thermodynamics\cite{aris_prolegomena_1965,kondepudi_modern_2014,polettini_irreversible_2014,aslyamov2023nonideal} applied to chemistry in a wealth of contexts, \cite{polettini_irreversible_2014,dal_cengio_geometry_2023,penocchio_nonequilibrium_2021,blokhuis_reaction_2018,poulton_nonequilibrium_2019,gaspard_fluctuating_2018},including its connections to complex behavior\cite{polettini_dissipation_2015,aslyamov2023nonideal,Oberreiter2022,Liang2024,liang2024thermodynamicspacechemicalreaction}. In their study, CRNs have been principally approached from a perspective where they are known \textit{a priori}. 
As the CRN deduction problem remains open, it seems fitting to reinvestigate the 1963 viewpoint of an observer interpreting chemical data from an unknown CRN. 

In this paper, we consider the \emph{dimensional indices} of chemical data a CRN would produce and how it manifests in a series of (e.g. spectroscopic) measurements. The first dimension we discuss is the number of discernible independent reactions, $d= \#$ measured variables $-\#$ constraints. First, we derive what we call the \emph{co-production law}, which captures emergent non-integer conservation laws and broken cycles due to collinear reactions. We demonstrate that it resolves a conundrum posed by an anomalous conservation law recently discovered\cite{Liu_PRE_2024}, and apply it to derive new expressions for mechanisms in chemistry and models in statistical physics. One major application, elucidating CRN structure, is briefly illustrated, and developed further in our companion paper \cite{blokhuis2024ejoc}.
We then derive deeper structural links between coproduction and other integer quantities in CRN theory, resulting in the coproduction-linkage law (CLL), which allows for quick visual elucidation of all instances of coproduction, and their structural causes.
Subsequently, we apply the same methodology to derive a dimensional law for "concealed" species i.e. species that are not measured directly and another one for indistinguishable (i.e. quantified, but not disentangled) species. 
We illustrate these results through examples using literature data and literature CRNs. In our companion paper\cite{blokhuis2024ejoc}, we describe how determining these dimensions from the experimental data relates portions of data to specific parts of the CRN structure.

Finally, since a large variety of models can be treated as CRNs or readily mapped to them, our results have bearing well beyond chemical data. We make this point explicit by looking at several models in statistical physics - where these types of results have so far rarely been used - and show how our results provide new tools to study them. We consider the random-sequential adsorption of n-mers on a discrete lattice of size L. Due to co-production, this process has $n$ conservation laws constraining the dynamics, and we provide exact expressions for these conservation laws for arbitrary $L,n$. Our new conservation laws let us explicitly and easily calculate the statistics of this class of 1d adsorbtion models. Further applications and extensions are provided in the appendix.

Our paper is organized as follows. Sec. II a motivating example of the phenomenology of emergent conservation laws due to pairs of linearly dependent reactions (co-production). In Sec. III we introduce the stoichiometric matrix formalism and procedures by which we can build a theory. Sec. IV formalizes co-production and emergent conservation laws through index laws. Sec. V introduces the dimensionality of chemical data and the effect of factors like concealed species and coproduction. Sec. VI discusses the experimental observation of data dimension. Sec. VII Applies insights from co-production and its index laws to models in statistical physics. Sec. VIII provides a broader discussion of our results and the literature.

An overview of important quantities can be found in the glossary in Appendix \ref{subsection:glossary}.



\section{Excess Conservation: an example}

We will here consider a motivating example of a system that exhibits a conservation law in excess of the number predicted by current CRN theories. 

\begin{figure}[tbhp!]
\centering
\includegraphics[width=1.0\linewidth]{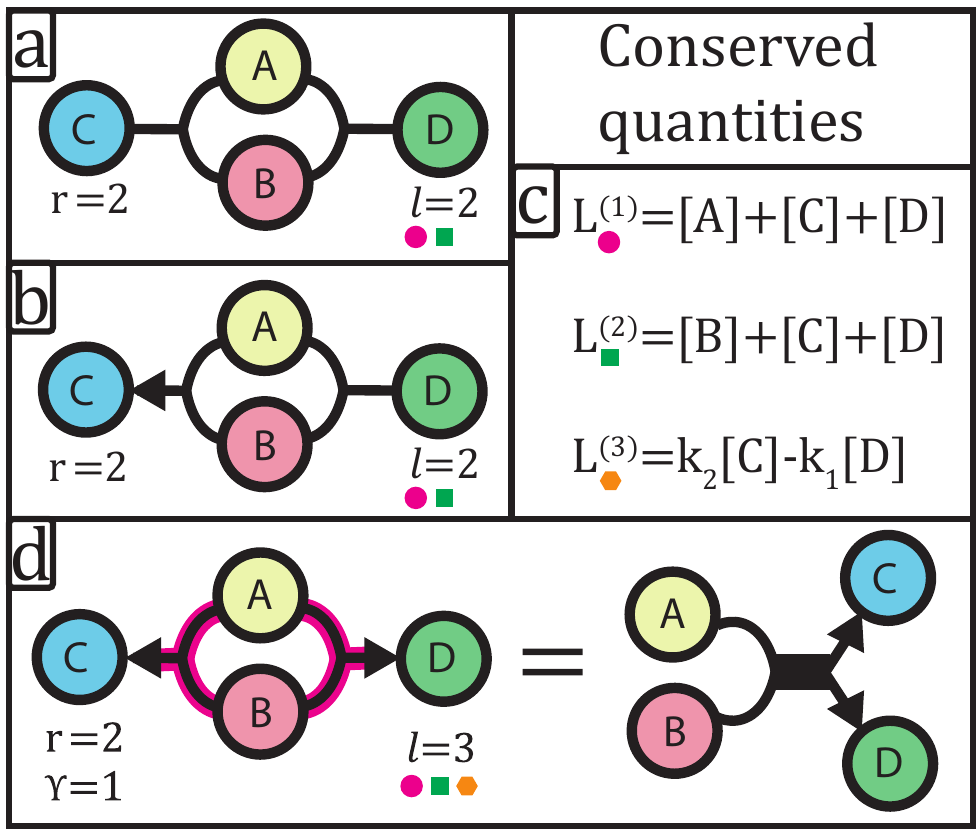}
\caption{a-b) CRNs whose conservation laws follow from integer stoichiometry. c) conservation laws. Making reactions $r_1,r_2$ irreversible, an additional non-integer conservation conserved quantity $L^{(3)}$ emerges. d) CRN with an emergent non-integer conserved quantity. For integer stoichiometry, such quantities do not follow from the first structural law (Eq.\ref{equation:SL1}). $r: \#$ reactions, $\ell: \#$ conserved quantities, $\text{\footnotesize{$\Upsilon$}}$: co-production index, $\#$ collinear reactions. Species are represented by nodes. Reactants (resp. products) in a reaction are joined by round arcs. The networks depicted are equal '=' with respect to their description of a dynamical system on a macroscopic level: they represent the exact same ODEs.}
\label{fig:Motivation}
\end{figure}

As an insightful example in which a new conservation law emerges, we will consider the CRNs in Fig. \ref{fig:Motivation}. We consider $s=4$ species, and $r=2$ bimolecular reactions
\be
\ce{C} \overset{1}{\leftrightarrows} \ce{A} + \ce{B} \overset{2}{\leftrightarrows} \ce{D}. \label{equation:CRN1}
\ee

For mass-action kinetics (Sec. \ref{subsec:timeev}), we obtain ordinary differential equations (ODEs)
\hspace*{-0.5cm}\vbox{\bea
d_t [\ce{A}] &=& - \left(\kappa_1 + \kappa_2\right) [\ce{A}][\ce{B}] + \kappa_3 [\ce{C}] + \kappa_4 [\ce{D}],  \nonumber \\
d_t [\ce{B}] &=& - \left(\kappa_1 + \kappa_2\right) [\ce{A}][\ce{B}] + \kappa_3 [\ce{C}] + \kappa_4 [\ce{D}], \nonumber \\
d_t [\ce{C}] &=& \kappa_1 [\ce{A}][\ce{B}] - \kappa_3 [\ce{C}], \nonumber \\
d_t [\ce{D}] &=& \kappa_2 [\ce{A}][\ce{B}] - \kappa_4 [\ce{C}]. \label{equation:ODEex1}
\eea}
We're looking for conserved quantities $L$ that are linear combinations of concentrations  
\bea
L &=& \sum_i \ell_i [\ce{X}_i], \ \ \\
d_t L &=& 0.
\eea
Assuming conservation laws are due to stoichiometry only, CRN theory would predict (Sec. \ref{subsection:SL1}) that there are exactly $\ell = s - r = 2$ linear conservation laws,
\bea
L^{(1)} &=& [{\ce{A}}] + [{\ce{C}}] + [{\ce{D}}], \nonumber \\ 
L^{(2)} &=& [{\ce{B}}] + [{\ce{C}}] + [{\ce{D}}]. \nonumber 
\eea
which is correct. 

However, when  both reactions are rendered irreversible, that is
\be
\ce{C} \overset{1}{\leftarrow} \ce{A} + \ce{B} \overset{2}{\rightarrow} \ce{D}. \label{equation:CRN2}
\ee
an additional conserved quantity appears.
Since now $\kappa_3=\kappa_4=0$, it follows from inspection that
\bea
d_t [\ce{C}] = \frac{\kappa_1}{\kappa_2} d_t [\ce{D}] = \kappa_1 [\ce{A}][\ce{B}]. \ \ \ 
\eea
Thus $\ell = 3$, i.e., a novel type of conservation law with non-integer coefficients emerged 
\be
L^{(3)} = k_2 [\ce{C}] - k_1 [\ce{D}] ,
\ee
which does not follow from stoichiometry. As a shorthand, such emergent invariants will be referred to as \textit{emanants}\footnote{Lat. \textit{emanatus}. To emanate: To flow out. Figuratively: to arise from. We consider that conservation laws beyond the usual stoichiometric ones are 'emergent' in that they arise from a special set of conditions. An analogous concept already exists for cycles, where   emergent cycles are also referred to as affinities \cite{polettini_irreversible_2014}}.

\begin{figure}[tbhp!]
\centering
\includegraphics[width=1.0\linewidth]{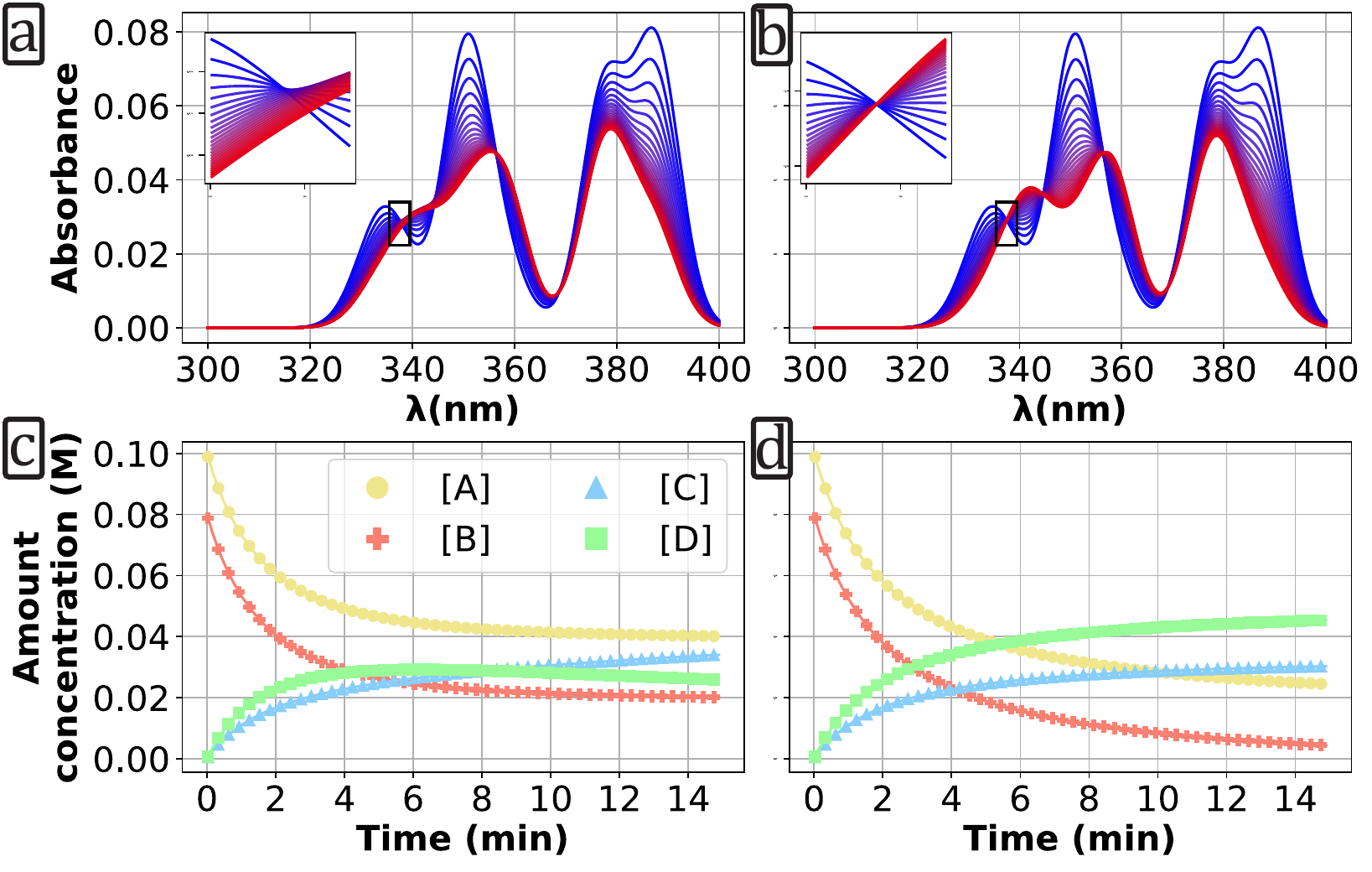}
\caption{(a) Simulated successive absorption spectra, during the time the CRN $\ce{C} \leftrightarrows \ce{A} + \ce{B} \leftrightarrows \ce{D}$ undergoes chemical change (i.e., dynamics in c). A close-up reveals no isosbestic points  ($d(\Lambda)=2$), i.e., no fixed total absorbance. b) Simulated successive absorption spectra, during the time the CRN $\ce{C} \leftarrow \ce{A} + \ce{B} \rightarrow \ce{D}$, undergoes chemical change (i.e., dynamics in D) a close-up reveals an isosbestic point ($d(\Lambda)=1$), i.e., a total absorbance that remains fixed over time. The underlying dynamics being simulated is given by system of ODEs in Eq. \eqref{equation:ODEex1}. (c) Underlying dynamics for a) $\pmb{\kappa}=(2.0, 0.03, 3.0, 0.11)$, d) underlying dynamics for b). $\pmb{\kappa}=(2.0, 0, 3.0, 0)$. For both, $[\pmb{\ce{X}}]_0=(0.1,0.08,0,0)$. Absorption spectra for each species were generated randomly once and used for both simulations. }
\label{fig:spec1}
\end{figure}

Importantly, if we monitored both the reversible CRN \eqref{equation:CRN1} and the irreversible CRN \eqref{equation:CRN2} spectroscopically (Fig. \ref{fig:spec1}), we would - provided spectral overlap - observe distinct spectral features. For irreversible CRN \eqref{equation:CRN2}, in Fig. \ref{fig:spec1}b we zoom in on a region where at a specific wavelength $\lambda^*$ the total absorbance remains perfectly fixed. Such a feature is known as an isosbestic\cite{braslavsky_glossary_2007,kaspar_quality_2023,cohen_588_1962,scheibe_uber_1937} point, usually interpreted as a feature of $1d$ data (but see Sec. \ref{subsec:isosbesticpts} for a more complete treatment).

Here, an isosbestic point is an immediate observable consequence of an additional conservation law lowering the dimension (number of linearly independent variables) from $d=2$ for CRN \eqref{equation:CRN1} to $d=1$ for CRN \eqref{equation:CRN2} (Sec. \ref{section:datadimension}). \footnote{Although no mention is made of conserved quantities, Cohen and Fischer have reported on this phenomenon in their work on conditions for the occurrence of isosbestic points\cite{cohen_588_1962}, by arguing that while $\ce{B} \leftarrow \ce{A} \rightarrow \ce{C}$ has 2 reactions, their simultaneity still enables an isosbestic point. Indeed, due to the introduction of reverse reactions, CRN \eqref{equation:CRN1} has no isosbestic points.}  


Thus, conserved quantities need not derive from (classical) stoichiometry alone\footnote{That is to say, in chemistry there are 'natural' choices for which reactions are described and treated as distinct. The stoichiometry adopted in this classical reference frame may not explain all conserved quantities if further ones arise from co-production. After merging reactions, however, one can also furnish a (nonclassical) stoichiometric interpretation for these laws.}. This forms the motivation to reinspect how conservation laws in CRNs are derived, and extend this formalism to count chemically meaningful quantities that are missed via standard approaches.

\section{Stoichiometric matrices}

We will now reinspect CRN theory from the point of view of stoichiometry, and subsequently identify where this analysis can overlook conservation laws.
A reversible chemical reaction $r_{i}^{\circ}$ is represented as

\be
\sum_k \nu^{\ominus}_{k,i} \ce{X_k} \overset{i}{\rightleftarrows} \sum_k \nu^{\oplus}_{k,i} \ce{X_k}
\ee

where $\nu^{\ominus}_{k,i}, \nu^{\oplus}_{k,i}$ are integer stoichiometric coefficients.
Similarly, an irreversible reaction $r_{i}^{\triangleright}$ is represented as

\be
\sum_k \nu^{\ominus}_{k,i} \ce{X_k} \overset{i}{\rightarrow} \sum_k \nu^{\oplus}_{k,i} \ce{X_k}
\ee

We can then define a stoichiometric matrix by taking the difference in the stoichiometric coefficients

\bea
\mathbb{S} = \nu^{\oplus} - \nu^{\ominus}, \label{equation:decompositionSirr}
\eea

and we refer to $\nu^{\ominus}$ (resp. $\nu^{\oplus}$) as the stoichiometric reactant (resp. product) matrix. We endow $\mathbb{S}$ with a suffix to denote further context or a submatrix of $\mathbb{S}$. For reversible and irreversible reactions, we now define stoichiometric matrices that respectively contain: \\
$\mathbb{S}_{\circ}$: all reactions counted once \\
$\mathbb{S}_{\bowtie}$: only reversible reactions \\
$\mathbb{S}_{\triangleright}$: only irreversible reactions \\
$\mathbb{S}_{\ffor}$: all reactions, reversible reactions included twice (once for each direction) \\

In general we can write:

\bea 
\mathbb{S}_\circ &=& (\mathbb{S}_\triangleright, \mathbb{S}_{\bowtie} ) ,  \label{equation:SDEF} \\
\mathbb{S}_{\ffor} &=& (\mathbb{S}_\circ, -\mathbb{S}_{\bowtie} ) = (\mathbb{S}_\triangleright, \mathbb{S}_{\bowtie}, -\mathbb{S}_{\bowtie})  \label{equation:SDEF_2}.
\eea

Formally, a reversible reaction can be built up from a pair of opposing irreversible reactions, 

\bea
&\ce{X_1} \overset{1}{\leftrightarrows} \ce{X_2} + \ce{X_3},  \\
&\ce{X_1} \overset{1}{\rightarrow} \ce{X_2} + \ce{X_3} \overset{2}{\rightarrow} \ce{X_1} \eea

For the truly reversible case, we may then write:

\bea
&\ce{X_1} \overset{1}{\leftrightarrows} \ce{X_2} + \ce{X_3}, \label{equation:stoichmatrixrevirrev} \\
&\mathbb{S}_{\circ} = \mathbb{S}_{\bowtie} = \begin{pmatrix} -1 \\ 1 \\ 1 \end{pmatrix}, \ \ \mathbb{S}_{\ffor} = \begin{pmatrix} -1 & 1 \\ 1 & -1  \\ 1 & -1 \end{pmatrix}, \nonumber 
\eea
where succesive rows act on $\ce{X_1}, \ce{X2},\ce{X3}$. As there are no irreversible reactions, the matrix $\mathbb{S}_\triangleright $ is 'empty' \footnote{One can treat $\mathbb{S}_\triangleright $ here as an 'empty' 3-by-0 matrix $()$, s.t.  $(\mathbb{S}_{\triangleright}, \mathbb{S}_{\bowtie}) = (\mathbb{S}_{\bowtie}) $ .}.

For a fully reversible chemical reaction network (CRN), we can always represent the same CRN in terms of pairs of irreversible reaction steps:

\bea
\mathbb{S}_{\bowtie} = \mathbb{S}_{\circ} \leftrightarrow  \mathbb{S}_{\ffor} = (\mathbb{S}_\circ, -\mathbb{S}_\circ). 
\eea

Conversely, if at least some reactions do not have a reverse, we can only fully represent the CRN (including distinctions between reversible and irreversible) using $\mathbb{S}_{\ffor}$, for instance: 

\bea
&\ce{X_4} \overset{1}{\leftarrow}  \ce{X_1} \underset{3}{\overset{2}{\rightleftarrows}} \ce{X_2} + \ce{X_3}, \nonumber \\
& \mathbb{S}_{\ffor} = \begin{pmatrix} -1 & -1 & 1 \\ 0 & 1 & -1  \\ 0 & 1 & -1 \\ 1 & 0 & 0 \end{pmatrix}, \ \ \  \mathbb{S}_{\triangleright} = \begin{pmatrix} -1  \\ 0   \\ 0  \\ 1  \end{pmatrix}, \nonumber \\
& \mathbb{S}_{\circ} = \begin{pmatrix} -1 & -1  \\ 0 & 1   \\ 0 & 1  \\ 1 & 0  \end{pmatrix}, \ \ \ 
 \mathbb{S}_{\bowtie} = \begin{pmatrix} -1  \\  1   \\  1  \\  0 \end{pmatrix}. \nonumber
\label{equation:stoichmatrixirrex}
\eea
where succesive rows act on $\ce{X_1}, \ce{X2},\ce{X3}, \ce{X4}$. Thermodynamically, reactions have a microscopic reverse. Irreversible reactions can parsimoniously describe kinetics when this reverse becomes negligible enough \footnote{Another scenario where irreversible reactions can be used is for reactions with driving forces fixed (e.g., phase reactions, reactions with chemostatted concentrations) in such a way that their rates remain fixed. A reversible and irreversible description of the reaction would yield the same fixed rate (but only the former can be linked to thermodynamics).}. This can for instance occur when products disappear from the reaction medium (e.g., as gas or precipitate), on short timescales (Sec. \ref{subsection:res-dep}) or for reactions with high driving forces. The latter is of special importance for the production and subsequent reaction of highly reactive species, as usually occurs in fields such as radiochemistry, plasma chemistry, and photochemistry. Irreversible reactions are thus of chemistry-wide importance, but particularly prevalent in some branches of chemistry such as astrochemistry \cite{Oberg2016,Millar2024,Indriolo2013}, photochemical motors \cite{boursalian_all-photochemical_2020,GarciaLopez2020} and atmospheric chemistry \cite{Cox2003,sturm_conservation_2022,Herrmann2015}.

\subsection{The first structural law (SL1)}
\label{subsection:SL1}

A matrix can be characterized by four fundamental subspaces:
The span of its column vectors (image), the vectors spanning the left nullspace orthogonal to these (cokernel), the span of row vectors (coimage) and - orthogonally - the vectors spanning the right nullspace (kernel) (Sec. \ref{subsection:stoich_int_sl1}). Dimensions of these fundamental subspaces for $\mathbb{S}$ count quantities with a chemical interpretation, through the fundamental theorem of linear algebra (FTLA).
We refer to the following relation \cite{aris_prolegomena_1965,polettini_irreversible_2014} as the first structural law (SL1):

\be
\text{rk}(\mathbb{S}) = s - \ell = r - c \label{equation:SL1}
\ee

Letting $\#$ denote 'the number of', we have: \\
{$\text{rk}(\mathbb{S})$}: rank of matrix $\mathbb{S}$ \\
{$s: \ \#$} species ({$\ce{X_1}, ... , \ce{X_s}$})\\
{$\ell: \ \#$} conserved quantities {$(\text{dim}(\text{coker}(\mathbb{S}))$} \\
{$r: \ \#$} reactions ({$r_1, ..., r_r$}) \\
{$c: \ \#$} cycles {$(\text{dim}(\text{ker}(\mathbb{S})))$} 

The number of conserved quantities (resp. cycles) are thus directly related to the dimension of the left (resp. right) nullspace, i.e., of the cokernel (resp. kernel) of $\mathbb{S}$. The fundamental theorem of linear algebra (FTLA) provides two ways to count the rank in terms of chemical quantities, thus giving us a combinatorial identity (SL1) (Eq. \eqref{equation:SL1}) that relates them. SL1 thereby relates fundamental CRN quantitities based on structure. This technique has notably been used to find index relations for open CRNs \cite{polettini_irreversible_2014}, flow reactors, serial transfer \cite{blokhuis_reaction_2018}, and autocatalytic networks \cite{Despons2024auto}.

We furthermore adopt the convention that indices counting the size of subspaces ($s,\ell,r,c$) adopt the suffixes from the matrix they are applied to, for instance $s_{\ffor}, \ell_{\ffor}, r_{\ffor}, c_{\ffor}$ are respectively $\#$ species, conserved quantities, reactions, and cycles for $\mathbb{S}_{\ffor}$.

\subsection{Stoichiometric interpretation of SL1}
\label{subsection:stoich_int_sl1}

We will now take a closer look at the number of cycles ($c$) and conserved quantities ($\ell$). These correspond respectively to $c$ right nullvectors and $\ell$ left nullvectors: 

\bea
\mathbb{S} \ \pmb{c}^{(i)} = \pmb{0}, \ \ \ (i \in \{1, ..., c\}) \\
\mathbb{S}^T \pmb{\ell}^{(i)}  = \pmb{0} \ \ \ (i \in \{1, ..., \ell\}). 
\eea

Where  $\pmb{c}^{(1)},..,\pmb{c}^{(c)}$, (resp. $\pmb{\ell}^{(1)},..,\pmb{\ell}^{(\ell)} $) denote basis vectors that span the right (resp. left) nullspace. 

\begin{figure}[tbhp!]
\centering
\includegraphics[width=1.0\linewidth]{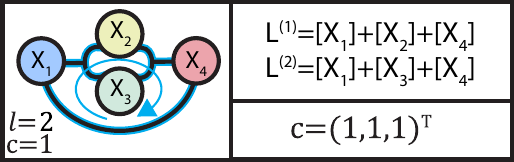}
\caption{A small CRN with a cycle. $\ell: \#$ conserved quantities. $c: \#$ cycles.}
\label{fig:X1X2X3}
\end{figure}

Usually, cycles are interpreted as combinations of reactions that leave the system unchanged and the constraints are interpreted as integer combinations of species that remain unchanged. For instance, the network below (Fig. \ref{fig:X1X2X3}) has $s=4, r=3$, $\ell=2, c=1$:

\bea 
\ce{X_1} \overset{1}{\leftrightarrows} \ce{X_2} + \ce{X_3} \overset{2}{\leftrightarrows} \ce{X_4}  \overset{3}{\leftrightarrows} \ce{X_1} \\
\mathbb{S}_\circ = \left(\begin{smallmatrix} -1 & 0 & 1  \\ 
1 & -1 & 0  \\
1 & -1 & 0  \\ 
0 & 1 & -1   
\end{smallmatrix}\right).
\eea

Performing reactions 1, 2, and 3 in equal amounts leaves the system unchanged: 

\be 
\begin{pmatrix} \Delta [\ce{X}_1] \\ \Delta [\ce{X}_2] \\ \Delta [\ce{X}_3] \\  \Delta [\ce{X}_4]\end{pmatrix} = \mathbb{S} \begin{pmatrix} 1 \\ 1 \\ 1 \end{pmatrix}  = \pmb{0},
\ee

and the following integer combinations of species are conserved:

\bea
L^{(1)} &=& [\ce{X_1}] + [\ce{X_2}] + [\ce{X_4}], \\
L^{(2)} &=& [\ce{X_1}] + [\ce{X_3}] + [\ce{X_4}].
\eea

Notably, stoichiometric constraints are known by many names \footnote{There exist further decompositions of $\ell_\circ$ for open systems. For instance, mass-like conservation laws with purely positive entries are used to describe moiety conservation in biochemistry \cite{hofmeyr_metabolic_1986,avanzini_thermodynamics_2022}. Charge-like conservation laws - with coefficients of mixed sign - can always be eliminated in a closed system \cite{blokhuis:tel-02566386}, in favor of a basis with purely positive conservation laws \cite{Schuster1991,Muller2022}. This is not true for an open system, e.g., $\ce{A} + \ce{B} \leftrightarrows \emptyset$ has as its $\ell_\circ=1$ unique conservation law the charge-like quantity $[\ce{A}]-[\ce{B}]=L$}, for instance, conserved charges \cite{hirono_structural_2021}, components, invariants \cite{aris_prolegomena_1965}, conserved quantities \cite{polettini_irreversible_2014}, or conservation laws\cite{Liu_PRE_2024}. 

As foreshadowed by our motivating example, SL1 need not count all linear conservation laws (as is typically assumed) that constrain the ODEs of a system. We will show that this is because reaction rates may be linearly dependent. Accounting for this fact correctly produces extensions of SL1 that require extensions of aforementioned interpretations of cycles and conservation laws.

In anticipation of this extension - in which cycles may vanish and additional conservation laws emerge - we will henceforth add a suffix to the conserved quantities and cycles we counted without this extension, and count their number as $\ell_\circ, c_\circ$, which correspond to the number of conservation laws and cycles one would have when all reactions proceeded at independent rates (which would, for example, occur for reversible reactions).

For the stoichiometric matrix $\mathbb{S}_{\ffor}$, all reactions are encoded as one-way reactions. We denote the number of reactions that have a reverse by $r_{\bowtie}$. Performing a pair of conjugate reactions leaves the system unchanged, yielding a trivial 2-membered cycle\footnote{Ordinarily, one does not seek to count 2-membered cycles (in the sense of right nullvectors). They are excluded naturally describing forward and reverse reactions by the same entity. Certain notions of cycle explicitly require them to have three members, as for 2-membered ones one cannot disentangle the orientations.} 

All reactions in $\mathbb{S}_{\ffor}$ in excess of those in $\mathbb{S}_\circ$ yield 2-cycles, we thus always have $\text{rk}(\mathbb{S}_{\ffor}) = \text{rk}(\mathbb{S}_\circ)$.

We will now split the established counting of reactions $r$ and cycles $c_\circ$ into their reversible and irreversible components:
\bea
\! \! \! \! \text{rk}(\mathbb{S}_{\ffor}) &=& s - \ell_{\ffor}  = r_{\ffor} - c_{\ffor} \\ 
&=& s - \ell_\circ = r_\circ - c_\circ  \label{SL1irrb} \\ c_\circ &=& c_{\ffor} - r_{\bowtie}. \\
\ell_\circ &=& \ell_{\ffor}. 
\label{equation:reactrevirr} 
\eea

Conventionally, $r,c$ are used to denote nontrivial reactions and cycles:

\be
r = r_o, \ \ \ c = c_\circ.
\ee

The necessity of further decomposing constraints and cycles emerges from kinetics, which we  will  introduce next.

\subsection{Time evolution}
\label{subsec:timeev}

The time evolution of species concentrations $[\ce{X_1}], ..., [\ce{X_s}]$ follows the kinetic equations

\be
d_t [\pmb{\ce{X}}] = \mathbb{S} \ \pmb{J} , \label{equation:timeevo}
\ee

where $\pmb{J}=(J_1, ..., J_r)^T$ is a vector of $r$ reaction currents. We will consider these currents to follow mass-action kinetics \footnote{Some authors investigate CRNs with current functions that are not mass-action\cite{feinberg_foundations_2019,Vassena2024}, allowing to generalize results to larger classes of dynamical systems. For instance, N. Vassena defines current functions as \emph{chemical} when they are nonnegative and monotonically increasing functions depending only on reactant concentration(s). Co-production can occur for more general classes of current functions, but need then no longer be implied by reactant stoichiometry.} so that their expression follows from stoichiometry as

\be
J_{i,\ffor} = \kappa_i \prod_{k=1}^s \nu_{\ffor,k,i}^\ominus [\ce{X_k}]^{\nu_{\ffor,k,i}^\ominus} . \label{equation:currentlawmassaction}
\ee

From Eq. \eqref{equation:currentlawmassaction} it follows directly that, under mass-action, a pair of reactions $r_i,r_j$ with the same reactant stoichiometry will have linearly proportional rates

\be
\nu_{\ffor,k,i}^{\ominus} = \nu_{\ffor,k,j}^{\ominus} \leftrightarrows J_{i,\ffor} \propto J_{j,\ffor}. \label{equation:proprate}
\ee

Other kinetic schemes than mass-action (e.g., Michaelis-Menten, Lindemann-Haldane, Eigen-Wilkins kinetics) are normally derived by evaluating mass-action under certain limits or approximations. In Sec. \ref{appendix:co-production beyond mass action}, we show that these approximations tend to preserve the proportional rate property in Eq. \eqref{equation:proprate}.

We are ultimately interested in finding linear conservation laws $L^{(i)}$ with coefficients $\pmb{\ell}^{(i)}$: 

\bea
&\ell_1^{(i)} [\ce{X}_1] + ... + \ell_s^{(i)} [\ce{X}_s] = L^{(i)}, \\
&d_t L^{(i)} = 0.   
\eea

Rewriting this equation using the above notation yields

\be
d_t \left(\pmb{\ell} \cdot [\pmb{X}] \right) = d_t \ \left(\pmb{\ell}^T  [\pmb{\ce{X}}]\right) = \pmb{\ell}^T \ \mathbb{S} \ \pmb{J} = 0 ,\label{equation:dyneq} 
\ee
where $\cdot$ denotes an inproduct. This formulation shows that the left nullvectors of $\mathbb{S}$ form a solution, as for a left nullvector $\pmb{\ell}$ we have 

\be
\pmb{\ell}^T \ \mathbb{S} = \pmb{0}^T .\label{equation:leftnull} 
\ee


However, these are not necessarily the only solutions. We are ultimately interested in the constant solutions $\pmb{\ell}^T$ of Eq. \eqref{equation:dyneq}, where we would like to remark that $\pmb{J}$ depends on the concentrations $[\ce{X_k}]$. These solutions then give us the coefficients of the conserved quantities. To see that Eq. \eqref{equation:dyneq} can have further solutions, we will now revisit our initial examples of reversible CRN Eq. \eqref{equation:CRN1} and irreversible CRN Eq. \eqref{equation:CRN2}. We then derive a network law for the additional chemical quantities SL1 needs to count when applied to the time evolution of species. Sec.\ref{section:atmosphericexample} of the Appendix applies this law to verify and explain the recent algorithmic detection of a non-integer conservation law \cite{Liu_PRE_2024} that was not anticipated, due to not being a solution to Eq. \eqref{equation:leftnull}.

\subsection{Hidden irreversible conservation law}
We can now  address the emergence of further conservation laws by irreversibility. We  will first revisit the case of our motivating example, in which a procedure is introduced that we generalize in Sec. \ref{subsection:genprocmerg}. This in turn allows the derivation of index laws governing emergent quantities due to irreversible reactions in Sec. \ref{section:emconslaw}:

\be
\ce{C} \overset{1}{\leftrightarrows} \ce{A} + \ce{B} \overset{2}{\leftrightarrows} \ce{D}.
\ee

Using mass-action (or a rate law that derives from it, see Sec. \ref{appendix:co-production beyond mass action}), we have $r_\circ=2$ linearly independent currents

\bea
J_{1,\circ} &=& \kappa_1 [\ce{A}] [\ce{B}] - \kappa_3 [\ce{C}] = J_{1,\ffor} - J_{3,\ffor} \nonumber \\ 
J_{2,\circ} &=& \kappa_2 [\ce{A}] [\ce{B}] - \kappa_4 [\ce{D}] = J_{2,\ffor} - J_{4,\ffor} \nonumber
\eea

and stoichiometric matrices $\mathbb{S}_\circ$, $\mathbb{S}_{\ffor} $

\bea
\mathbb{S}_\circ &=& \left(\begin{smallmatrix} -1 & -1 \\ 
-1 & -1 \\
1 & 0 \\
0 & 1 \end{smallmatrix}\right), \ \mathbb{S}_{\ffor}  = \left(\begin{smallmatrix} -1 & -1 & 1 & 1 \\ 
-1 & -1 & 1 & 1 \\
1 & 0 & -1 & 0 \\
0 & 1 & 0 & -1 \end{smallmatrix}\right) \nonumber 
\eea
And $\mathbb{S}_{\triangleright}$ is 'empty'. i.e., $\mathbb{S}_{\bowtie} = \mathbb{S}_\circ$, so that

\bea
d_t [\pmb{\ce{X}}] = \mathbb{S}_\circ \pmb{J}_\circ = \mathbb{S}_{\ffor}^\circ  \pmb{J}_{\ffor}.
\eea

Hence, we can represent the same species dynamics using a distinct number of reactions $r$. Since we have $r_\circ = 2$ linearly independent reactions, we need at least $r=2$.

We now modify the above network by rendering both bimolecular reactions irreversible

\be
\ce{C} \overset{1}{\leftarrow} \ce{A} + \ce{B} \overset{2}{\rightarrow} \ce{D},
\ee

so that $r_{\ffor} = 0$, and $\mathbb{S}_{\ffor}$ loses its last two columns compared to $\mathbb{S}_{\ffor}^\circ$.
We obtain the irreversible currents by removing the reverse reaction from $J_1, J_2$:
\bea
J_{1,\ffor} &=& \kappa_1 [\ce{A}] [\ce{B}],  \\ 
J_{2,\ffor} &=& \kappa_2 [\ce{A}] [\ce{B}] = \frac{\kappa_2}{\kappa_1} J_{1,\ffor} . \label{equation:colin}
\eea

Dynamically, the reactions are not independent since the currents are linearly proportional (cf. Eq. \eqref{equation:colin}). Thus, we only need 1 current and 1 reaction for a deterministic description \footnote{Note that in a stochastic description where deterministic $[\ce{A}], [\ce{B}]$ are replaced with fluctuating $n_{\ce{A}}, n_{\ce{B}}$, the collinear reactions still occur independently. Hence, the stochastic analogue of the conserved quantity is not conserved intrinsically, $\kappa_2 n_{\ce{A}} - \kappa_1 n_{\ce{B}} \neq 0$. Instead it holds on average, $\kappa_2 \langle n_{\ce{A}} \rangle - \kappa_1 \langle n_{\ce{B}} \rangle = 0$, and, thus, imposes a constraint on statistical momenta.} 
Hence, we can merge reactions with linearly proportional rates

\hspace*{-0.75cm}\vbox{\bea
\ce{A} + \ce{B} \overset{\blackffor}{\rightarrow} p \ce{C} + \left(1-p\right) \ce{D}, \label{equation:mergedratesfirst} \\ 
p=\frac{\kappa_1^+}{\kappa_1^+ + \kappa_2^+} , \ \ \ J_{\blackffor} = J_1 + J_2 
\eea}

We now construct a new stoichiometric matrix $\mathbb{S}_{\blackffor}$ that admits merged reactions 

\be
\mathbb{S}_{\ffor} = \left(\begin{smallmatrix} -1 & -1 \\ 
-1 & -1 \\
1 & 0 \\
0 & 1 \end{smallmatrix}\right), \ \ \mathbb{S}_{\blackffor} = \left(\begin{smallmatrix} -1  \\ 
-1  \\
p  \\
1 - p  \end{smallmatrix}\right). \label{equation:matrixmerge}
\ee
where succesive rows act on $\ce{A}, \ce{B},\ce{C}, \ce{D}$. 
The time evolution of the system is then fully described by

\be
d_t [\pmb{\ce{X}}] = \mathbb{S}_{\ffor} \pmb{J}_{\ffor} =  \mathbb{S}_{\blackffor} \pmb{J}_{\blackffor}. 
\ee

Since $\text{rk}(\mathbb{S}_{\blackffor}) = 1$, we now correctly predict the existence of $\ell_{\blackffor} = 3$ conservation laws. Denoting the number of emenants (emergent conservation laws) as $\rotatebox[origin=c]{180}{e}_{\blackwhiteffor}$, we have

\be
\rotatebox[origin=c]{180}{e}_{\blackwhiteffor} = \ell_{\blackffor} - \ell_{\circ}.
\ee

The left nullvector $\pmb{\ell}_{\blackffor}^{(3)} = (0, 0, \kappa_2, -\kappa_1)$ can be shown to only live in the nullspace of $\mathbb{S}_{\blackffor}$:

\be
\pmb{\ell}_{\blackffor} \mathbb{S}_{\blackffor} = \pmb{0}, \ \ \pmb{\ell}_{\blackffor} \mathbb{S}_{\ffor} = (\kappa_2, -\kappa_1) \neq \pmb{0} .\label{equation:co-production_a}
\ee

Rendering reactions irreversible can thus lead to the emergence of additional conserved quantities, because currents can become linearly dependent. However, not every pair of reactions with proportional reaction rates supplies us with a conservation law. In the following section (Sec. \ref{subsection:genprocmerg}), we will establish the general procedure . From there, we can derive laws (Sec. \ref{section:emconslaw}) detailing the additional quantities we need to count.

\subsection{General procedure for merging reactions}
\label{subsection:genprocmerg}

Let us now establish a more general merging procedure. We write all reactions as individual irreversible reactions (splitting up reversible reactions into irreversible pairs, as in $\mathbb{S}_{\ffor}$) and merge all reactions with proportional rates, so that no pair of currents that remains is pairwise proportional (Fig. \ref{fig:copro_derivation}a-c). We denote $\Phi$ the number of pairwise mergers needed to this end and $\mathbb{S}_{\blackffor}$ as the resulting stoichiometric matrix. In the derivation of most kinetic schemes (mass-action, Michaelis-Menten, Lindemann-Haldane \ref{appendix:co-production beyond mass action}) pairwise proportional rates directly follow from having the same reactant stoichiometry, i.e. the same reactant complex. In matrix form, these reactions have identical columns in $\nu_{\ffor}^{\ominus}$.

More formally, we use $\Phi$ to denote the number of pairwise mergers of reactions. Starting from $\mathbb{S}_{\ffor}$, we first partition all $r_{\ffor}$ reactions in $r_{\ffor}-\Phi$ disjoint sets $P^{(q)}$, with the condition that reactions $r_i,r_j$ in the same set have pairwise proportional rates, and pairwise proportional reactions are always in the same single set, i.e., $J_i \propto J_j \implies \exists q \ \ s.t. \ \ i,j \in P^{(q)}, i,j \notin P^{(k)}, (k \neq q)$.

For each set $P^{(q)}$, we can now introduce a single total current $J_{(q)}$, and express the current of each underlying reaction as a fraction of rate constants:

\bea
J_{(q)} &=& \sum_{i \in P^{(q)}} J_i \\
J_i &=& p_i J_{(q)}  \\
p_i &=&  \frac{\kappa_i }{ \sum_{j \in P^{(q)}} \kappa_j } . 
\eea

We can, by the same token, now express the reaction stoichiometry as a single reaction, through the weighted sum:
\bea
\sum_{i \in P^{(q)}} p_i \sum_{k=1}^s  \nu_{\ffor,k,i}^{\ominus} \ce{X}_k \rightarrow \\
\sum_{i \in P^{(q)}} p_i
\sum_{k=1}^s  \nu_{\ffor,k,i}^{\oplus} \ce{X}_k . \label{equation:reactionq}
\eea
The resulting stoichiometric coefficients make up the stoichiometry of a single merged reaction.

We now define $\mathbb{S}_{\blackffor}$ as the matrix obtained by this process, and Eq. \eqref{equation:reactionq} provides the coefficients in the q\textsuperscript{th} column of $\mathbb{S}_{\blackffor}$.


\section{Emergent conservation laws}
\label{section:emconslaw}

\begin{figure}[tbhp!]
\centering
\includegraphics[width=1.0\linewidth]{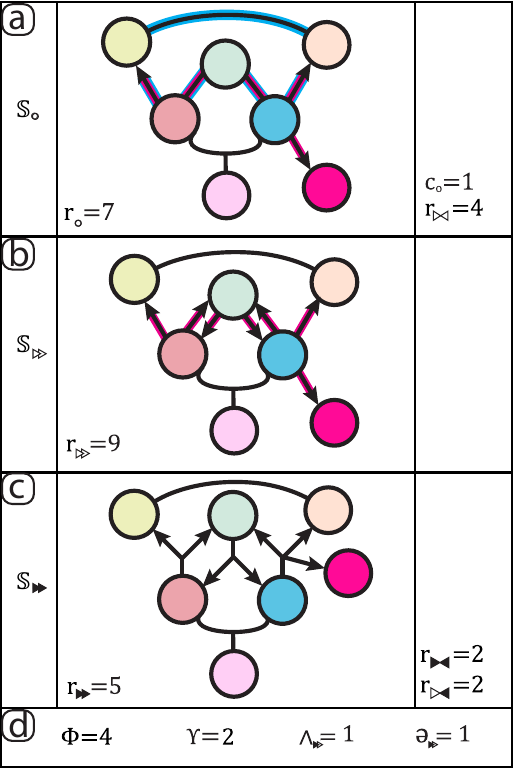}
\caption{a-c) Illustration of the general merging procedure. a) Representation with each reaction represented once ($\mathbb{S}_{\circ}$), a cycle is highlighted in blue. b) Representation with reversible reactions written as pairs of irreversible reactions ($\mathbb{S}_{\ffor}$). c) Representation after merging pairwise proportional reactions ($\mathbb{S}_{\ffor}$). d) $\Phi = 4$ reactions were merged with respect to $\mathbb{S}_{\ffor}$, whereas compared to $\mathbb{S}_{\circ}$ there are only $\copro = 2$ less reactions after mergers. Here, $\copro = 2$ results in one emanant and one broken cycle.}
\label{fig:copro_derivation}
\end{figure}

To correctly account for conserved quantities due to pairwise proportional reactions, we need to construct a stoichiometric matrix in which all pairs of reaction currents are linearly independent. Under mass-action (and extensions covered in Sec. \ref{appendix:co-production beyond mass action}), a pair of irreversible reactions is collinear if they have the same stoichiometry in their reactants (discounting reactants fixed in concentration, i.e., chemostatted reactants \cite{polettini_irreversible_2014}, cf. Sec \ref{subsubsec:obscure_vs_chemostat}). Collinear reactions thus have the same column entries in the stoichiometric reactant matrix $\nu^{\ominus}$ (cf. Eq. \eqref{equation:decompositionSirr}). Starting from $\mathbb{S}_{\ffor}$, we merge every pair of collinear reactions in $\mathbb{S}_{\ffor}$  until none remain to obtain $\mathbb{S}_{\blackffor}$ (Fig. \ref{fig:copro_derivation}a-c).

To characterize the consequences of merged reactions, we need to distinguish between reactions that came in reversible pairs and those that did not. We use $\mathbb{S}_{\blackbowtie}$ to denote the stoichiometric matrix containing all $r_{\blackbowtie}$ reversible reactions (written as single reaction), whose 
pair of irreversible reactions was retained in $\mathbb{S}_{\blackffor}$ (written as single reaction), i.e., the  reactions that are left untouched by mergers. Then $r_{\halfblackbowtie}$  denotes the number of reversible reactions involved in mergers, or equivalently the difference (highlighted by the half-filled symbol) in reversible reactions in going from $\mathbb{S}_{\bowtie}$ to $\mathbb{S}_{\blackbowtie}$ : 

\bea
\mathbb{S}_{\bowtie} &=& (\mathbb{S}_{\blackbowtie},\mathbb{S}_{\halfblackbowtie}) \\
r_{\halfblackbowtie}  &=& r_{\bowtie} -r_{\blackbowtie} .
\eea

Using $\Phi$ (total number of pairwise reaction mergers), we let the co-production index $\text{\footnotesize{$\Upsilon$}}$ be $\Phi$ after
discounting the $r_{\halfblackbowtie}$ merged reversible reactions (these inherently lead to lost trivial 2-cycles):

\bea
\Phi = r_{\ffor} - r_{\blackffor} \label{equation:phi_react} \\ 
\text{\footnotesize{$\Upsilon$}} =  \Phi - r_{\halfblackbowtie} \label{equation:upsdef}
\eea

We can now apply the fundamental theorem of linear algebra on the difference

\be
\! \! \! \!  \! \!  \text{rk}(\mathbb{S}_{\blackffor}) - \text{rk}(\mathbb{S}_{\ffor})  = \Delta s - \Delta \ell  = \Delta r - \Delta c .
\ee

By definition,  $\Delta r = - \Phi$. $\Delta s = 0$, as merging columns leaves $\#$ species in the description untouched. Mergers thus lead to more conservation laws or less cycles:

\be
\Phi = \Delta \ell - \Delta c \geq 0 .
\ee

Every merged 2-cycle trivially decreases $\Delta c$ and is not of interest for our further discussion, we will hence seek to eliminate these. Again discounting the $r_{\halfblackbowtie}$ merged reversible reactions, the number of nontrivial lost cycles, symbolized by $\wedge_{\blackwhiteffor}$ is then

\bea
\wedge_{\blackwhiteffor} &=&   c_{\ffor} - c_{\blackffor} - r_{\halfblackbowtie} \\
&=& (c_{\ffor} - r_{\bowtie}) - (c_{\blackffor}-r_{\blackbowtie}) \nonumber
\eea

We introduce emanants $\rotatebox[origin=c]{180}{e}$ \footnote{The symbol $\rotatebox[origin=c]{180}{e}$ is called 'schwa'} to refer to emergent conservation laws. The number of emanants is simply 

\be
\rotatebox[origin=c]{180}{e}_{\blackwhiteffor} = \ell_{\blackffor} - \ell_{\ffor}
\ee

\subsection{The co-production law}
We can now state the co-production law as

\be
\text{\footnotesize{$\Upsilon$}} = \rotatebox[origin=c]{180}{e}_{\blackwhiteffor} + \wedge_{\blackwhiteffor}, \label{Delta_SL1b}
\ee

where \\ %
$\text{\footnotesize{$\Upsilon$}}$: co-production index \\ %
$ \rotatebox[origin=c]{180}{e}_{\blackwhiteffor}$: $\#$ co-production emanants (emergent conservation laws due to global co-production) , \\ %
$\wedge_{\blackwhiteffor}$: $\#$ broken cycles due to global co-production. \footnote{To allude to the splitting up of a cycle, we have adopted a (splitting) wedge symbol.}

From the co-production law it follows that each instance of co-production must either produce an irreversible conservation law or lead to the loss of a previously present (but directionally impossible to perform) cycle upon merging reactions, an example for the latter is shown in Fig. \ref{fig:a broken cycle}, which is elaborated on in Sec. \ref{appendix:wedge}.

We now want to revisit SL1, for which we will first again construct quantities to count reactions and cycles, without double counting and trivial cycles:

\bea
r_{\bullet} = r_{\blackffor} - r_{\blackbowtie} \\
c_{\bullet} = c_{\blackffor} - r_{\blackbowtie} .
\eea

Upon substitution, we obtain a more complete SL1 (where $\ell$ includes co-production):

\hspace*{-0.5cm}\vbox{\bea
\! \! s - \ell_{\blackffor} &=& r_{\bullet} - c_{\bullet} ,\\
\! \! s - (\ell_\circ + \rotatebox[origin=c]{180}{e}_{\blackwhiteffor}) &=& (r-  \text{\footnotesize{$\Upsilon$}}) - (c_\circ - \wedge_{\blackwhiteffor}). \label{equation:modifiedSL1} 
\eea}

However, we recall that these indices are governed by two independent relations:

\bea
s - \ell_\circ &=& r - c_\circ, \label{equation:SL1a}\\
\text{\footnotesize{$\Upsilon$}} &=& \rotatebox[origin=c]{180}{e}_{\blackwhiteffor} + \wedge_{\blackwhiteffor}, \label{equation:SL1b}
\eea

We will now consider several applications and examples of these relations. Let us first remark that CRNs are not restricted to chemistry, but capture all Markov Models (Examples in Appendix \ref{appendix:markov-processes}) and many models in, e.g., statistical physics. As such, we can use co-production to find conserved quantities in, e.g., models for n-mer adsorption, asset-exchange, diffusion, and fragmentation.  
One major application - inference of network structure from data - is the subject of a companion article \cite{blokhuis2024ejoc} that develops this subject in its own right. We provide a brief illustration in Sec. \ref{subsection:SVDspec}.

\subsection{Example: a dimensional conundrum in atmospheric chemistry}
\label{subsection:dimensional-conundrum}

With the co-production law, we are now equipped to characterize the candidate for a non-integer conservation law in Ref. \cite{Liu_PRE_2024}. This is illustrated in Fig. \ref{fig:atmosphericnetwork} and analyzed in detail in Sec. \ref{section:atmosphericexample} of the Appendix. A co-production index $\copro = 1$ follows from visual inspection of the highlighted reactions in Fig. \ref{fig:atmosphericnetwork}. 

\begin{figure}[tbhp!]
\centering
\includegraphics[width=1.0\linewidth]{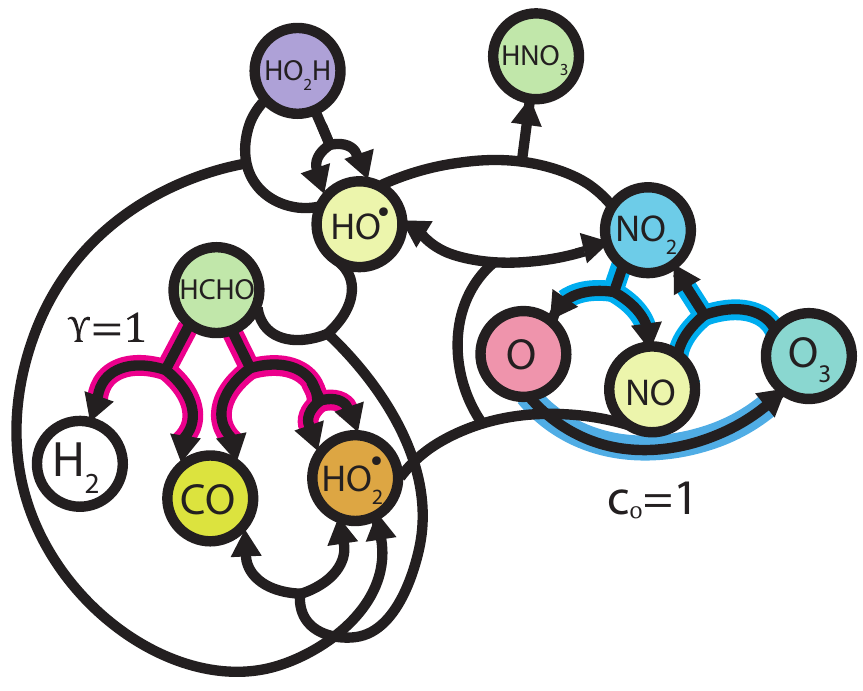}
\caption{An Atmospheric CRN model with co-production index $\text{\footnotesize{$\Upsilon$}}=1$. Collinear reactions are highlighted in pink. These do not include cycle reactions -  highlighted in blue - and  hence no cycles break upon merging collinear reactions: $\wedge_{\blackwhiteffor} = 0$, thus  $\rotatebox[origin=c]{180}{e}_{\blackwhiteffor} = 1$, i.e. a co-production conservation law then results from the co-production law $\text{\footnotesize{$\Upsilon$}}= \rotatebox[origin=c]{180}{e}_{\blackwhiteffor} + \wedge_{\blackwhiteffor}$, confirming the newly found law $\text{CQ}_3$ marks the detection of a genuine conservation law.}
\label{fig:atmosphericnetwork}
\end{figure}

The collinear reactions correspond to 
$r_4$, $r_5$ in

\hspace*{-0.75cm}\vbox{\be
\mathbb{S}_{\triangleright} = \left(\begin{smallmatrix} 0 & 1 & -1 & 0 & 0 & 0 & 0 & 0 & 0 & 0 \\
1 & 0 & -1 & 0 & 0 & 0 & -1 & 0 & 0 & 0  \\
-1 & 0 & 1 & 0 & 0 & 0 & 1 & -1 & 0 & 0 \\
0 & 0 & 0 & \colorbox{pink}{-1} & \colorbox{pink}{-1} & -1 & 0 & 0 & 0 & 0 \\
0 & 0 & 0 & \colorbox{pink}{2} & 0 & 1 & -1 & 0  & 0 & 1 \\
0 & 0 & 0 & 0 & 0 & 0 & 0 & 0 & -1 & -1 \\
0 & 0 & 0 & 0 & 0 & -1 & 1 & -1 & 2 & -1 \\
1 & -1 & 0 & 0 & 0 & 0 & 0 & 0 & 0 & 0 \\
0 & 0 & 0 & 0 & 0 & 0 & 0 & 1 & 0 & 0 \\
0 & 0 & 0 & \colorbox{pink}{1} & \colorbox{pink}{1} & 1 & 0 & 0 & 0 & 0 \\
0 & 0 & 0 & 0 & \colorbox{pink}{1} & 0 & 0 & 0 & 0 & 0
\end{smallmatrix}\right) . 
\ee}
With successive rows corresponding to \ce{O3},  \ce{NO, NO2, HCHO, HO2, HO2H, OH, O,HNO3}, \ce{CO , H2}. By merging $r_4$ and $r_5$, a  stoichiometric matrix $\mathbb{S}_{\blackffor}$ of independent reactions is obtained:

\hspace*{-0.75cm}\vbox{\be
\mathbb{S}_{\blackffor} = \left(\begin{smallmatrix} 
0 & 1 & -1 & 0 & 0 & 0 & 0 & 0 & 0 \\
1 & 0 & -1 & 0 & 0 & -1 & 0 & 0 & 0  \\
-1 & 0 & 1 & 0 & 0 & 1 & -1 & 0 & 0 \\
0 & 0 & 0 & \colorbox{pink}{-1} & -1 & 0 & 0 & 0 & 0 \\
0 & 0 & 0 & \colorbox{pink}{$2 p$} & 1 & -1 & 0  & 0 & 1 \\
0 & 0 & 0 & 0  & 0 & 0 & 0 & -1 & -1 \\
0 & 0 & 0 & 0  & -1 & 1 & -1 & 2 & -1 \\
1 & -1 & 0 & 0  & 0 & 0 & 0 & 0 & 0 \\
0 & 0 & 0 & 0  & 0 & 0 & 1 & 0 & 0 \\
0 & 0 & 0 & \colorbox{pink}{1}  & 1 & 0 & 0 & 0 & 0 \\
0 & 0 & 0 & \colorbox{pink}{1-$p$} & 0 & 0 & 0 & 0 & 0 \end{smallmatrix}\right).
\ee}

This leads to the emergence of a new co-production law as left nullvector:

\hspace*{-0.75cm}\vbox{\be
\! \! \! \! \pmb{\ell}_{\blackffor} = \left(6,-5,1,3,9,6,3,6,4,-3,\frac{6 - 18 p}{1-p} \right)^T.
\ee}

We show in Sec. \ref{section:atmosphericexample} that this is the quantity found by the SID algorithm in Ref. \cite{Liu_PRE_2024}, and that Eq. \eqref{equation:modifiedSL1} accounts for all true \footnote{There are other sources of emergent conserved quantities that SID may detect. For instance, Hirono et al have listed 3 types of systems exhibiting emergent conserved quantities \cite{hirono_structural_2021} characterized as 'pathological': CRNs i) requiring extra parameters to specify steady-state, ii) having no steady-state, iii) having unphysical kinetics. Bazhin and Baez et al have discussed conservation laws underlying biochemical coupling \cite{bazhin_essence_2012,baez2018biochemical}, which are conservation laws that are valid for a subnetwork of fast reactions (and hence respected on short time approximations of the time evolution of species abundances).} conserved quantities SID may find for a CRN. Co-production thus confirms the validity of this heretofore unexplained numerical result found through Machine Learning, and provides the missing theoretical foundation to interpret it.

\begin{figure}[tbhp!]
\centering
\includegraphics[width=1.0\linewidth]{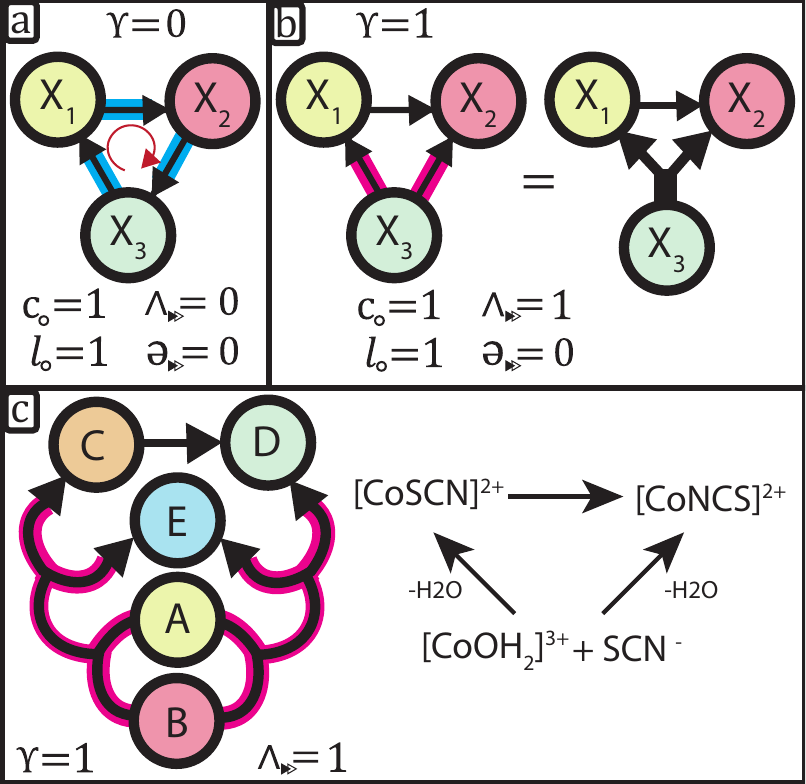}
\caption{a) A cyclic CRN composed of irreversible reactions. Cycle reactions are highlighted in blue. Reversing any of the reaction arrows in this cyclic CRN affords CRN b) with co-production index $\copro = 1$. After merging collinear reactions, the original reaction cycle can no longer be performed, and co-production manifests as one lost cycle ($ \wedge_{\blackwhiteffor} = 1$), zero emanants ($\rotatebox[origin=c]{180}{e}_{\blackwhiteffor}=0$). Using CLL Eq.\eqref{equation:copro-linkage-law}, we find for a) $r^+ = 3, \lambda_{\bowtie} = 3, \lambda_{\bowtie}^0 = 0$,  $\copro = 3 - (3-0) = 0$ and for b)
$r^+=3, \lambda_{\bowtie} = 3, \lambda_{\bowtie}^0 = 1$, $\copro = 3 - (3-1) = 1$. c) Literature example \cite{Jackson1985} of a CRN with a broken cycle $ \wedge_{\blackwhiteffor} = 1$, addition of thiocyanate to $[\ce{CoOH}_2]^{3+}$.}
\label{fig:a broken cycle}
\end{figure}

\subsection{Generalizing beyond the Curtin-Hammett principle}
\label{subsection:Beyond-Curtin-Hammett}

Through (nonlocal) coproduction, we can derive an expression that provides a generalization of the Curtin-Hammett principle\cite{IUPAC_Curtin_Hammett} beyond the fast equilibration limit. The Curtin-Hammett principle prescribes the interpretation of chemical experiments in which two irreversible competing reactions for the same isomer occur~\cite{IUPAC_Curtin_Hammett}, proceeding through different conformers. In the limit of fast reactions, the relative concentrations of $\ce{X},\ce{Y}$ is only controlled by the difference in standard free energies $\delta \Delta^{\ddagger} G^\circ$ of respective transition states.


The CRN used as foundation for the Curtin-Hammett principle involves interconverting reactants $\ce{A}, \ce{B}$ that form products $\ce{C}, \ce{D}$

\be
\ce{C} \overset{k_1}{\leftarrow} \ce{A} \ \underset{k_2^-}{\overset{k_2^+}{\leftrightarrows}} \ce{B} \overset{k_3}{\rightarrow} \ce{D} ,  \label{equation:curhamex}
\ee
and the Curtin-Hamett principle now states that if $k_2^+, k_2^- \gg k_1,k_3$, then the ratio of produced $\ce{C}$ over $\ce{D}$ follows
\hspace*{-0.75cm}\vbox{\be
\frac{\Delta [\ce{C}]}{\Delta [\ce{D}]} = \frac{k_1 k_2^+}{k_3 k_2^-} =  \frac{k_1}{k_3 K_2} = e^{-\delta \Delta^\ddagger G^\circ / RT},
\ee}
where the latter follows from transition state theory\footnote{in transition state theory, a rate $k$ depends on the Gibbs energy of activation $\Delta^\ddagger G^\circ$ according to $k = (k_B T/h) \ e^{-\Delta^\ddagger G^\circ/RT}$.}, $R$ is the gas constant, $T$ absolute temperature.

More generally we can write currents for CRN \eqref{equation:curhamex}
\bea
J_1 &=& k_1 [\ce{A}]  \ , \nonumber \\
J_2 &=& k_2^+ [\ce{A}] - k_2^- [\ce{B}] \ , \\ 
J_3 &=& k_3 [\ce{B}] \ \nonumber .
\eea

As we have three currents, expressed as linear combination of only 2 variables, a collinearity necessarily occurs. We have coproduction  $\copro = 1$,
\be
J_2 = \frac{k_2^+}{k_1} J_1 - \frac{k_2^-}{k_3} J_3 .
\ee

We note that

\bea
d_t [\ce{A}] &=& - J_1 - J_2 \nonumber \\
&=& - \frac{k_2^+ + k_1}{k_1} J_1 + \frac{k_2^-}{k_3} J_3  , \nonumber \\
d_t [\ce{B}] &=& - J_3 + J_2 \nonumber , \\
d_t [\ce{C}] &=&  J_1   , \label{equation:odescurham} \\
d_t [\ce{D}] &=&  J_3 , \nonumber
\eea

There are no cycles, and our coproduction manifests as an emanant $\rotatebox[origin=c]{180}{e}_{\blackwhiteffor} = 1$. Inspection of the ODEs \eqref{equation:odescurham} allows us to write one co-production conservation law $L_2$:

\bea
L_1 &=&  [\ce{A}] + [\ce{B}] + [\ce{C}] + [\ce{D}] ,  \\
L_2 &=& [\ce{A}] + \frac{k_2^+ + k_1}{k_1} [\ce{C}] - \frac{k_2^-}{k_3} [\ce{D}] .
\eea

We then consider an initial state with only reactants $[\ce{C}]^0 = [\ce{D}]^0 = 0$, $[\ce{A}]^0, [\ce{B}]^0 > 0$. For $t \rightarrow \infty$, only $[\ce{C}], [\ce{D}]$ are populated: 

\hspace*{-0.75cm}\vbox{\bea
\ [\ce{X}]^\infty &=& \lim_{t\rightarrow \infty} [\ce{X}](t) , \\
\ [\ce{A}]^\infty &=& [\ce{B}]^\infty = 0 , \ [\ce{C}]^\infty, [\ce{D}]^\infty > 0 . 
\eea}

We can then write:

\hspace*{-0.75cm}\vbox{\bea
L_1 &=& [\ce{C}]^{\infty} + [\ce{D}]^{\infty} = [\ce{A}]^{0} + [\ce{B}]^{0} , \\
L_2 &=&  \frac{k_2^+ + k_1}{k_1} [\ce{C}]^{\infty} - \frac{k_2^-}{k_3} [\ce{D}]^{\infty} \nonumber \\
&=& [\ce{A}]^0 .
\eea}

\begin{figure}[tbhp!]
\centering
\includegraphics[width=1.0\linewidth]{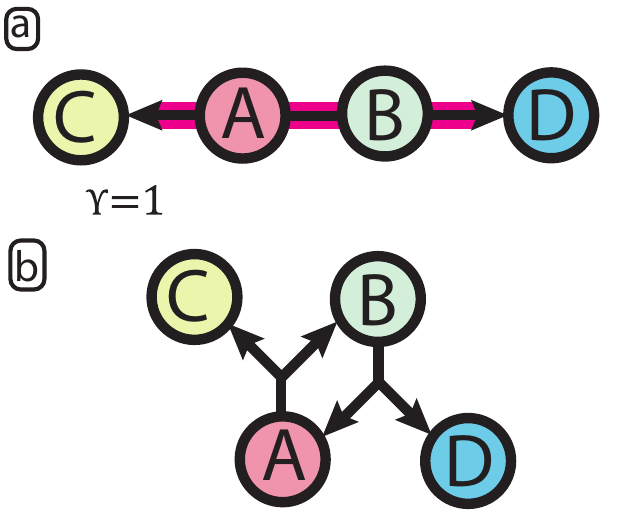}
\caption{a) A CRN $\ce{C} \leftarrow \ce{A} \leftrightarrows \ce{B} \rightarrow \ce{D}$ with nonlocal coproduction. Using Eq.\eqref{equation:copro-linkage-law}, we find $r^+=2, \lambda_{\bowtie} = 1, \lambda_{\bowtie}^0 = 0$, hence $\text{\footnotesize{$\Upsilon$}} = 2 - (1-0) = 1$. b) Upon merging reactions, we obtain the representation $\mathbb{S}_{\blackffor}$ with one less reaction.} 
\label{fig:CH}
\end{figure}

Following our procedure, one can alternatively directly merge reactions $R_1 + R_2^+$ and $R_2^- + R_3$ (see Fig. \ref{fig:CH}) to obtain the reactions
\bea
\ce{A} \rightarrow \frac{k_1}{k_1 + k_2^+} \ce{C} + \frac{k_2^+}{k_1 + k_2^+} \ce{B} , \\
\ce{B} \rightarrow \frac{k_3}{k_3 + k_2^-} \ce{D} + \frac{k_2^-}{k_3 + k_2^-} \ce{A} ,
\eea
after which $r$ decreases by 1, hence $\Phi=1$. One can readily check that either reaction conserves both $L_1$ and $L_2$.

We can rearrange the expressions above to express the final composition directly:

\bea
\ [\ce{C}]^{\infty} &=& k_1 \frac{ \left(k_2^- + k_3\right) [\ce{A}]^0 + k_2^- [\ce{B}]^0 }{k_3 k_2^+ + k_3 k_1 + k_1 k_2^-} , \nonumber \\
\ [\ce{D}]^{\infty} &=& k_3 \frac{k_2^+ [\ce{A}]^0 + \left( k_2^+ + k_1 \right) [\ce{B}]^0 }{k_1 k_2^- + k_3 k_2^+ + k_3 k_1}  .
\eea

Then:

\bea
\frac{\ [\ce{C}]^{\infty}}{\ [\ce{D}]^{\infty}} &=& \frac{k_1}{k_3} \left(\frac{\left(k_2^- + k_3\right) [\ce{A}]^0 + k_2^- [\ce{B}]^0}{k_2^+ [\ce{A}]^0 + \left( k_2^+ + k_1 \right) [\ce{B}]^0}\right)  \nonumber \\
&=& \frac{k_1}{k_3 K_2}  \left(\frac{\left(1 + \frac{k_3}{k_2^-} \right) [\ce{A}]^0 +  [\ce{B}]^0}{[\ce{A}]^0 + \left( 1 + \frac{k_1}{ k_2^+} \right) [\ce{B}]^0}\right) \nonumber \\
&=& e^{-\delta \Delta^{\ddagger} G^{\circ}/RT } Q \label{equation:Curtin-hammett-extend} , \\
Q &=& \left(\frac{\left(1 + \frac{k_3}{k_2^{-}} \right) [\ce{A}]^0 +  [\ce{B}]^0}{[\ce{A}]^0 + \left( 1 + \frac{k_1}{k_2^{+}} \right) [\ce{B}]^0}\right) . \nonumber
\eea

The ratios of rate constants appearing in $Q$ are dimensionless quantities that compare the timescales of irreversible reaction with isomerization. In the limit where $k_2^+, k_2^-  \gg k_1, k_3, \ \ Q \rightarrow 1$, we recover the Curtin-Hammett limit:

\bea
\ [\ce{C}]^{\infty} &\approx& k_1 \frac{  [\ce{A}]^0 +  [\ce{B}]^0 }{k_3 K_2 + k_1} ,\\
\ [\ce{D}]^{\infty} &\approx& k_3 K_2 \frac{ [\ce{A}]^0 + [\ce{B}]^0 }{k_3 K_2 + k_1  }  ,\\
\frac{\ [\ce{C}]^{\infty}}{\ [\ce{D}]^{\infty}} &\approx& \frac{k_1}{k_3 K_2 } = e^{-\delta \Delta^\ddagger G^{\circ}/RT} .
\eea

Outside this limit, the initial conditions and isomerization rates become pertinent, and 
our result Eq.\eqref{equation:Curtin-hammett-extend} replaces the Curtin-Hammett formula. A related extension for this CRN was reported by Zefirov \cite{Zefirov1977}.

Importantly, via co-production conservation laws, far greater generalizations of the principle are possible: one can always derive formulas of the form of Eq. \eqref{equation:Curtin-hammett-extend} for elaborate CRNs using coproduction. See Appendix \ref{appendix:chextension}.






\subsection{Structural aspects of coproduction: linkage classes and complexes}

\begin{figure}[tbhp!]
\centering
\includegraphics[width=1.0\linewidth]{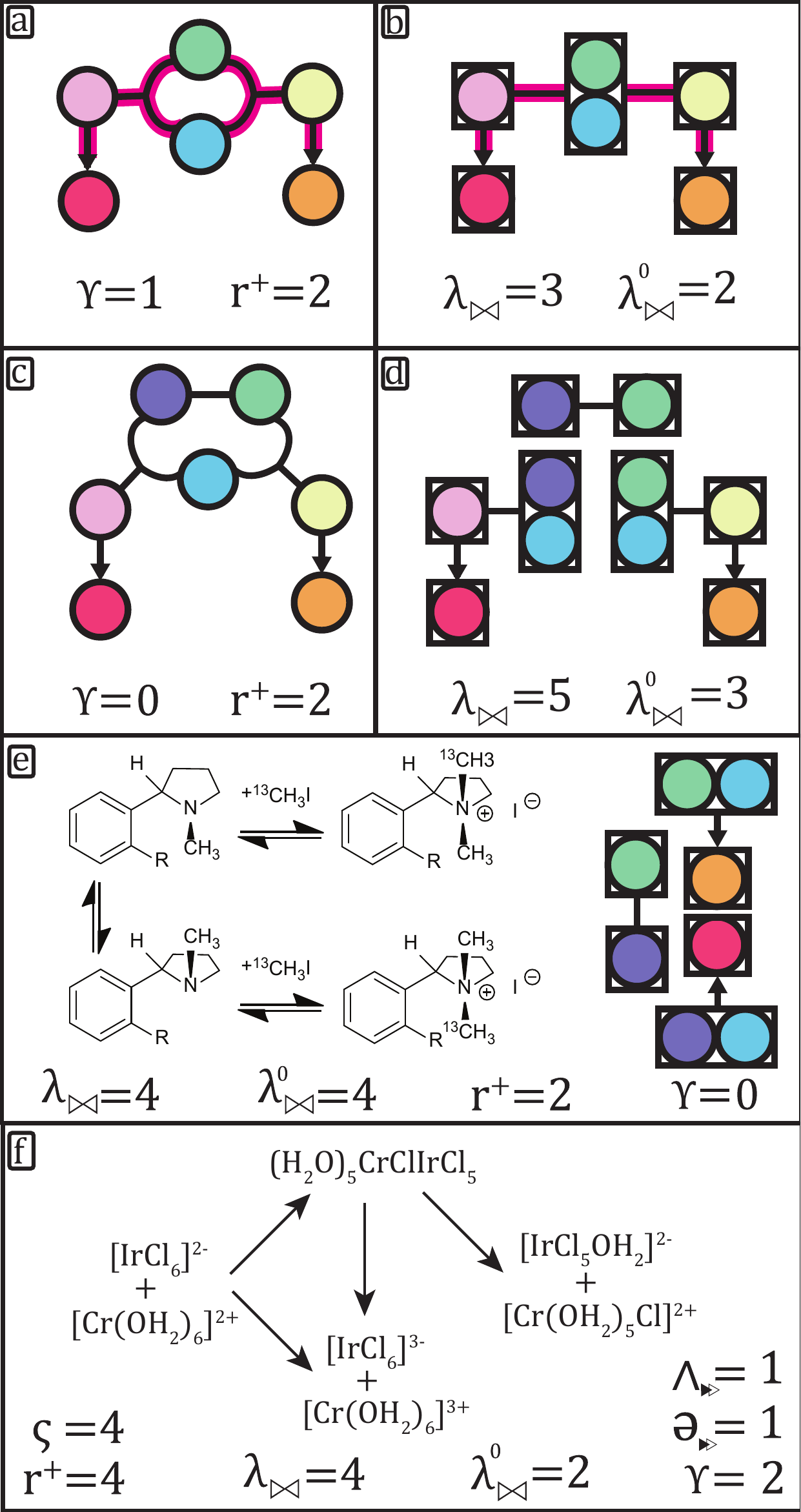}
\caption{A nonrobust conservation law: upon adding an additional intermediate to the CRN in a) to obtain c), co-production is lost, and hence a conservation law is lost. Graph representations of a) $\mathbb{S}_{\circ}$ for CRN \eqref{equation:CRN_nonrob} b) $\mathbb{\partial}_{\circ}$ for CRN \eqref{equation:CRN_nonrob}  c) $\mathbb{S}_{\circ}$ for CRN \eqref{equation:crn_nonrob_b} d) $\mathbb{\partial}_{\circ}$ for CRN \eqref{equation:crn_nonrob_b}. e) Literature example \cite{Seeman1980}: a CRN without coproduction analogous to c). If isomerization is rapid, coproduction emerges as in a). f) Literature example \cite{Haim1983}: concurrent inner sphere and outer sphere reactions with binuclear intermediate of a CRN with $\textit{\textvarsigma} = 4$ complexes, $\lambda_{\bowtie} = 4,  \lambda_{\bowtie}^0 = 2, r^+ = 4, \copro = 2, \rotatebox[origin=c]{180}{e}_{\blackwhiteffor} = 1, \wedge_{\blackwhiteffor} = 1$.}
\label{fig:robustness}
\end{figure}

In CRN theory, the combination of species occurring in either side of a reaction is referred to as a complex \cite{feinberg_foundations_2019} (counted by \textit{\textvarsigma}). One can draw a graph representation of complexes (nodes) connected by reactions (edges), and each connected graph in this representation is then called a linkage class. The counterpart to the stoichiometric matrix $\mathbb{S}$ is the incidence matrix $\partial$. As an example, consider the following CRN depicted in Fig.~\ref{fig:robustness}a,b:

\bea
\! \! \! \! \mathbb{S}_{\circ}&:& \ \ \ce{X}_1 \overset{1}{\leftarrow} \ce{X}_2 \overset{2}{\leftrightarrows} \ce{X}_3  + \ce{X}_4 \overset{3}{\leftrightarrows} \ce{X}_5 \overset{4}{\rightarrow} \ce{X}_6 , \label{equation:CRN_nonrob} \nonumber \\
\! \! \! \! \mathbb{S}_{\bowtie} &:& \ \ \ce{X}_1, \ \ \  \ce{X}_2 \overset{2}{\leftrightarrows} \ce{X}_3  + \ce{X}_4 \overset{3}{\leftrightarrows} \ce{X}_5,  \ \ \ \ce{X}_6  \nonumber 
\eea

which can be written as the following complexes:

\bea
\partial_{\circ}&:& \ \ \Gamma_1 \overset{1}{\leftarrow} \Gamma_2 \overset{2}{\leftrightarrows} \Gamma_3  \overset{3}{\leftrightarrows} \Gamma_4 \overset{4}{\rightarrow} \Gamma_5 , \nonumber \\
\partial_{\bowtie}&:& \ \ \ \ \Gamma_1 \ \ \   \Gamma_2 \overset{2}{\leftrightarrows} \Gamma_3  \overset{3}{\leftrightarrows} \Gamma_4,  \ \ \ \Gamma_5  \nonumber
\eea

From inspection, it can be seen that $\text{\footnotesize{$\Upsilon$}} = 1$.  As $c=0$, only conserved quantities can emerge, $\rotatebox[origin=c]{180}{e}_{\blackwhiteffor}=1$.

In order for it to be possible for a pair of reactions to be merged, the reactions need to have the same stoichiometry. By definition, they then act on the same complex. To merge successive reactions, they all need to be reversible, and occur in the same linkage class (See Fig. \ref{fig:robustness}a-d). We will define a set of complexes that can reach each other by reversible reactions within a static linkage class, and denote their number as $\lambda_{\bowtie}$.

Representing $\mathbb{S}$ in terms of its complexes, we obtain $\partial$, and the FTLA yields a well-known structural relation \cite{feinberg_foundations_2019,blokhuis:tel-02566386}, which we will refer to as the second structural law (SL2)

\be
\text{rk}(\partial_{\circ}) = \textit{\textvarsigma} - \lambda_{\circ} = r - \sqsupset_{\circ} ,\label{equation:SL2} 
\ee

where \\
$\textit{\textvarsigma}$
: $\#$ complexes in $\partial_{\circ}$ \\
$\lambda_{\circ}$: $\#$ linkage classes $(\text{dim}(\text{coker}(\mathbb{\partial_{\circ}}))$ \\
$r$: $\#$ reactions \\
$\sqsupset_{\circ}$: $\#$ cycles within linkage classes $(\text{dim}(\text{ker}(\mathbb{\partial_{\circ}}))$ \\

To count pertinent quantities for coproduction such as $\lambda_{\bowtie}$, we represent $\mathbb{S}_{\bowtie}$ in terms of its complexes, to obtain $\partial_{\bowtie}$, and the appropriate form of SL2 (Eq. \eqref{equation:SL2}) then becomes

\be
\text{rk}(\partial_{\bowtie}) = \textit{\textvarsigma}_{\bowtie}
 - \lambda_{\bowtie} = r_{\bowtie} - \sqsupset_{\bowtie} \label{equation:FTLAcomplexes} ,
\ee

where \\
$\textit{\textvarsigma}
_{\bowtie}$: $\#$ complexes in $\partial_{\bowtie}$ ($\textit{\textvarsigma}_{\bowtie} = \textit{\textvarsigma}$) \\
$\lambda_{\bowtie}$: $\#$ static linkage classes $(\text{dim}(\text{coker}(\mathbb{\partial_{\bowtie}}))$ \\
$r_{\bowtie}$: $\#$ reversible reactions \\
$\sqsupset_{\bowtie}$: $\#$ cycles contained within static linkage classes $(\text{dim}(\text{ker}(\mathbb{\partial_{\bowtie}}))$ \\

Cycles of $\partial_{\bowtie}$ are simple graph cycles. All cycles of $\partial_{\bowtie}$ are cycles of $\mathbb{S}_{\bowtie}$ but the converse is not true, i.e., $c_{\bowtie} \geq \sqsupset_{\bowtie}$.

For our example, we have
\be
\partial_{\bowtie} = \begin{pmatrix} 
0 & 0 \\ 
-1 & 0 \\
1 & -1 \\
0 & 1 \\
0 & 0  
\end{pmatrix} 
\ee
which acts on $\Gamma_1, \Gamma_2,\Gamma_3,\Gamma_4,\Gamma_5$, so that  
 $\textit{\textvarsigma} = 5, \lambda_{\bowtie}=3, \lambda_{\bowtie}^{0}=2, r_{\bowtie} = 2, \sqsupset_{\bowtie}=0$.

Irreversible reactions  decompose into those whose description includes reactant and product ($\pm$), only reactant ($+$), and only product ($-$):

\be
\ce{X}_i \underset{\pm}{\rightarrow} \ce{X}_j, \ \ \ \ce{X}_i \underset{+}{\rightarrow} \emptyset, \ \ \  \emptyset \underset{-}{\rightarrow} \ce{X}_k, \label{equation:threereactions}
\ee

where $\emptyset$ denotes an absence of species (in the description of the system), e.g. because of exchange with an environment. 
Eq. \eqref{equation:threereactions} lets us decompose the number of irreversible reactions as 

\bea
r_{\triangleright} &=& r_{\triangleright}^{\pm} + r_{\triangleright}^{+} + r_{\triangleright}^{-} , \\
r^+ &=& r_{\triangleright}^{\pm} + r_{\triangleright}^{+} . \label{equation:rplus}
\eea

The pairs of irreversible reactions that can be merged over a distance are exactly the outgoing irreversible reactions attached to the same static linkage class. Through the pigeonhole principle we then count them and thereby also count co-production $\text{\footnotesize{$\Upsilon$}}$.

Let us use $\lambda_{\bowtie}^0$ to denote the number of static linkage classes with no outgoing irreversible reactions attached to them ($r^+$ cancels locally). Then the remaining linkage classes contain at least one outgoing irreversible reaction, and each reaction in excess of the first contributes to co-production $\text{\footnotesize{$\Upsilon$}}$, which results in the coproduction-linkage law (CLL)
\be
\text{\footnotesize{$\Upsilon$}} = r^+ - (\lambda_{\bowtie} - \lambda^0_{\bowtie}), \label{equation:copro-linkage-law}
\ee
where \\
$\text{\footnotesize{$\Upsilon$}}:$ Co-production \\
$r^+: \#$ irreversible reactions with reactant in description (outgoing irreversible reactions) \\
$\lambda_{\bowtie}: \#$ static linkage classes \\
$\lambda_{\bowtie}^0: \#$ static linkage classes with no outgoing irreversible reactions.

In Appendix  \ref{appendix:coproductionFTLA}, we present an alternative derivation of the coproduction-linkage law Eq.\eqref{equation:copro-linkage-law} using FTLA, wherein the indices correspond to fundamental subspaces.

The coproduction $\text{\footnotesize{$\Upsilon$}}=1$ in all our examples can be assessed rapidly through the coproduction-linkage law (Eq.\eqref{equation:copro-linkage-law}). For instance, in Fig. \ref{fig:robustness}a,b, it can readily be seen that $r^+=2,\lambda_{\bowtie} = 3, \lambda^0_{\bowtie} = 2$.

\subsection{Resolution-dependent co-production}

Now suppose there exists a more fine-grained representation of the system in Fig. \ref{fig:robustness}a,b, namely \ref{fig:robustness}c,d, where $\ce{X}_4$ is replaced by two interconverting species $\ce{X}_{4a}$ and $\ce{X}_{4b}$, whose abundance we suppose to be variables accessible by a higher-resolution measurement:
\bea
\ce{X}_1 \overset{1}{\leftarrow} \ce{X}_2 &\overset{2}{\leftrightarrows}& \ce{X}_3  + \ce{X}_{4a}, \ \ \ \ce{X}_{4a} \overset{3}{\leftrightarrows} \ce{X}_{4b}, \nonumber \\
\ce{X}_3 + \ce{X}_{4b} &\overset{4}{\leftrightarrows}& \ce{X}_5 \overset{5}{\rightarrow} \ce{X}_6 . \label{equation:crn_nonrob_b}
\eea
The addition of this level of detail breaks the system up in more static linkage classes $\lambda_{\bowtie}=3, \lambda_{\bowtie}^0=1$, $r^+=2$, whereby the pair of irreversible reactions has become isolated, $\text{\footnotesize{$\Upsilon$}}= 2 - (3-1) = 0$. Consequently, we no longer have an emergent conservation law. Intuitively, adding a species and an isomerization reaction may be expected to generally add one degree of freedom, and hence increase $d$ by 1. As we have now shown, this is not true when doing so breaks coproduction, and here the data dimension increases by 2 ($d=5$). A corollary is that co-production may emerge due to coarse-graining.

Several powerful tools in CRN theory rely on dimensionality (of the ODEs) always increasing by 1 for certain network modifications\cite{Banaji2018, Banaji2023, Banaji2024}, in order for behavior (e.g., multistability) to be inherited by the modified CRN. Nonrobust co-production provides a mechanism by which exceptions to these assumptions arise, and by which they can be understood and structurally characterized (Eq.\eqref{equation:crn_nonrob_b}). Emanants can more broadly be linked to nonlinear dynamics (Appendix \ref{appendix:fixedpointstab}).

\section{Beyond isosbestic points: data dimension}
\label{section:datadimension}

To generalize the utility of isosbestic points, we introduce the notion of a data dimension $d$. 
For a CRN modeled through a set of variables $V$, we define the model dimension $d(V)$ as

\be
d(V) = v(V) - \ell (V) ,\label{equation:dimfromvar}
\ee

where
$v(V)$: $\#$ assessed variables ($v(V)=|V|$) \\
$\ell (V)$: $\#$ constraints on these variables, i.e., number of conserved quantities for these variables.

In the context of a CRN described using $s$ species variables, we have a set of variables $\ce{X}=\{[\ce{X_1}],..,[\ce{X_s}]\}$, so that $V=\ce{X}$. We retain the convention that $s=|\ce{X}|$, $\ell=\ell(\ce{X})$ and, thus:

\be
d(\ce{X})  = s - \ell(\ce{X}) .
\ee 

Suppose we monitor a system spectroscopically, assuming a fixed linear relation between species and absorbance throughout the experiment (e.g., temperature is fixed and we are within the linear regime of the detector). Thus, we suppose that the Lambert-Beer law is valid

\be
A_\lambda = L_{p}^{-1} \sum_{k=1}^s [\ce{X_k}] \epsilon_k (\lambda), \label{equation:lambert-beer}
\ee

Where $A_\lambda$ is the total absorbance at wavelength $\lambda$, $\epsilon_k(\lambda)$ the molar extinction coefficient of $\ce{X}_k$ at wavelength $\lambda$, and $L_p$ is the light path length. We monitor in bins around $v(\Lambda)=n$ wavelengths, $\Lambda=\{ \lambda_1,\lambda_2,...,\lambda_n \}$, and, now, the absorbance $A_{\lambda_k}$ at each binned wavelength $\lambda_k$ is an assessed variable ($V=\{A(\Lambda)$). Denoting $\pmb{A}(\Lambda)=(A_{\lambda_1}, ... , A_{\lambda_n})^T$, we have 

\bea
d_t \pmb{A}(\Lambda) &=&  L_p^{-1} \mathbb{E}^T d_t [\pmb{\ce{X}}] \\
&=& L_{p}^{-1} \ \mathbb{E}^T(\Lambda) \ \mathbb{S} \ \pmb{J} ,
\eea

where $\pmb{J}$ is a vector of reaction currents as introduced in Eq.\eqref{equation:timeevo} and $\mathbb{E}$ is an $s$-by-$n$ matrix with each row corresponding to the spectrum of a species:

\be
\mathbb{E} = \left(\begin{smallmatrix} \epsilon_1 (\lambda_1) & \epsilon_1 (\lambda_2) & ... & \epsilon_1 (\lambda_n)  \\ 
\epsilon_2 (\lambda_1) & \epsilon_2 (\lambda_2) & ... & \epsilon_2 (\lambda_n)  \\ 
\vdots & \vdots & \ddots & \vdots \\
\epsilon_s (\lambda_1) & \epsilon_s (\lambda_2) & ... & \epsilon_s (\lambda_n)   \end{smallmatrix} \right) . \label{equation:matrixE}
\ee

The rank of $\mathbb{E}$ is at most $s$, when all species are spectroscopically distinct and $n$ is sufficiently large to discern them, i.e.,
\be
\text{rk}(\mathbb{E}) \leq \text{min}(|\Lambda|,|\ce{X}|).
\ee

The rank of $\mathbb{E}$ can be lower than the  number of species for a variety of reasons. We will here consider two contributions and partition of $\mathbb{E}$ accordingly. We have

\be
\mathbb{E} = \begin{pmatrix} \mathbb{E}_x \\ \mathbb{E}_{\blacksquare} \\ \mathbb{E}_{\text{\textsection}} \end{pmatrix} ,
\ee

namely species that are not measurably absorbing in $\Lambda$. We refer to species in the original system that cannot be accessed by the observer as concealed (i.e., "black-boxed").

We consider the following contributions that lower the rank

\be
\text{rk}(\mathbb{E} (\Lambda)) = s - s_{\blacksquare}(\Lambda) - \text{\textsection}(\Lambda) - \ell_{||}(\Lambda), \label{equation:specrank}
\ee

where
$s_{\blacksquare}(\Lambda)$ : $\#$ concealed  species, species not measurably absorbing in $\Lambda$, modeled as row of zeros in $\mathbb{E}$. \\
$\text{\textsection}(\Lambda)$ : indistinguishability index, $\#$ indistinguishable isomer species wrt. $\Lambda$. $\#$ independent pairs of proportional rows in $\mathbb{E}$, here with the added assumption that the underlying species are isomers \footnote{The assumption that isospectral species are taken to be isomeric (i.e. in principle 1-to-1 interconvertible}  enables us to merge them here. For a more general theory this assumption will need to be relaxed. .
$\ell_{||}(\Lambda)$ : $\#$ further collinearities among absorption spectra in $\Lambda$ \\

Further collinearities ($\ell_{||}(\Lambda)$) can e.g. occur when $n$ becomes comparable to the dimension. For instance, when computer vision is used for reaction monitoring \cite{barrington_computer_2022,yan_computer_2023,daglish_determining_2023}, the RGB images have $n=3$, which may set an upper bound on the dimension that may be resolved. More elaborate collinearities may exist among reactants and products and (approximately) isospectral species in practice need not be isomers. A more complete treatment of further collinearities (e.g. isospectral reactions) requires an extension of formalism which will be developed elsewhere.

The data dimension for $n=|\Lambda|$ spectral variables can be written Eq.\eqref{equation:dimfromvar} as

\be
d(\Lambda) = n - \ell(\Lambda) \leq d(\ce{X}) .
\ee


\subsection{The concealed species law versus chemostatting}
\label{subsubsec:obscure_vs_chemostat}

Next, we consider the subnetwork we see effectively due to the first two constraints in Eq.\ref{equation:specrank} relating species to their spectral observables.
First, we partition $\mathbb{S}$ into a matrix $\mathbb{S}_\square$ of species that are observable (e.g., give rise to detectable absorbance), and a matrix $\mathbb{S}_\blacksquare$ of species in the system that are not observable (e.g., they do not give rise to detectable absorbance in the assessed spectral range):

\be
\mathbb{S} = \begin{pmatrix} \mathbb{S}_\blacksquare \\ \mathbb{S}_\square \end{pmatrix} .\label{equation:concealdecomposition}
\ee

From applying the FTLA / SL1 to $\mathbb{S}_\blacksquare$, we obtain

\bea
s_{\blacksquare}(\Lambda) &=& b_{\textifsym{\SquareShadowA}}(\Lambda) + a_{\textifsym{\SquareShadowA}}(\Lambda) , 
\label{equation:concealedspecieslaw} \\
b_{\textifsym{\SquareShadowA}}(\Lambda) &=& \ell - \ell_{\square}  , \\
a_{\textifsym{\SquareShadowA}}(\Lambda) &=& c_{\square} - c ,
\eea

where
$b_{\textifsym{\SquareShadowA}}(\Lambda)$: $\#$ concealed conservation laws in visible subnetwork ($\mathbb{S}_\square$) \\
$a_{\textifsym{\SquareShadowA}}(\Lambda)$: $\#$ apparent cycles in visible subnetwork \\

We borrow this partitioning procedure \cite{polettini_irreversible_2014} from a different context, where $s_{\ce{Y}}$ species are chemostatted (i.e., fixed in concentration),

\be
\mathbb{S} = \begin{pmatrix} \mathbb{S}_{\ce{Y}} \\ \mathbb{S}_{\ce{X}}\end{pmatrix} , \label{equation:chemostatdecomposition}
\ee

and then can write

\bea
s_{\ce{Y}} &=& b_{\ce{X}} + a_{\ce{X}}, \\
b_{\ce{X}} &=& \ell - \ell_{\ce{X}}, \\
a_{\ce{X}} &=& c_{\ce{X}} - c , 
\label{equation:chemostatlaw}
\eea
where
$b_{\ce{X}}$: $\#$ broken conservation laws in $\mathbb{S}_{\ce{X}}$, \\
$a_{\ce{X}}$: $\#$ emergent cycles in internal network. \\

An emergent cycle $\pmb{a}$ denotes a right nullvector of the subnetwork with fewer species that does not extend to the full network:

\bea
\mathbb{S} \ \pmb{a}_{\textifsym{\SquareShadowA}} \neq \pmb{0}, \ \ \mathbb{S}_\square \ \pmb{a}_{\textifsym{\SquareShadowA}}  = \pmb{0} , \\
\mathbb{S} \ \pmb{a}_{\ce{X}} \neq \pmb{0}, \ \ \mathbb{S}_{\ce{X}} \ \pmb{a}_{\ce{X}}  = \pmb{0} . 
\eea

However, the net effect of chemostatting species is not equivalent to concealing them: previously independent irreversible reactions can become collinear by fixing variables (chemostatting), but not by concealing variables. A minimal example is illustrated in Fig. \ref{fig:csvsconceal}. The same phenomenon occurs in the example provided in Appendix \ref{section:atmosphericexample}. 

Concealment of a species from direct observation need not make its quantification inaccessible. Provided we have knowledge of underlying (concealed) conservation laws, hidden species may still be quantified indirectly. Unambiguous quantification of all species is possible as long as no apparent cycles are generated, i.e., $a_{\textifsym{\SquareShadowA}} = 0$.
We will now look at an example of concealment using literature data where all concealed species can still be quantified.

\begin{figure}
\centering
\includegraphics[width=1.0\linewidth]{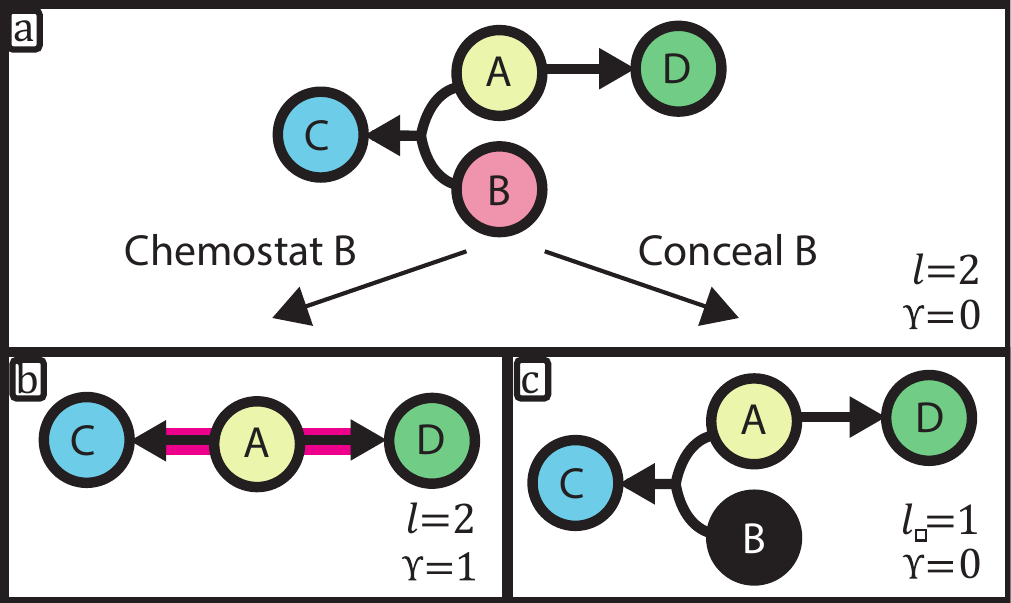}
\caption{Contrasting chemostatting and concealing for the reference CRN in shown a). b) Upon fixing concentration $[\ce{B}]=cte$, the reactions become collinear, and we obtain a co-production index $\copro = 1$. Upon concealing $[\ce{B}]$, it remains a variable, but we lose access to it. The reactions then remain independent. Through CLL (Eq.\eqref{equation:copro-linkage-law}) we find for b) $r^+=2,\lambda_{\bowtie} = 3,\lambda_{\bowtie}^0 = 2$, $\copro = 2 - (3-2) =1$, c) $r^+=2,\lambda_{\bowtie} = 4,\lambda_{\bowtie}^0 = 2$, $\copro = 2 - (4-2) =0$. }
\label{fig:csvsconceal}
\end{figure}

\subsection{Example: concealment for dynamic covalent macrocycles}

\begin{figure}
\centering
\includegraphics[width=1.0\linewidth]{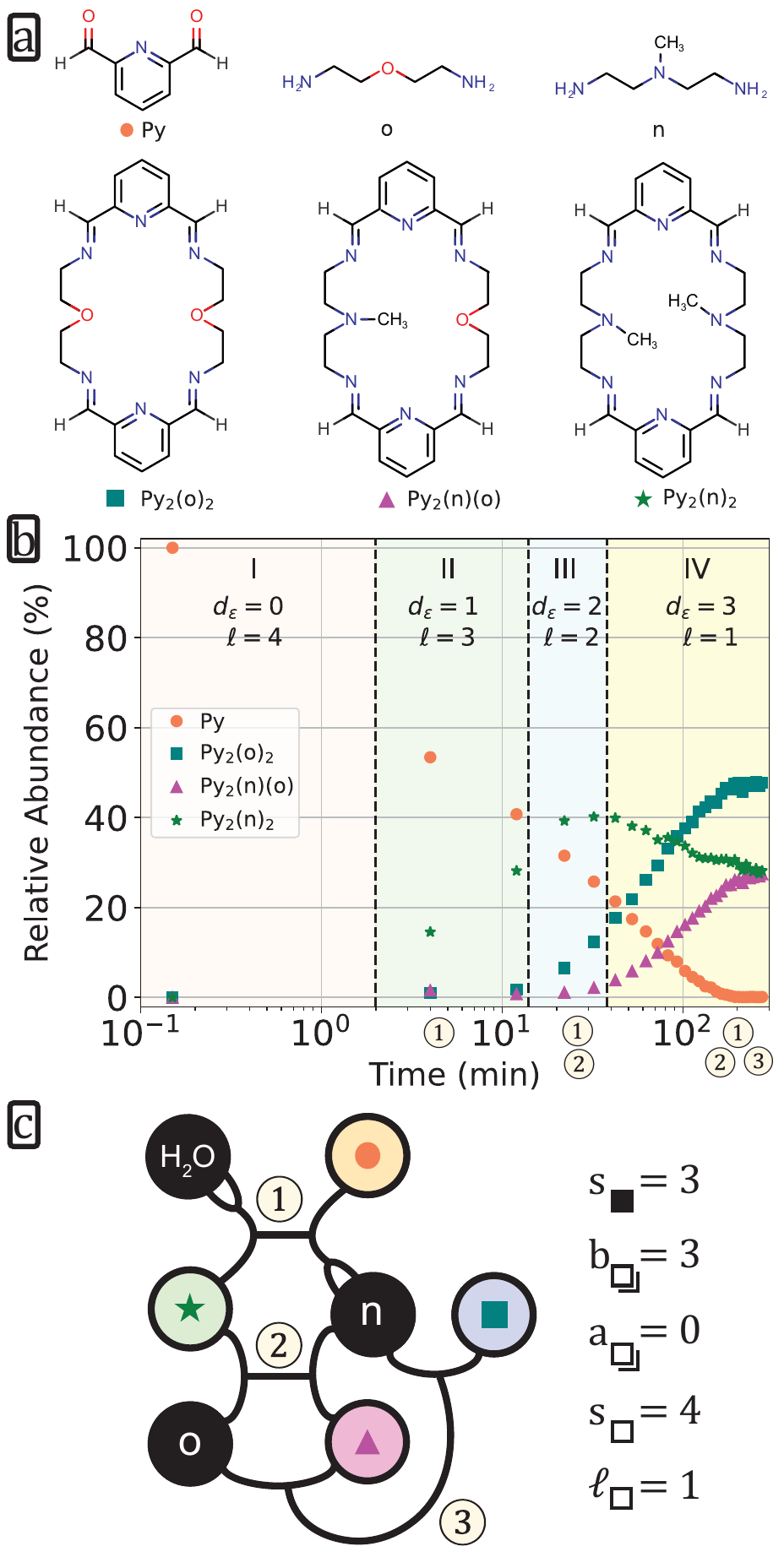}
\caption{a) Structures of organic molecules used and formed in Ref. \cite{Yang2024}, b) the monitored concentrations of Py and its cyclic products, progressively, all $d_0=3$ dimensions become discernible (regime IV) (see Sec: \ref{subsection:res-dep}). c) CRN with concealed species. $s_\blacksquare = 3$ out of $s=7$ species are not monitored. Since they give rise to $b_{\textifsym{\SquareShadowA}}=3$ concealed conservation laws ($a_{\textifsym{\SquareShadowA}}=0$), their values can readily be reconstructed indirectly.} 
\label{fig:yang2024}
\end{figure}

Fig.~\ref{fig:yang2024} shows a literature example of concealed species in chemical data from Ref. \cite{Yang2024}, where cyclic species are formed through successive condensation reactions
\bea
\ce{Py} + 2 \ce{n} \leftrightarrows \ce{Py}(\ce{n})_2 + 2 \ce{H2O} \nonumber \\
\ce{Py}(\ce{n})_2 + \ce{o} \leftrightarrows  \ce{Py}(\ce{n})(\ce{o}) + \ce{n} \nonumber \\
\ce{Py}(\ce{n})(\ce{o}) + \ce{o} \leftrightarrows \ce{Py}(\ce{o})_2  + \ce{n} . \nonumber
\eea
At this level of description, we have $s=7$ species, $r=3$ reactions, $c=0$ cycles, and thus $\ell=4$ conservation laws. One natural choice of basis is

\hspace*{-1.5cm}\vbox{\bea
L_1 &=& [\ce{Py}] + [\ce{Py(n)_2}] + [\ce{Py}(\ce{n})(\ce{o})]  + [\ce{Py(o)_2}] , \nonumber \\
L_2 &=& [\ce{n}] + 2 [\ce{Py(n)_2}] + [\ce{Py}(\ce{n})(\ce{o})]  , \nonumber  \\
L_3 &=& [\ce{o}] + [\ce{Py}(\ce{n})(\ce{o})]  + 2 [\ce{Py(o)_2}]   , \\
L_4 &=& 2 [\ce{H2O}] + [\ce{Py}] .\nonumber
\eea}

Fig.~\ref{fig:yang2024}b shows data monitoring species abundance versus time, with $s_\blacksquare=3$ species ($\ce{n},\ce{o},\ce{H2O}$) concealed, effectively yielding

\bea
\ce{Py} &+& \blacksquare \leftrightarrows \ce{Py}(\ce{n})_2 + \blacksquare  , \nonumber \\
\ce{Py}(\ce{n})_2 &+& \blacksquare  \leftrightarrows  \ce{Py}(\ce{n})(\ce{o}) + \blacksquare ,   \\
\ce{Py}(\ce{n})(\ce{o}) &+& \blacksquare \leftrightarrows \ce{Py}(\ce{o})_2  + \blacksquare . \nonumber
\eea

Concealment either conceals conservation laws or makes apparent cycles appear (Eq. \eqref{equation:concealedspecieslaw}). Here, all concealed species result in concealed conservation laws:  $b_{\textifsym{\SquareShadowA}} = 3$, and only $L_1$ remains valid for the observed species. 

Indirectly, we can still monitor all concealed species: provided we know the values of $L_2$ to $L_4$, we can quantify $[\ce{H2O}], [\ce{n}]$ and $[\ce{o}]$ using the values of our observables. In Ref~\cite{Yang2024}, the values of $L_1$ to $L_4$ are controlled by the experimentalist preparing the initial composition of the system ($\ce{Py}, \ce{n}, \ce{o}$) and, ipso facto, their values are known.

\subsection{The isospectral species law}

In the observable subnetwork $\mathbb{S}_\square$, we now merge all pairs of isospectral isomer species that measurably absorb, to describe the effective subnetwork due to their indistinguishibility. We denote as $\text{\textsection}(\Lambda)$ the number of mergers thus performed. We obtain
\hspace*{-0.25cm}\vbox{\be
\text{rk}(\mathbb{S}_\square)-\text{rk}(\mathbb{S}_\textifsym{\RightDiamond}) = \Delta s - \Delta \ell = \Delta r - \Delta c , \label{equation:towardsisospectral}
\ee}

for which $\Delta s = \text{\textsection}(\Lambda)$. Reactions that vanish by merging a pair of isomeric species are reactions appearing as direct isomerization reactions in $\mathbb{S}_\square$, $\Delta r = r_\sim$. Merging isospectral species can either break conservation laws or create apparant cycles, affording the isospectral species law

\be
\text{\textsection}(\Lambda) = b_{\text{\textsection}}(\Lambda) + a_{\text{\textsection}}(\Lambda) + r_\sim (\Lambda) ,\label{equation:isospectralspecieslaw}
\ee

where \\
$b_{\text{\textsection}}(\Lambda): \#$ lost conservation laws \\
$a_{\text{\textsection}}(\Lambda): \#$ apparent cycles (combinations of reactions that leave discernable variables unchanged) \\
$r_\sim (\Lambda): \#$ direct isomerization reactions among isospectral species pairs

We can then express the loss of dimension between the full CRN and what is discerned spectroscopically as

\bea
\Delta d_\Lambda &=& \text{\textsection}(\Lambda) + s_{\blacksquare}(\Lambda) - r_\sim(\Lambda)  \nonumber \\ &\ & - b_{\textifsym{\SquareShadowA}}(\Lambda) - b_{\text{\textsection}}(\Lambda)  ,\label{equation:lostdimension}\\
&=& a_{\textifsym{\SquareShadowA}}(\Lambda) + a_{\text{\textsection}}(\Lambda) .
\eea

\begin{figure}[tbhp!]
\centering
\includegraphics[width=1.0\linewidth]{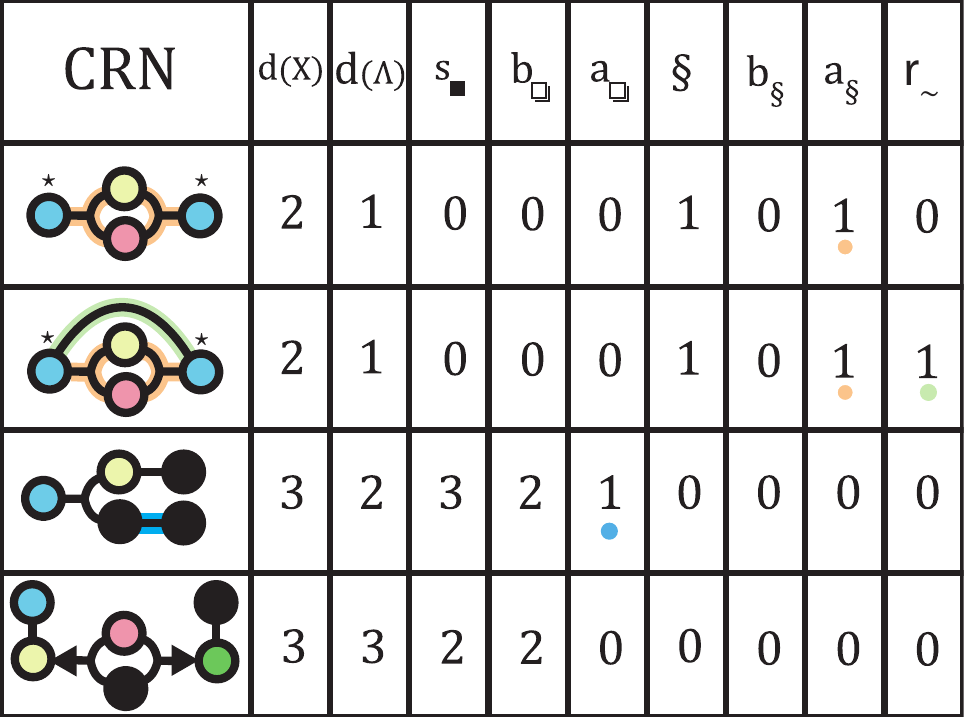}
\caption{Illustration of laws for isospectral (Eq. \eqref{equation:isospectralspecieslaw}) and concealed (Eq. \eqref{equation:concealedspecieslaw}) species for several CRNs. Dark nodes correspond to spectroscopically inactive species. Species that are indistinguishable are marked with a star. New cycles - leading to loss of data dimension - are highlighted in orange and blue. Data dimensions: $d(\ce{X}), d(\Lambda)$ (spectral), $d(\ce{X})$ (species). $d(\Lambda)$ spectral data dimension, $s_{\blacksquare}(\Lambda): \#$ concealed species, $b_{\textifsym{\SquareShadowA}} (\Lambda): \#$ concealed conservation laws, $a_{\textifsym{\SquareShadowA}} (\Lambda): \#$ apparent cycles, $\text{\textsection}(\Lambda):$ isospectral index, $r_\sim (\Lambda): \#$ isospectral isomerization reactions, $b_{\text{\textsection}}(\Lambda): \#$ isospectral broken conservation laws, $a_{\text{\textsection}}(\Lambda) \#$, isospectral emergent cycles.}
\label{fig:effectiveCRN}
\end{figure}

We can represent these spectral properties in network representations, and thereby graphically determine $d(\ce{X})$ and $d(\Lambda)$. Some examples are provided in Fig. \ref{fig:effectiveCRN}. More examples with elaboration are provided in Sec. \ref{appendix:isospectral_concealed}.

The dimension of spectroscopic data can match the original data dimension $d(\ce{X})$ $(\Delta d_\Lambda = 0)$ even if not all species are discernable. Up to $\ell$ variables can be lost (Eq.\eqref{equation:lostdimension}), provided this only leads to unobservable conservation laws or isospectral isomerization reactions. Conversely, each missing dimension corresponds to an apparent cycle, i.e., combinations of reactions that leave the discernable variables unchanged (this includes reactions which have become unobservable $\blacksquare \leftrightarrows \blacksquare$). For nonabsorbing species, such a cycle requires two chemostats ($s_{\blacksquare}(\Lambda) \geq 1$) for the effective CRN. Hence, for $s_{\blacksquare}(\Lambda) = 1$, no loss of dimension occurs. 

\section{Measuring data dimension}
\label{sec:measuring_data_dimension}

We will now formalize the process of measuring the data dimension. 
We define a spectral data matrix $\mathbb{A}$ as the matrix containing spectra measured at successive times $t_1, t_2, ..., t_m$

\be
\mathbb{A} = \left(\begin{smallmatrix} A_{\lambda_1}(t_1) & A_{\lambda_2}(t_1) & ... & A_{\lambda_n}(t_1) \\
A_{\lambda_1}(t_2) & A_{\lambda_2}(t_2) & ... & A_{\lambda_n}(t_2) \\
\vdots & \vdots & \ddots & \vdots \\
A_{\lambda_1}(t_m) & A_{\lambda_2}(t_m) & ... & A_{\lambda_n}(t_m) \\
\end{smallmatrix}\right) .
\ee

We denote by $\Delta A_{\lambda_n}$ a mean subtracted absorption

\hspace*{-0.5cm}\vbox{\be
\Delta A_{\lambda_n}(t_q) = A_{\lambda_n}(t_q) - \frac{1}{m} \sum_{k=1}^m A_{\lambda_n}(t_k),
\ee}
and we analogously define 

\hspace*{-0.75cm}\vbox{\be
\Delta \mathbb{A} = \left(\begin{smallmatrix} \Delta A_{\lambda_1}(t_1) & \Delta A_{\lambda_2}(t_1) & ... & \Delta A_{\lambda_n}(t_1) \\
\Delta A_{\lambda_1}(t_2) & \Delta A_{\lambda_2}(t_2) & ... & \Delta A_{\lambda_n}(t_2) \\
\vdots & \vdots & \ddots & \vdots \\
\Delta A_{\lambda_1}(t_m) & \Delta A_{\lambda_2}(t_m) & ... & \Delta A_{\lambda_n}(t_m) 
\end{smallmatrix}\right).
\ee}

To contrast with genuine (noisy) data, we let $\Delta \tilde{\mathbb{A}}$ derive from an exact (noise-free) solution $[\pmb{\ce{X}}](t)$ to the dynamics multiplied by exact theoretical spectra. Whereas we expect a true data matrix to be of full rank, noise-free $\Delta \tilde{\mathbb{A}}$ can have a nonzero kernel, and (starting away from equilibrium) its rank is given by the spectroscopic data dimension $d(\Lambda)$
\be
\text{rk}(\Delta \tilde{\mathbb{A}}) = d(\Lambda).
\ee
We can then model noisy data by addition of a random matrix $\eta$ whose entries are independent identically distributed random variables with mean 0 and variance $\epsilon^2$

\be
\Delta \mathbb{A} = \Delta \tilde{\mathbb{A}} + \eta (\epsilon), 
\ee
so that almost always
\be
\text{rk}(\mathbb{A}) = \text{rk}(\eta) = \text{max}(n,m).
\ee

While we cannot directly asses $d(\Lambda)$, we can adopt a decomposition procedure to attempt to separate signal-rich dimensions from noisy ones, and estimate a dimension $d_\epsilon$ from that by some criterion. A simple criterion introduced in Sec. \ref{subsection:SVDspec} considers the spectral properties of random matrix \cite{marcenko_distribution_1967} $\eta(\epsilon)$ and will suffice for our examples (see Refs.\cite{venturi_proper_2006,epps_singular_2019,epps_error_2010} for more thorough considerations). We will call the estimate $d_\epsilon(\Lambda)$ hereby obtained the discernable dimension. 

The procedure we will use for our illustration is singular value decomposition (SVD), which decomposes the data in successive components that maximally explain remaining variance (alternative decompositions to estimate data dimension exist \cite{aris_independence_1963}). The SVD has several desirable mathematical properties and is widely implemented. A disadvantage in our context is that the decomposition is not tailored to the structure of our problem, and is prone to underestimating $d(\Lambda)$ when dimensions become small.

\subsection{Isosbestic points}
\label{subsec:isosbesticpts}

An isosbestic point is defined \cite{braslavsky_glossary_2007} as a wavelength $\lambda^*$ where the total absorbance $A$ remains fixed during chemical or physical change of a sample

\be
d_t A_{\lambda^*} = 0. \label{equation:isosbesticpointdef}
\ee
The expression

\be
A_{\lambda^*} = \sum_{k=1}^{s_{\square}} [X_k] \epsilon_k (\lambda^*)
\ee
contains $s_{\square}$ variables $[\ce{X}_k]$ and coefficients $\epsilon_k$. For the expression to reduce to a constant, we then require that same number of constraints on species and coefficients (which depend on $\lambda$). To have an isosbestic point be confined to a point $\lambda^*$, we thus require at least one constraint on coefficients $\epsilon_k$ and that absorbance $A_\lambda$ is nonzero around $A_{\lambda^*}$. In the absence of special symmetries, it is normally assumed that, at $\lambda^*$, only one further constraint on coefficients is introduced. Under this condition, we get $d_\Lambda = 1$ if the assessed set of wavelengths $\Lambda$ is confined to a small enough neighborhood around the isosbestic point.

For $s_{\square}=1$ visible species, no local isosbestic point can exist if we require $\epsilon_1(\lambda)>0$, as then

\be
d_t A_\lambda = \epsilon_1(\lambda) d_t [\ce{X_1}] ,
\ee
which only vanishes if it vanishes everywhere due to $d_t[\ce{X_1}]=0$. We thus require $s_{\square} \geq 2$. 

Furthermore, we require the visible (sub)network to at least possess one conservation law with positive coefficients. This follows since the expression of stationary absorbance is a conservation law:

\bea
L &=& \frac{1}{\epsilon^\circ} \sum_{k=1}^{s_{\square}} [X_k] \epsilon_k (\lambda^*), \ \ \ \ (\epsilon_k > 0), \nonumber \\
d_t L &=& 0 ,
\eea
where $\epsilon^\circ$ is a constant introduced to render coefficients dimensionless. In this sense, an isosbestic point \textit{is} itself a conservation law \cite{blokhuis2024ejoc}.

An isosbestic point provides local information about the visible subnetwork. When different species absorb appreciably at different wavelengths and not at others, a CRN can have several independent sets of isosbestic points, each pointing to local low-dimensional substructures. For instance, in the CRN

\bea
\ce{X_1} + \ce{X_2} \leftrightarrows \ce{X}_4 , \nonumber \\
\ce{X_2} + \ce{X_3} \leftrightarrows \ce{X}_5 ,
\eea
we have two independent substructures that can give rise to isosbestic points. Denoting 
$\blacksquare$ an obscured species, we have
\hspace*{-0.0cm}\vbox{\bea
\ce{X_1} + \blacksquare \leftrightarrows \ce{X}_4 , \ \ (L_1 = [\ce{X}_1] + [\ce{X}_4]) ,
\eea}
if $\epsilon_1(\lambda^*) = \epsilon_4(\lambda^*)$, $\epsilon_k(\lambda^*)=0 \ \ (k\neq 1,4)$, and
\hspace*{-0.0cm}\vbox{\bea
\ce{X_3} + \blacksquare \leftrightarrows \ce{X}_5 , \ \ (L_3 = [\ce{X}_3] + [\ce{X}_5]) ,
\eea}
if $\epsilon_3(\lambda^{**}) = \epsilon_5(\lambda^{**})$, $\epsilon_k(\lambda^*)=0 \ \ (k\neq 3,5)$.
This multiplicity of independent isosbestic points is a nontrivial conclusion: previous work on isosbestic points \cite{cohen_588_1962} worked with the assumption that all species are visible $s_\square=s$ and that they appreciably absorb at all wavelengths being assessed, which is the exception \footnote{The broad absorbance is first and foremost encountered for spectroscopy in the visible and ultraviolet range (UV-VIS).} rather than the rule. Under these hypotheses, isosbestic points become global features of low dimension.

It may furthermore be noted that there are other - more common - spectral features with the same structural implications as isosbestic points. The use of isosbestic points and 'equivalent' dimensional features \footnote{For instance, a pair of independent absorbance peaks $A_i, A_j$ for which the absorbances satisfy a conservation law $L = a_i A_i + a_j A_j$, $a_i,a_j >0$ similarly imply $d_\Lambda = 1$ and positive conservation locally.} as structural probe can thus be greatly extended to characterize higher-dimensional system.

The exact structural interpretation of isosbestic points has not been cleared up yet. Foundations for such a theory are contained in this section and the dimensional laws developed in this paper.
Developing a full structural theory of isosbestic points would go beyond the scope of the current work and will be developed further elsewhere.



\subsection{Resolution-dependence of dimension}
\label{subsection:res-dep}

The need for $d_\epsilon(\Lambda)$ becomes apparent by revisiting the CRN $\ce{C} \leftrightarrows \ce{A} + \ce{B} \leftrightarrows \ce{D}$, starting with only $\ce{A},\ce{B}$ ($[\ce{A}]_0, [\ce{B}]_0 > 0, [\ce{C}]_0,[\ce{D}]_0=0$). Up to a linear approximation, we then have
\bea
\Delta [\ce{A}] &=& - \left(k_1 + k_2\right) [\ce{A}]_0 [\ce{B}]_0 t + \mathcal{O}(t^2) , \nonumber \\
\Delta [\ce{C}] &=& - k_1 [\ce{A}]_0 [\ce{B}]_0 t + \mathcal{O}(t^2) , \\
\Delta [\ce{D}] &=& - k_2 [\ce{A}]_0 [\ce{B}]_0 t + \mathcal{O}(t^2) .\nonumber
\eea
For a signal-to-noise ratio $\epsilon>0$, we can then choose a sufficiently small time $t^*$ such that higher order corrections starting at $\mathcal{O}(t^2)$ become comparable to noise. For data recorded on shorter timescales, we cannot yet distinguish reverse reactions, and then $\ce{C} \leftrightarrows \ce{A} + \ce{B} \leftrightarrows \ce{D}$ behaves as $\ce{C} \leftarrow \ce{A} + \ce{B} \rightarrow \ce{D}$, i.e., we have a discernable dimension $d_\epsilon (\ce{X}) = 1$ for $t \ll t^*$, $d_\epsilon (\ce{X}) = 2$ for $t \gg t^*$. Under such circumstances, a co-production law $\rotatebox[origin=c]{180}{e}_{\blackwhiteffor}$ may be transient and conservation laws more generally may be transient.

\begin{figure}[tbhp!]
\centering
\includegraphics[width=1.0\linewidth]{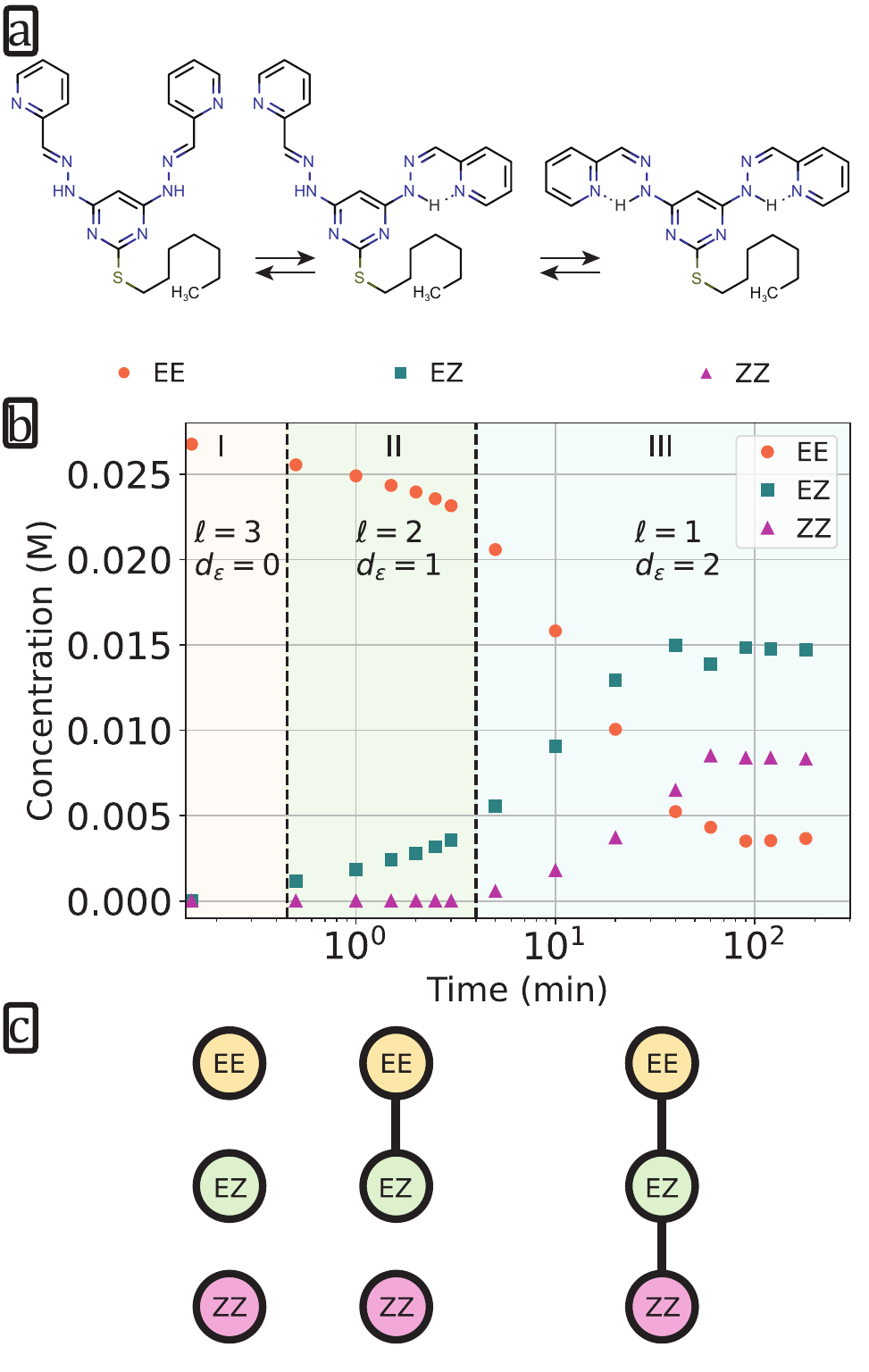}
\caption{a) Photochemically interconvertible stereoisomers from Ref \cite{Gordillo2017} b) The observable dimensionality is time scale-dependent ($\epsilon = 3 \cdot 10^{-5}$): Reaction monitoring data (via \textsuperscript{1}H-NMR) quantifying isomers $\ce{EE}$, $\ce{EZ}$ to $\ce{ZZ}$. On very short times (regime I) no change is discernable, thus $d=0, \ell=3$. Subsequently (regime II), consequences of the first isomerization become observable, but not the second ($d=1, \ell=2$). Finally (regime III), both reactions yield observable consequences ($d=2,\ell=1$). c) effective CRNs in regime I), II) and III).}
\label{fig:Gordillo}
\end{figure}

We provide a small  example of the phenomenon of resolution-dependent conservation laws using literature data of a small CRN in Fig. \ref{fig:Gordillo}, setting $\epsilon = 3 \cdot 10^{-5}$. We have used data from Ref. \cite{Gordillo2017}, where a photochemical isomerization reaction is monitored for a CRN of the form $\ce{A} \leftrightarrows \ce{B} \leftrightarrows \ce{C}$. On short timescales, changes are not yet discernable (either by being too small $<\epsilon$ or due to the range not including a second datapoint), i.e., all species appear conserved ($\ell=3, d_{\epsilon}=0$):
\bea
L^{(1)}_{\text{I}} &=& [\ce{A}] = [\ce{A}]^{0}, \ \ \ L^{(2)}_{\text{I}} =  [\ce{B}] = 0, \nonumber \\   
L^{(3)}_{\text{I}} &=&  [\ce{C}] = 0. 
\eea
Shortly after, the reaction $\ce{A} \leftrightarrow \ce{B}$ becomes noticeable, $\ell=2, d=1$:
\bea
L^{(1)}_{\text{II}} &=& [\ce{A}] + [\ce{B}] = [\ce{A}]^{0}, \\ L^{(2)}_{\text{II}} &=&  [\ce{C}] = 0. 
\eea
On longer times, the reaction $\ce{B} \leftrightarrows \ce{C}$ also becomes noticeable, and now all $d_0=2$ dimensions are resolved, $\ell=1$ conservation law remains:
\be
L^{(1)}_{\text{III}} = [\ce{A}]  + [\ce{B}] + [\ce{C}].
\ee

Resolution and chemical interpretations of all missing dimensions $d_0  - d_\epsilon $ is outside the scope of this paper but is developed further in our companion paper\cite{blokhuis2024ejoc} and will be the subject of follow-up work. In the current example, the dimension could intuitively be deduced by eye. However, this is not generally possible, especially for higher-dimensional data, and more formal procedures such as SVD can be used to establish $d_\epsilon$, which we will discuss next.


\subsection{SVD for spectral data}
\label{subsection:SVDspec}

A singular value decomposition of the $m$-by-$n$ matrix $\Delta \mathbb{A}$ represents it as a product of three matrices
\be
\Delta \mathbb{A} = U \Sigma V^T.
\ee
In our context, $V$ is an $n$-by-$n$ matrix with rows $v_1(\lambda), v_2(\lambda), ...,v_n(\lambda)$ (right singular vectors) corresponding to spectra
\be
V = \left(\begin{smallmatrix} v_1 (\lambda_1) & v_2 (\lambda_1) & .. &  v_n (\lambda_1) \\ 
v_1 (\lambda_2) & v_2 (\lambda_2) & .. &  v_n (\lambda_2) \\ 
\vdots & \vdots & \ddots & \vdots \\
v_1 (\lambda_n) & v_2 (\lambda_n) & .. &  v_n (\lambda_n) \end{smallmatrix}\right)
\ee
$U$ is an m-by-m matrix with rows $u_1(t), u_2(t), ...,u_n(t)$ (left singular vectors) corresponding to time trajectories, i.e., time-dependent spectral contributions
\be
U = \left(\begin{smallmatrix} u_1 (t_1) & u_2 (t_1) & .. &  u_m (t_1) \\ 
u_1 (t_2) & u_2 (t_2) & .. &  u_m (t_2) \\ 
\vdots & \vdots & \ddots & \vdots \\
u_1 (t_m) & u_2 (t_m) & .. &  u_m (t_m) \end{smallmatrix}\right).
\ee
Singular vectors thus produced are orthonormal 
\be
U U^T = I, \ \ \ V V^T = I.
\ee
The contributions of the singular vectors to the data are weighed by singular values $\sigma_1,..\sigma_n$, which occur diagonally in the $n-$by$-m$ matrix $\Sigma$,
\bea
\Sigma_{ki} &=& \delta_i^k \sigma_k, \\
\sigma_i &=& \Sigma_{ii} \geq 0, \ \ \ \sigma_i \geq \sigma_{i+1}.
\eea
Singular values are non-negative and placed in decreasing order. Based on the results of SVD, a single spectrum at time $t_q$ can now be reconstituted as
\be
A_\lambda(t_q) = \sum_i^{min(n,m)} \sigma_i u_i (t_q) \pmb{v}_i, 
\ee
which, on the level of the data matrix $\Delta \mathbb{A}$, can be written as a sum of rank-1 matrices
\bea
\Delta \mathbb{A} = \sum_i \Delta \mathbb{A}^{(i)} , \\
\Delta \mathbb{A}^{(i)} = \sigma_i \pmb{u}_i \pmb{v}_i^T , \\
\text{rk}(\Delta \mathbb{A}^{(i)}) = 1.
\eea
By truncating at $k$ we can now approximate the data with data of rank $k$.

Since $\sigma_i \geq \sigma_{i+1}$, the reconstructive contribution of successive $\Delta \mathbb{A}^{(i)}$ diminish with increasing $i$. Informed by the spectral properties of a random matrix \cite{marcenko_distribution_1967} $\eta(\epsilon)$, and supposing $n>m$, we will consider a threshold 
\be
\sigma^* = \epsilon \left(1+\sqrt{\frac{m}{n}}\right)^2  \sqrt{m}. \label{equation:threshold}
\ee
 We then let the discernable dimension $d_\epsilon$ be the number of singular values above this threshold. For our examples, $d_\epsilon(\Lambda)$ will also be qualitatively evident from visual inspection.

\begin{figure}[tbhp!]
\centering
\includegraphics[width=1.0\linewidth]{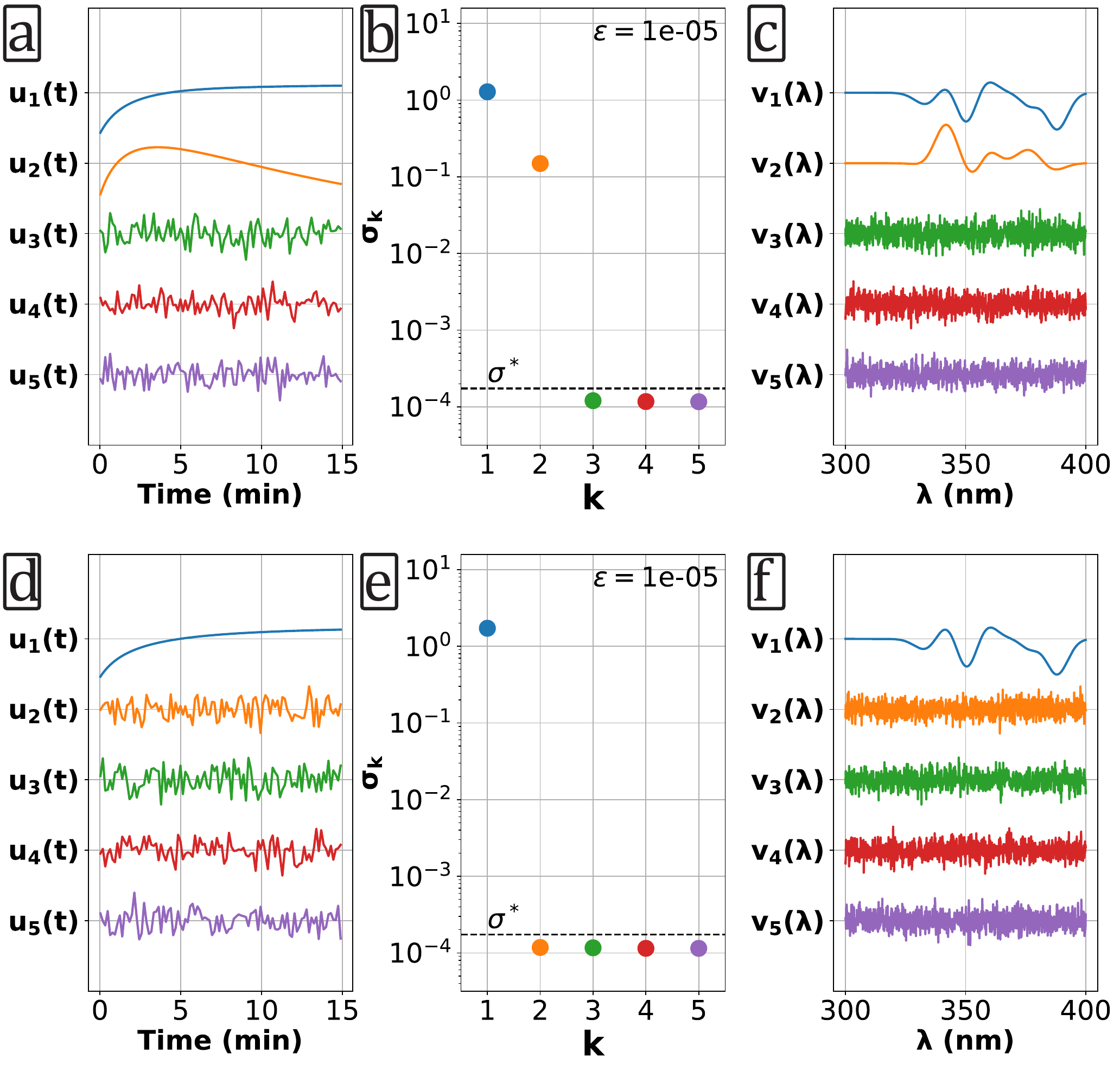}
\caption{(a-c) First 5 components of singular value decomposition of $\Delta \mathbb{A}$ for CRN $\ce{C} \rightleftarrows \ce{A} +\ce{B} \rightleftarrows \ce{D}$. By visual inspection, only 2 non-noisy a) trajectories and c) spectra are discerned, consistent with $d_\epsilon(\ce{\Lambda})=2$. (d-f) First 5 components of Singular Value Decomposition of $\Delta \mathbb{A}$ for CRN $\ce{C} \leftarrow \ce{A} +\ce{B} \rightarrow \ce{D}$. This CRN has $d_\epsilon(\ce{\Lambda})=1$, and its only f) spectral component $v_1(\lambda)$ crosses the origin 4 times. At these points, the change in absorbtion $\Delta A_\lambda = \sigma_1 u_1(t) \pmb{v}_1$ is zero, which (Eq. \eqref{equation:isosbesticpointdef}) correspond to isosbestic points. These isosbestic points can also be seen in Fig. \ref{fig:Motivation}b. For both, we have noise level $\epsilon = 10^{-5}$, $m=100$ successive spectra, $n=1000$ wavelengths. }
\label{fig:Fig7_SVDmotivate}
\end{figure}


Provided small enough $\epsilon$, signal is captured within the first $d(\Lambda)$ components, and further dimensions should look qualitatively distinct. Fig. \ref{fig:Fig7_SVDmotivate} shows the first 5 components for the spectral data for our initial example in Fig. \ref{fig:Motivation}a,b ($d(\Lambda)=2$) and Fig. \ref{fig:Motivation}c,d $(d(\Lambda)=1)$. In these examples, $d_\epsilon (\Lambda)=d(\Lambda)$ and a clear transition from signal to noise can be discerned.   

An analysis that makes use of the whole spectrum (such as SVD) allows for a more robust identification of a dimension compared to the use of special features such as isosbestic points or isosbestic lines. 


\section{Further applications: statistical physics}

A large diversity of models that are not necessarily treated as CRNs can readily be mapped to CRNs and our results consequently apply to a wide variety of models. The features (e.g. irreversible reactions) that give rise to our results are very common - even in simple and linear models - and enable the use of structural CRN considerations to furnish tools for classes of models where CRNs have traditionally seen limited application, e.g., many models studied in statistical physics (diffusion, fragmentation, scission collision, adsorption \cite{Krapivsky2010}).

Any Markov process can always be mapped to a CRN by mapping states to species and, consequently, we can apply our results such as the co-production law and the coproduction-linkage law to the study of these processes and derive co-production conservation laws for them. Our approach provides a tool to derive dynamic constraints and exact solutions for a broad class of models in statistical physics. We will here illustrate this point by the study of n-mer adsorption and provide two further examples of co-production in diffusion models in Appendix \ref{appendix:markov-processes}. In Appendix \ref{appendix:fixedpointstab}, we briefly illustrate how co-production pertains to the stability of certain fixed points and system behavior.

\subsection{Random sequential adsorption (RSA) of n-mers}
\label{subsec:nmer_abs}
\begin{figure}[tbhp!]
\centering
\includegraphics[width=1.0\linewidth]{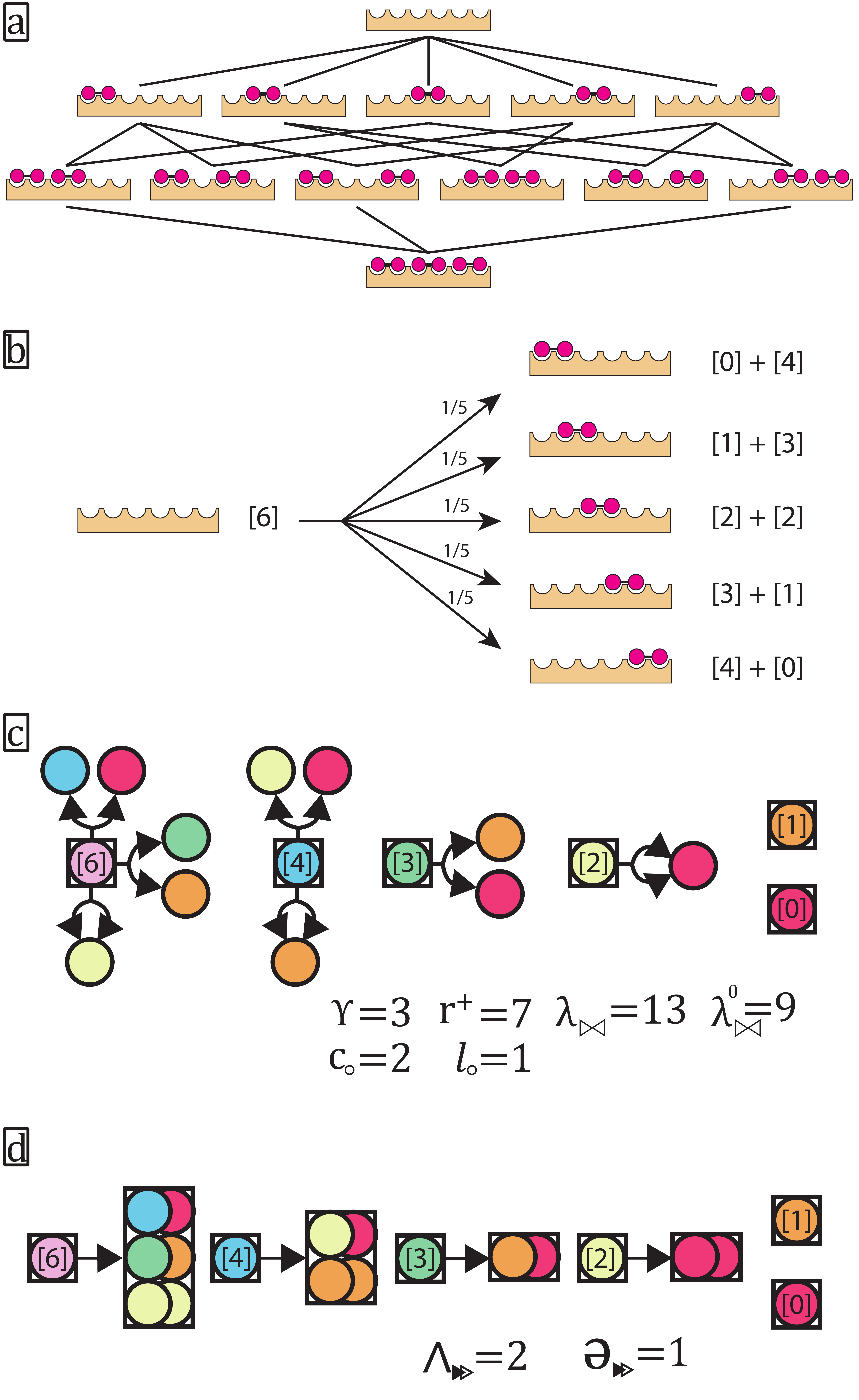}
\caption{a) Successive dimer ($n=2$) adsorption on a lattice of size $L=6$. Gaps sizes go from 0 to $n-1$, (here: 0,1). b) Merged dimer adsorption reactions for a gap  $[6]$. c) CRN representation of reactions on gap populations for $L=6, n=2$. $[5]$ is not included, as such a gap cannot be formed starting from $[6]$. The unique conservation law corresponding to $\ell^\circ=1$ is $L= 4 \langle [6]_t \rangle + 3 \langle [4]_t \rangle + \frac{5}{2} \langle [3]_t \rangle + 2 \langle [2]_t \rangle + \frac{3}{2} \langle [1]_t \rangle + \langle [0]_t \rangle$. d) Representation - after merging reactions - in terms of complexes.}
\label{fig:orbtion}
\end{figure}

We consider the irreversible random sequential adsorption\cite{Evans1993,Krapivsky2010} (RSA\footnote{See Ref.\cite{Evans1993} for a review on discrete RSA models and Ref.\cite{Talbot2000} for a review on continuous RSA models}) of n-mers on an initially empty discrete lattice of L sites (illustrated in Fig.\ref{fig:orbtion}a,b). We denote with $[k]$ the population of gaps of k successive empty sites between adsorbed n-mers and/or lattice extremities. We denote with $[k]_t$ the value of $[k]$ after $t$ reactions (but choosing a continuous time instead leaves our results unchanged), so that, initially ($t=0$), $[L]_0=1,[k]_0=0 \ \ (k<L)$.

In terms of gap populations, the adsorption process resembles a fragmentation process:

\hspace*{-0.5cm}\vbox{\be
\ \! \! \! [m+n] \rightarrow [m-k] + [k] \ \ \ (0 \leq k \leq m) .
\ee}
When merging collinear reactions, we have

\bea
\ \! \! \! [m+n] &\rightarrow& \frac{1}{m+1} \sum_{k=0}^{m} \left( [m-k] + [k] \right) \nonumber \\
\ \! \! \! [m+n] &\rightarrow& \frac{2}{m+1} \sum_{k=0}^{m} [k] ,
\eea
which in a stoichiometric matrix yield columns of the form $(-1, [0]_n, [\frac{2}{m+1}]_{m+1}$). For instance, for $L=9$, $n=3$ we have the stoichiometric matrix 

\hspace*{-0.75cm}\vbox{\be
\mathbb{S}_{\blackffor} = \begin{pmatrix}
\frac{2}{7} & \frac{2}{6} & \frac{2}{5} & \frac{2}{4} & \frac{2}{3} & 1 & 2 \\

\frac{2}{7} & \frac{2}{6} & \frac{2}{5} & \frac{2}{4} & \frac{2}{3} & 1 & 0  \\		  

\frac{2}{7} & \frac{2}{6} & \frac{2}{5} & \frac{2}{4} & \frac{2}{3} & 0 & 0  \\

\frac{2}{7} & \frac{2}{6} & \frac{2}{5} & \frac{2}{4} & 0 & 0 & -1 \\
          
\frac{2}{7} & \frac{2}{6} & \frac{2}{5} & 0 & 0 & -1 & 0 \\ 

\frac{2}{7} & \frac{2}{6} & 0 & 0 & -1 & 0 & 0 \\

\frac{2}{7} & 0 & 0 & -1 & 0 & 0 & 0 \\

0  & 0 & -1 & 0 & 0 & 0 & 0  \\  

0  & -1 & 0 & 0 & 0 & 0 & 0  \\  

-1  & 0 & 0 & 0 & 0 & 0 & 0  
      \end{pmatrix}   . 
\ee}

With successive rows corresponding to $[0], [1] , .. , [9]$. To show the highly regular structure of $\mathbb{S}_{\blackffor}$, we have included columns for $[8]$, $[7]$. These columns can be freely omitted when starting from $L=9, n=3$ as no pathway then exists to form these gap sizes.

Gaps of size smaller than $n$ cannot be reduced further, whereas gaps of size $n$ or greater always have an outgoing reaction. We thus have $n$ coproduction conservation laws $\pmb{\ell}^{(1)},..,\pmb{\ell}^{(n)}$ governing state probabilities (see Eq. \eqref{equation:gensol_nmer} for a general solution),  satisfying $\pmb{\ell}  \mathbb{S}_{\blackffor} = \pmb{0}^T$. A convenient basis for these conservation laws is one where each vector has a nonzero argument for only a single of the irreducible gaps:

\bea
\pmb{\ell}^{(1)} &=& \left( 1, 0, 0, 2, 1, \frac{2}{3}, \frac{3}{2}, \frac{8}{5}, \frac{14}{9}, \frac{37}{21} \right)^T , \nonumber \\
\pmb{\ell}^{(2)} &=& \left( 0, 1, 0, 0, 1, \frac{6}{9}, \frac{1}{2}, \frac{4}{5}, \frac{8}{9}, \frac{19}{21} \right)^T ,
\nonumber \\ 
\pmb{\ell}^{(3)} &=& \left(0,0,1,0,0,\frac{6}{9},\frac{1}{2},\frac{2}{5},\frac{5}{9},\frac{13}{21} \right)^T. \label{equation:coeffcons_law_nmer}
\eea
The value of conserved quantity  $L^{(1)},L^{(2)},L^{(3)}$ follows from evaluation of the initial condition
\be
L^{(k)} = \pmb{\ell}^{(k)} \cdot \begin{pmatrix} \langle [0]_t \rangle \\
\langle [1]_t \rangle \\
\langle [2]_t \rangle \\
\vdots \\
\langle [L-1]_t \rangle \\
\langle [L]_t \rangle  \end{pmatrix} . \label{equation:cons_avg_nmer}
\ee

\begin{figure}[tbhp!]
\centering
\includegraphics[width=1.0\linewidth]{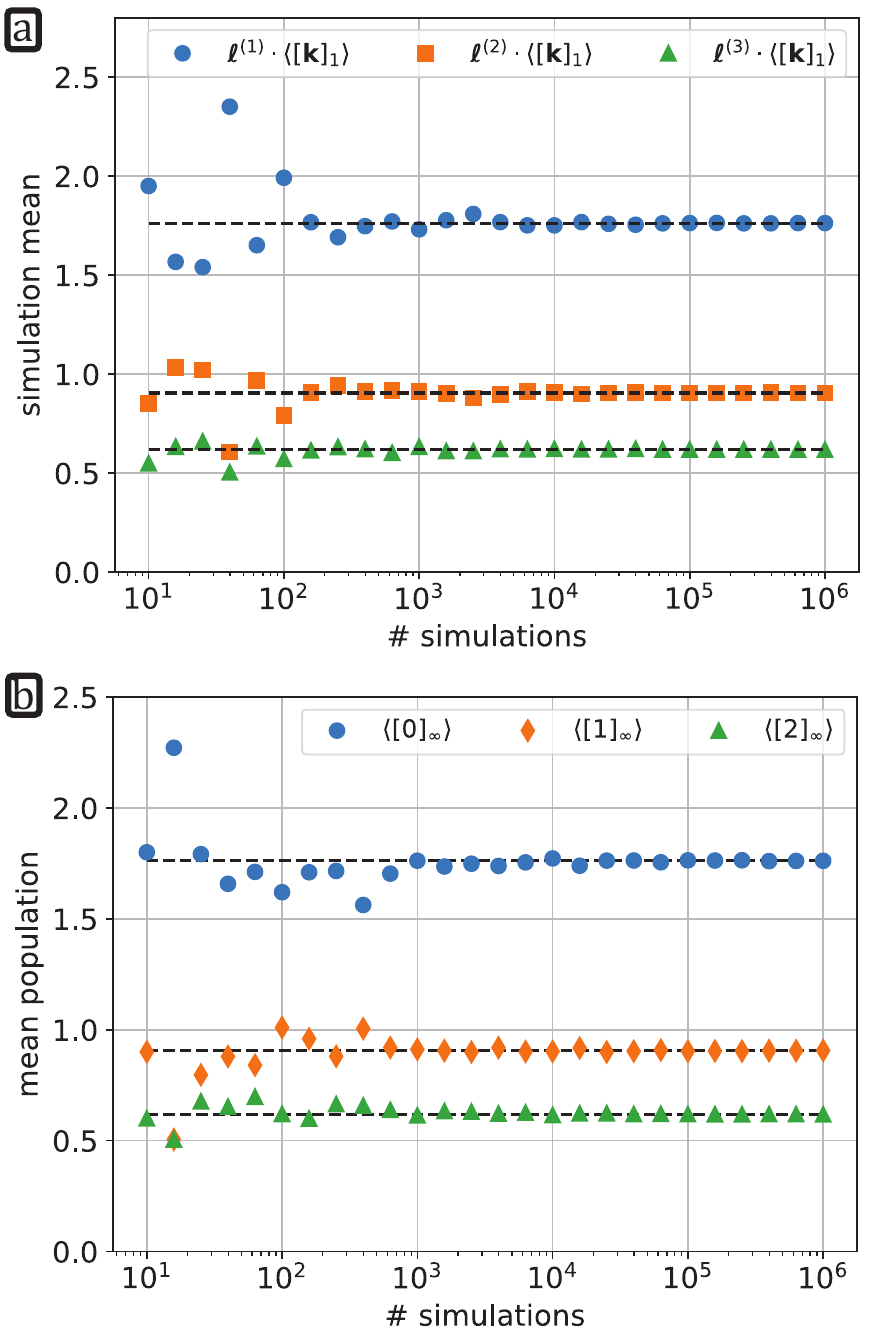}
\caption{a) Evaluation of conserved quantities over averages after 1 reaction, by averaging over an increasing number of simulations. As the number of simulations increases, convergence occurs to the theoretical values (dotted line) of $L^{(1)}, L^{(2)}, L^{(3)}$. b) Averaged number of gaps $[0],[1],[2]$ as a function of the number of individual simulated realizations, for trimer ($n=3$) adsorption on a lattice of size $L=9$. Dotted lines correspond to theoretical expectation values $L^{(1)}$, $L^{(2)}$, $L^{(3)}$, to which the simulated averages converge. Note that the conserved quantities here behave as regular conservation laws on the level of deterministic ODEs. Co-production conservation laws do not constrain individual realizations of a stochastic process, but do constrain the ensemble of realizations as a linear law on expectation values (Eq. \eqref{equation:cons_avg_nmer}).}
\label{fig:n-merplot}
\end{figure}

The conserved quantity pertains to averages and not individual trajectories. This is illustrated in Fig. \ref{fig:n-merplot}a, showing an evaluation of the conserved quantity after $t=1$ reaction, averaging over an increasing number of simulations. Convergence to the $L^{(1)},L^{(2)},L^{(3)}$ occurs as the number of simulations is increased to approach the theoretical expectation value.

Initially starting with an empty lattice of size $L=9$, ($[k](t=0)=\delta_k^L$, i.e., $[9]_0=1, [k]_0=0 \ (k<9)$), at long times $t\rightarrow \infty$, we then attain as stationary expectation value for the gap population $\langle [0]_\infty \rangle$

\bea
\langle [0]_\infty \rangle &=& \langle [\pmb{k}]_\infty \rangle \cdot \pmb{\ell}^{(1)}  =  L^{(1)}_{t \rightarrow \infty} = L^{(1)}_{t=0} \nonumber \\
&=&  \langle [\pmb{k}]_0 \rangle \cdot \pmb{\ell}^{(1)} = \langle [9]_0\rangle \ell^{(1)}_{10} = L^{(1)} \nonumber \\
&=&  \frac{37}{21} 
\eea
The same argument for $\langle [1]_\infty \rangle, \langle [2]_\infty \rangle$ results in
\bea
\langle [1]_\infty \rangle &=& \langle [9]_0\rangle \ell^{(2)}_{10} = L^{(2)} =  \frac{19}{21} \\
\langle [2]_\infty \rangle &=&  \langle [9]_0\rangle \ell^{(3)}_{10} = L^{(3)} = \frac{13}{21} , 
\eea
where $\langle [k]_\infty \rangle$ denotes the expectation value for the number of gaps of size $k$ for $t \rightarrow \infty$. Simulations of the n-mer model for $L=9, n=3$ are consistent with this result, and simulated mean gap size populations are shown in Fig. \ref{fig:n-merplot}b. The jamming coverage $\rho_{n,L}$ can then be written as
\be
\rho_{n,L} = 1 - \sum_{k=1}^{n} \frac{k L^{(k)}}{L}
\ee
For our example, we obtain $\rho_{3,9} = \frac{144}{189}\approx 0.76190$, which is noticably smaller than the $L\rightarrow \infty$ limit $\rho_{3,\infty}=0.82365$

\subsection{General solution for arbitrary \textit{L,n}}

The coefficients of a co-production $\pmb{\ell}^{(q)}$ conservation law must be such that $\pmb{\ell}$ is a left null vector $\mathbb{S}_{\blackffor}$. For the adsorption of n-mers, we then have

\be
\ell^{(q)}_{j+n} = \frac{2}{j} \sum_{k=1}^{j} \ell^{(q)}_{k} .
\ee
Rather than summing over $k=1$ to $k=j$, we can confine our sum from $k=j-n+1$ to $k=j$,

\hspace*{-0.75cm}\vbox{\bea
\ell^{(q)}_{j+n} &=& \frac{2}{j} \left(\sum_{k=j-n+1}^{j} \ell^{(q)}_{k} +  \sum_{k=1}^{j-n} \ell^{(q)}_{k} \right) \nonumber \\
&=& \frac{2}{j} \sum_{k=j-n+1}^{j} \ell^{(q)}_{k} + \frac{j - n }{j} \ell^{(q)}_{j} . \nonumber 
\eea}
We may similarly express $\ell^{(q)}_{j+n-p} $, $0<p<n$ in terms of $\ell^{(q)}_{j-n+1}$ to $\ell^{(q)}_{j}$:

\hspace*{-1.0cm}\vbox{\bea
\ell^{(q)}_{j+n-p} &=& \frac{2}{j-p} \left(\sum_{k=j-n+1}^{j-p} \ell^{(q)}_{k} +  \sum_{k=1}^{j-n} \ell^{(q)}_{k} \right) \nonumber \\
&=& \frac{2}{j-p} \sum_{k=j-n+1}^{j-p} \ell^{(q)}_{k} + \frac{j - n}{j-p} \ell^{(q)}_{j} \nonumber 
\eea}

From these expression, we can then calculate any coefficient in a conservation law for arbitrary oligomer size $n$ and lattice size $L$:
\hspace*{-1.0cm}\vbox{\bea
&\begin{pmatrix} \ell^{(q)}_{mn-n+1} \\ \ell^{(q)}_{mn-n+2} \\ \vdots \\ \ell^{(q)}_{mn-2} \\ \ell^{(q)}_{mn-1}  \\ \ell^{(q)}_{mn}  \end{pmatrix} = \prod_{p=1}^{m-1} \mathbb{F}_{p,n} \begin{pmatrix}\ell^{(q)}_1 \\ \ell^{(q)}_{2} \\ \ell^{(q)}_{3} \\ \vdots \\ \ell^{(q)}_{n-1}  \\ \ell^{(q)}_{n} \end{pmatrix} \label{equation:gensol_nmer} \\ 
&\mathbb{F}_{p,n} = \left(\begin{smallmatrix}
\frac{2}{np-n+1} & 0 & 0  & \dots & 0 &    \frac{np-n}{np-n+1} \\
 \frac{2}{np-n+2} & \frac{2}{np-n+2} & 0 &  \dots & 0 & \frac{np-n}{np-n+2} \\
\vdots & \vdots & \vdots & \ddots & \vdots & \vdots \\
 \frac{2}{np-2} & \frac{2}{np-2} & \frac{2}{np-2}  & \dots & 0 & \frac{np-n}{np-2} \\
 \frac{2}{np-1}  & \frac{2}{np-1} & \frac{2}{np-1} & \dots & \frac{2}{np-1}  & \frac{np-n}{np-1} \\
 \frac{2}{np} & \frac{2}{np} &  \frac{2}{np}  & \dots  & \frac{2}{np} & \frac{np-n+2}{np} 
\end{smallmatrix} \right) . \nonumber
  \eea}
  
For oligomers of size $n$, a convenient base for nullvectors in Eq. \eqref{equation:gensol_nmer} follows from equating the first $n$ coefficients to a unit vector $\hat{\pmb{e}}_q$ of length $n$ for $\ell^{(1)}$ to $\ell^{(m)}$ and we get 

\hspace*{-0.5cm}\vbox{\be 
\begin{pmatrix} \ell^{(q)}_1 & \ell^{(q)}_{2} & \dots & \ell^{(q)}_{n} \end{pmatrix}^T = \hat{\pmb{e}}_q , \ \ \ (0 \leq k \leq n) , 
\ee}
which was the approach adopted in \ref{subsec:nmer_abs}, for which the matrices $\mathbb{F}_{1,3},\mathbb{F}_{2,3},\mathbb{F}_{3,3}$ can also be found in \ref{appendix:nmer_matrix}. With this choice, one can then obtain the stationary expectation values for gap populations for arbitrary $L,n$ as
\be
\langle [k]_\infty \rangle = \langle [L]_0 \rangle \ell_{L+1}^{(k+1)} = L^{(k+1)} .
\ee

\section{Discussion}

Machine learning (ML) approaches can learn representations of chemical data and make predictions based on this representation. Importantly, any ML representation that incorporates inherent symmetries and constraints will make more physically sound predictions and will need less training data \cite{batzner_e3-equivariant_2022}. The algorithmic detection of such symmetries has become an important development in its own right \cite{Liu_PRE_2024,sturm_conservation_2022,brunton_discovering_2016,schmidt_distilling_2009,Mototake2021,Wetzel2020}. Understandably, many ML approaches have adopted CRNs as the model representation to be learned, through a variety of learning strategies \cite{taylor_automated_2021, willis_inference_2016, bures_organic_2023,ji_autonomous_2021}. The challenges posed by ML and MCR thus bring us back to our original unresolved challenge, namely, how CRN structure follows from data, and \textit{vice versa}. 

Stochastic thermodynamics has recently started to unearth current-current relations, including affine\cite{Harunari2024} relations and their multilinear generalizations\cite{cengio2025} among stationary nonequilibrium currents in Markov Networks. Co-production provides a new body of results beyond Markov Networks and stationary states to the developing corpus of current-current relations. We surmise that these relations are connected and that there are deeper current-current relations yet to be elucidated, along with dimensional consequences.

%
In this paper we advanced a methodology to relate structural elements (cycles, conservation laws, reactions, species, co-production) in the CRN to the dimension of the data it can bring forth. We have by no means exhausted this strategy: further dimensional quantities (such as isosbestic points and its generalizations, data dimensions with added constraints) can be defined and measured, and related to structural elements by the fundamental theorem of linear algebra. Through the co-production linkage law (CLL), our framework fully characterizes the co-production in a system. However, other chemical mechanisms exist that can lead to emergent conservation laws, e.g. due to experimental resolution making some reactions unobservably slow (Sec. \ref{subsection:res-dep}) or fast (Fig. \ref{fig:robustness}c,d). In our companion paper\cite{blokhuis2024ejoc} we discuss further chemical sources of emanants. Further dimensional formalizations of chemistry and its data will form the object of future work.

The data dimension is useful for filtering competing hypotheses.
As more measured dimensional quantities (such as data dimension $d$) are combined, the identity of a CRN can be narrowed down further: the pool of CRNs that can obey all the observed constraints shrinks exponentially with the number of constraints.   

For our laws to pertain to as much data as possible recorded in experimental practice, we derived laws pertaining to the common chemical experimental context in which not all species can be observed or distinguished. Further consideration of the viewpoint of an observer \cite{Polletini2017} should be beneficial for CRNs in theory and experiment, as it has been for other physical theories.

\section{Conclusion}

We characterized effects of irreversible reactions and concealed species on the dimension of chemical data. This furnished several new CRN laws, which fundamentally derive from applying the first structural law (SL1) $s-\ell = r-c$ to new contexts: irreversible co-production and incomplete observation (through spectroscopy). The reinterpretation of underlying arguments in this new context in terms of countable chemically meaningful concepts thereby leads to laws of chemistry that govern it. 

The co-production law $\text{\footnotesize{$\Upsilon$}} = \rotatebox[origin=c]{180}{e}_{\blackwhiteffor} + \wedge_{\blackwhiteffor}$ formalizes one of the prominent topological differences that can arise in passing from CRNs with reactions that are reversible to ones that are irreversible. A set of common conservation laws can be found from an analysis of integer stoichiometry, but irreversbile CRNs can have further (non-integer) conservation laws that such an analysis would miss. Conversely, conservation laws are more informative of CRN structure than currently considered, as they pick up on more structural detail. This provides further motivation to extract them from chemical data and to elucidate how that can be done.

Critically, data collected in chemistry (e.g., by UV-VIS) often cannot 'see' all underlying CRN species. A dimensional theory for chemical data must thus account for the effects of what can and cannot be experimentally observed, and what cannot be distinguished. Within the scope of this work, these effects are governed by the concealed species law
$s_{\blacksquare}(\Lambda) = b_{\textifsym{\SquareShadowA}}(\Lambda) + a_{\textifsym{\SquareShadowA}}(\Lambda))$ and isospectral species law $\text{\textsection}(\Lambda) = b_{\text{\textsection}}(\Lambda)  + a_{\text{\textsection}}(\Lambda) + r_\sim(\Lambda)$. We thereby have started establishing bridges between data dimension and underlying CRN features, e.g., independent isosbestic points correspond to single reactions with associated conservation laws.

Our approach lays the foundation towards a CRN-level theory that characterizes the multifarious types of ambiguities and indeterminacies in chemical data and establishes chemical laws that govern them. By construction, such a theory provides deeper insights into the structure and chemical context of the ambiguity problem, the central problem in multivariate curve resolution (MCR). We surmise that this additional structure and context may be leveraged by a new generation of MCR algorithms and data collection protocols.

Importantly, our approach enables new inference methods by which portions of a reaction network are deduced from dimensions of portions of data, similar to how portions of NMR spectra illuminate portions of a molecular structure. We have illustrated that this is indeed possible through four case studies with experimental data from the literature in our companion paper \cite{blokhuis2024ejoc}. Understanding how structure in chemistry introduces structure in data thereby creates new tools allowing to address greater analytical challenges in chemistry. 

\begin{acknowledgments}
A.B. acknowledges support from the EU (Marie-Sklodowska-Curie grant 847675).
A.B. acknowledges fruitful discussions with Sijbren Otto.
We thank Troy Figiel for drawing our attention to the unexplained conserved quantity in Ref. \cite{Liu_PRE_2024}.

\end{acknowledgments}

\appendix

\section{Glossary}
\label{subsection:glossary}

In the context of a recently started effort to clarify and harmonize concepts and notation in stochastic thermodynamics, we clarify below the  notation and conventions that have been adopted.\cite{avanzini2023methods} Our notation closely follows the notation adopted in the framework of Ref \cite{polettini_irreversible_2014}, and the proposed unified notation from the (post)modern thermodynamics lecture notes \cite{avanzini2023methods}, with the following exceptions: \\
- Stoichiometric matrices are denoted by $\mathbb{S}$ \\
- Rate constants are denoted by $\kappa$ (reserving $k$ for indices). \\
We adopt the following additional conventions: \\
- For sub/superscripts not used as integer indices, non-alphanumeric symbols ($\circ, \blacksquare,\triangleright$) are preferred.  \\
- Non-alphanumeric sub/superscripts are chosen that have some relation to the context. E.g. equivalence $\sim$ for isomers, a black box $\blacksquare$ for a concealed subnetwork, a wedge $\wedge$ for quantities (cycles) that are broken / split up,  etc. \\
- Fillable sub/superscripts are preferred when decompositions will be taken (e.g. $\square, \triangleright, \circ, $)
\hspace*{-0.3cm}
\begin{tabular}{ | m{1.1cm} | m{4cm}| m{2cm} | } 
  \hline
  Symbol & Meaning & First appearance \\ 
  \hline
  \rowcolor{pink} \multicolumn{3}{| c |}{stoichiometric matrices ($\mathbb{S}, \nu, \mathbb{P}$)} \\
  \hline
  $\mathbb{S}$ & regular $\mathbb{S}$  & p.~\pageref{equation:decompositionSirr} Eq.\eqref{equation:decompositionSirr} \\ 
  \hline
  $\nu^\oplus$ & $\mathbb{S}$, positive part &  p.~\pageref{equation:decompositionSirr}, Eq.\eqref{equation:decompositionSirr}   \\
  \hline
  $\nu^\ominus$ & $\mathbb{S}$, negative part &  p.~\pageref{equation:decompositionSirr}, Eq.\eqref{equation:decompositionSirr}   \\
  \hline
  $\mathbb{S}_\circ$ & all reactions once & p.~\pageref{equation:stoichmatrixirrex} Eq.\eqref{equation:stoichmatrixrevirrev} \\ 
  \hline
    $\mathbb{S}_{\bowtie}$ & reversible reactions & p.~\pageref{equation:SDEF} Eq.\eqref{equation:SDEF} \\ 
  \hline
    $\mathbb{S}_{\triangleright}$ & irreversible reactions only & p.~\pageref{equation:SDEF} Eq.\eqref{equation:SDEF} \\ 
  \hline
  $\mathbb{S}_{\ffor}$ & all reactions, reversible reactions twice  & p.~\pageref{equation:SDEF_2} Eq.\eqref{equation:SDEF_2} \\ 
  \hline
  $\mathbb{S}_{\blackffor}$ & merged co-production &  p.~\pageref{equation:matrixmerge} Eq. \eqref{equation:matrixmerge} \\
  \hline
  $\mathbb{S}_{\ce{Y}}$ & external subnetwork & p.~\pageref{equation:chemostatdecomposition} Eq. \eqref{equation:chemostatdecomposition} \\
  \hline
  $\mathbb{S}_{\ce{X}}$ & internal subnetwork & p.~\pageref{equation:chemostatdecomposition} Eq. \eqref{equation:chemostatdecomposition} \\
  \hline
  $\mathbb{S}_\blacksquare$ & concealed subnetwork & p.~\pageref{equation:concealdecomposition} Eq. \eqref{equation:concealdecomposition} \\
  \hline
  $\mathbb{S}_\square$ & absorbing (visible) subnetwork & p.~\pageref{equation:concealdecomposition} Eq. \eqref{equation:concealdecomposition} \\
  \hline
  $\mathbb{S}_\textifsym{\RightDiamond}$  & merged isospectral species &  p.~\pageref{equation:towardsisospectral}, Eq. \eqref{equation:towardsisospectral} \\
  \hline
  \rowcolor{pink} \multicolumn{3}{| c |}{other matrices ($\partial, \Lmir$)} \\
\hline
$\partial$  & Incidence matrix for complexes &  p.~\pageref{equation:SL2}, Eq. \eqref{equation:SL2} \\
\hline
$\Lmir$  & Incidence matrix for linkage classes &  p.~\pageref{equation:ftlalmir}, Eq. \eqref{equation:ftlalmir} \\
\hline
   \rowcolor{pink}  \multicolumn{3}{| c |}{species, complexes ($s, \textit{\textvarsigma}$)} \\
  \hline
  $s$ & $\#$ species & p.~\pageref{equation:SL1} Eq.\eqref{equation:SL1} \\ 
  \hline
  $s_{\blacksquare}(\Lambda)$  & $\#$ concealed species & p.~\pageref{equation:specrank} Eq. \eqref{equation:specrank} \\
  \hline
  $s_{\ce{Y}}$ & $\#$ chemostats & p.~\pageref{equation:chemostatlaw}, Eq.\eqref{equation:chemostatlaw} \\
  \hline
  $\textit{\textvarsigma}$ & $\#$ complexes & p.~\pageref{equation:SL2}, Eq.\eqref{equation:SL2} \\
  \hline
  \rowcolor{pink} \multicolumn{3}{| c |}{reactions ($r, \text{\footnotesize{$\Upsilon$}}$)} \\
  \hline
  $r$ & $\#$ reactions & p.~\pageref{equation:SL1} Eq.\eqref{equation:SL1} \\ 
  \hline
    $r_{\triangleright}$ & $\#$ irreversible reactions & p.`~\pageref{SL1irrb} Eq.\eqref{SL1irrb} \\
  \hline
  $r_{\ffor}$ & $\#$ reactions, counting reversibility twice & p.~\pageref{SL1irrb} Eq.\eqref{SL1irrb} \\
  \hline
  $r_\circ$ & $\#$ reactions, counting reversibility once & p.~\pageref{SL1irrb} Eq.\eqref{SL1irrb} \\
  \hline
$r_{\bowtie}$ & $\#$ reversible reactions & p.~\pageref{equation:reactrevirr} Eq.\eqref{equation:reactrevirr} \\
  \hline
  $\copro$ & Co-production index &  p.~\pageref{equation:co-production_a} Eq. \eqref{equation:co-production_a} \\
  \hline
    $\Phi$ & $\#$ merged reactions, incl. reversible reactions &  p.~\pageref{equation:phi_react} Eq. \eqref{equation:phi_react} \\
  \hline
  $r^+$ & $\#$ irreversible reactions with explicit reactant  &  p.\pageref{equation:rplus} Eq. \eqref{equation:rplus} \\
  \hline
  $r_\sim (\Lambda)$ & $\#$ independent direct isomerization reactions  among isospectral species & p.\pageref{equation:isospectralspecieslaw} Eq.\eqref{equation:isospectralspecieslaw} \\ 
   \hline
\end{tabular}

\hspace*{-1cm}
\begin{tabular}{ | m{1.1cm} | m{4cm}| m{2cm} | } 
\hline 
  Symbol & Meaning & First appearance \\ 
      \hline
  \rowcolor{pink} \multicolumn{3}{| c |}{conserved quantities, linkage classes ($\ell, b, \lambda$)} \\
  \hline
  $\ell$ & $\#$ conserved quantities & p.~\pageref{equation:SL1} Eq.\eqref{equation:SL1} \\ 
  \hline
  $\ell_\circ$ & $\#$ conserved quantities, $\mathbb{S}_\circ$ & p.~\pageref{SL1irrb} Eq.\eqref{SL1irrb} \\ 
  \hline
  $\rotatebox[origin=c]{180}{e}_{\blackwhiteffor}$ &  $\#$ Co-production emanants, co-production conservation laws &  p.~\pageref{equation:matrixmerge} Eq. \eqref{equation:matrixmerge} \\
  \hline
  $\ell_{||}(\Lambda)$ & $\#$ further collinearities in absorbtion spectra &  p.~\pageref{equation:specrank} Eq. \eqref{equation:specrank} \\
  \hline
  $b_{\textifsym{\SquareShadowA}}(\Lambda)$ & $\#$ concealed conservation laws &  p.~\pageref{equation:concealedspecieslaw} Eq. \eqref{equation:concealedspecieslaw} \\
    \hline
$\lambda$ & $\#$ linkage classes &  p.~\pageref{equation:SL2} Eq. \eqref{equation:SL2} \\
    \hline
$\lambda_{\bowtie}$ & $\#$ static linkage classes &  p.~\pageref{equation:FTLAcomplexes} Eq. \eqref{equation:FTLAcomplexes} \\
    \hline
$\lambda_{\bowtie}^0$ & $\#$ static linkage classes with no outgoing reactions &  p.~\pageref{equation:FTLAcomplexes} Eq. \eqref{equation:FTLAcomplexes} \\
    \hline
  \rowcolor{pink} \multicolumn{3}{| c |}{cycles ($c, \Lambda, a$)} \\
  \hline
  $c_\circ$ & $\#$ cycles, reversible CRN & p.~\pageref{SL1irrb} Eq.\eqref{SL1irrb} \\ 
  \hline
   $c$ & $\#$ cycles & p.~\pageref{equation:SL1} Eq.\eqref{equation:SL1} \\ 
  \hline 
  $\wedge_{\blackwhiteffor}$ &  $\#$ broken cycles due to coproduction &  p.~\pageref{Delta_SL1b} Eq. \eqref{Delta_SL1b} \\
  \hline 
  $\sqsupset$ &  $\#$ cycles within a linkage class &  p.~\pageref{equation:SL2} Eq. \eqref{equation:SL2} \\
  \hline  
  $a_{\textifsym{\SquareShadowA}}(\Lambda)$ & $\#$ apparent cycles &  p.~\pageref{equation:concealedspecieslaw} Eq. \eqref{equation:concealedspecieslaw} \\
  \hline
  \rowcolor{pink} \multicolumn{3}{| c |}{dimension indices ($d, I, \text{\textsection}$)} \\
  \hline
  $d$ &  data dimension &  p.~\pageref{equation:dimfromvar} Eq. \eqref{equation:dimfromvar} \\
  \hline
$\text{\textsection}(\Lambda)$ & Isospectral index & p.~\pageref{equation:specrank} Eq. \eqref{equation:specrank} \\
  \hline
  \rowcolor{pink} \multicolumn{3}{| c |}{other quantities } \\
  \hline
  $A_\lambda$ &  absorbance &  p.~\pageref{equation:lambert-beer} Eq. \eqref{equation:lambert-beer} \\  
  \hline
  $L_{p}$ &  path length &  p.~\pageref{equation:lambert-beer} Eq. \eqref{equation:lambert-beer} \\
  \hline
  $[\ce{X_k}]$ &  amount concentration &  p.~\pageref{equation:lambert-beer} Eq. \eqref{equation:lambert-beer} \\
  \hline
  $\epsilon_k (\lambda)$ & molar extinction coefficient &  p.~\pageref{equation:lambert-beer} Eq. \eqref{equation:lambert-beer} \\
  \hline
  $\mathbb{E}(\Lambda)$ & Molar Extinction Matrix & p.~\pageref{equation:matrixE} Eq. \eqref{equation:matrixE} \\
  \hline
\end{tabular}

  

\section{A hidden conserved quantity in an atmospheric chemistry model}
\label{section:atmosphericexample}

As a final application example, we revisit the atmospheric model considered in  recent studies\cite{sturm_conservation_2022,Liu_PRE_2024}, in which a non-integer quantity that appeared to be conserved was discovered numerically. It was unclear whether this was an approximate conservation law, or a genuine one, and no clear interpretation could be given for its non-integer nature. We address both questions below by showing that it is an instance of co-production conservation.

The model starts by considering the irreversible reactions
\bea
\ce{NO2} &+& h\nu \rightarrow \ce{NO} + \ce{O} \\
\ce{O} &+& \ce{O2} \rightarrow \ce{O3}  \\
\ce{O3} &+& \ce{NO} \rightarrow \ce{NO2} + \ce{O2}  \\
\ce{HCHO} &+& 2 \ce{O2} + h\nu \rightarrow 2 \ce{HO}^\bullet_2 + \ce{CO}  \\
\ce{HCHO} &+& h\nu \rightarrow \ce{H2} + \ce{CO} \\
\ce{HCHO} &+& \ce{HO}^\bullet \rightarrow \ce{HO}^\bullet_2 + \ce{CO} + \ce{H2O} \\
\ce{HO}^\bullet_2 &+& \ce{NO} \rightarrow \ce{HO}^\bullet + \ce{NO2} \\
\ce{HO}^\bullet &+& \ce{NO2} \rightarrow \ce{HNO3} \\
\ce{HO2H} &+& h\nu \rightarrow 2 \ce{HO}^\bullet \\
\ce{HO2H} &+& \ce{HO}^\bullet \rightarrow \ce{H2O} + \ce{HO}^\bullet_2 
\eea
It is further supposed that $\ce{H_2O}$ is not monitored, and that $\ce{O2}$ is a reservoir species (chemostatted). The subnetwork thus afforded is 
\bea
\ce{NO2} + h\nu &\rightarrow& \ce{NO} + \ce{O} \\
\ce{O} &\rightarrow& \ce{O3}  \\
\ce{O3} + \ce{NO} &\rightarrow& \ce{NO2}   \\
\ce{HCHO} + h\nu &\rightarrow& 2 \ce{HO}^\bullet_2 + \ce{CO}  \\
\ce{HCHO} + h\nu &\rightarrow& \ce{H2} + \ce{CO} \\
\ce{HCHO} + \ce{HO}^\bullet &\rightarrow& \ce{HO}^\bullet_2 + \ce{CO} \\
\ce{HO}^\bullet_2 + \ce{NO} &\rightarrow& \ce{HO}^\bullet + \ce{NO2} \\
\ce{HO}^\bullet + \ce{NO2} &\rightarrow& \ce{HNO3} \\
\ce{HO2H} + h\nu &\rightarrow& 2 \ce{HO}^\bullet \\
\ce{HO2H} + \ce{HO}^\bullet &\rightarrow&  \ce{HO}^\bullet_2 
\eea
Since $\ce{H2O}$ only occurs as a sink species, the effects of concealing it and chemostatting it are equivalent ($b_{\ce{X}}=1$ or $b_\square=1$). 

We will now see how coproduction emerges: we remove $\ce{O}_2$ and $\ce{H_2O}$ from the description by chemostatting. By the chemostatting of $\ce{O}_2$, reactions $r_4$ and $r_5$ have become collinear, hence $\copro = 1$. The chemostat law (i.e., Eq. \eqref{equation:chemostatlaw})  states that $s^{\ce{Y}}=2$ reservoirs are introduced, and here one can readily check that $b=2$ stoichiometric conservation laws are broken (i.e., hydrogen and oxygen conservation being lost). Since $\rotatebox[origin=c]{180}{e}_{\blackwhiteffor} = 1$, the number of conservation laws is only reduced by 1 rather than 2. The subnetwork with the highlighted collinear reactions is depicted in Fig. \ref{fig:atmosphericnetwork}.

\begin{figure}[tbhp!]
\centering
\includegraphics[width=1.0\linewidth]{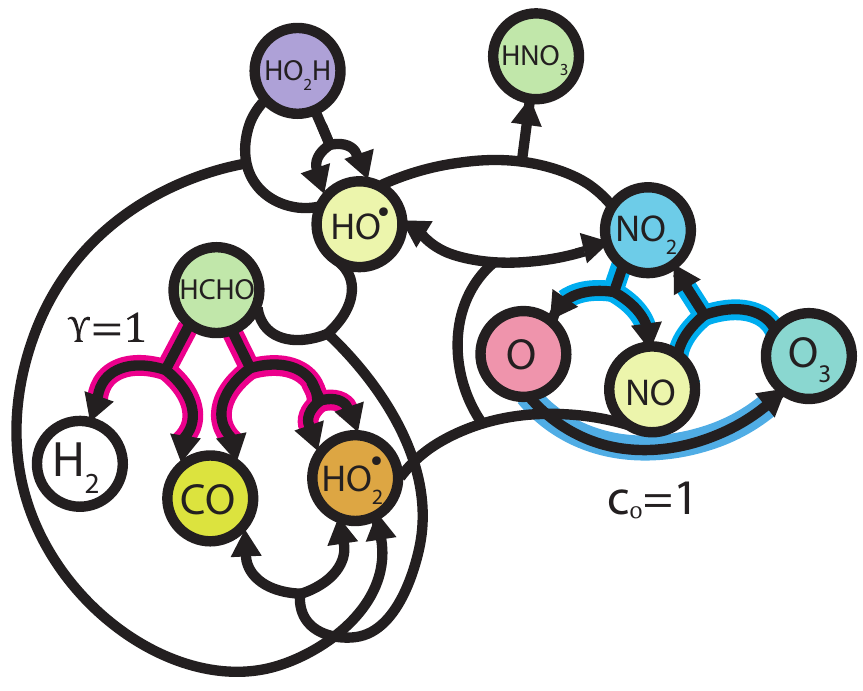}
\caption{Figure \ref{fig:atmosphericnetwork}, reproduced for convenience. An Atmospheric CRN model with co-production index $\copro = 1$. Collinear reactions are highlighted in pink. These do not include cycle reactions -  highlighted in blue - and, hence, no cycles break upon merging collinear reactions $\wedge_{\blackwhiteffor}=0$.  $\rotatebox[origin=c]{180}{e}_{\blackwhiteffor} = 1$  co-production conservation law then results from the co-production law $\copro= \rotatebox[origin=c]{180}{e}_{\blackwhiteffor} + \wedge_{\blackwhiteffor}$, confirming the newly found law $\text{CQ}_3$ marks the detection of a genuine conservation law. }
\label{fig:atmosphericnetworkb}
\end{figure}

The subnetwork satisfies $\mathbb{S}_{\ffor} = \mathbb{S}_{\triangleright}$, with stoichiometric matrix
\be
\mathbb{S}_{\triangleright} = \left(\begin{smallmatrix} 0 & 1 & -1 & 0 & 0 & 0 & 0 & 0 & 0 & 0 \\
1 & 0 & -1 & 0 & 0 & 0 & -1 & 0 & 0 & 0  \\
-1 & 0 & 1 & 0 & 0 & 0 & 1 & -1 & 0 & 0 \\
0 & 0 & 0 & -1 & -1 & -1 & 0 & 0 & 0 & 0 \\
0 & 0 & 0 & 2 & 0 & 1 & -1 & 0  & 0 & 1 \\
0 & 0 & 0 & 0 & 0 & 0 & 0 & 0 & -1 & -1 \\
0 & 0 & 0 & 0 & 0 & -1 & 1 & -1 & 2 & -1 \\
1 & -1 & 0 & 0 & 0 & 0 & 0 & 0 & 0 & 0 \\
0 & 0 & 0 & 0 & 0 & 0 & 0 & 1 & 0 & 0 \\
0 & 0 & 0 & 1 & 1 & 1 & 0 & 0 & 0 & 0 \\
0 & 0 & 0 & 0 & 1 & 0 & 0 & 0 & 0 & 0
\end{smallmatrix}\right)
\ee
Which has $\text{rk} \left( \mathbb{S}_\triangleright \right) = 9$. Since $r_{\triangleright}=10$, there is $c_\circ=1$ cycle, involving $r_1$ to $r_3$ 
\be
\pmb{c} = (1,1,1,0,0,0,0,0,0,0)
\ee
which is highlighted in Fig. \ref{fig:atmosphericnetworkb}.

As reactions $R_4$ and $R_5$ are now collinear, we merge them
\be
p R_4 + (1-p) R_5 , \ \ \ p = \frac{k_4}{k_4 + k_5}
\ee
which becomes
\be
\ce{HCHO} + hv \rightarrow 2 p \ce{HO2}^\bullet + \ce{CO} + \left(1-p\right) \ce{H2}
\ee
So that our stoichiometric matrix $\mathbb{S}_{\blackffor}$ in terms of independent reactions becomes
\be
\mathbb{S}_{\blackffor} = \left(\begin{smallmatrix} 
0 & 1 & -1 & 0 & 0 & 0 & 0 & 0 & 0 \\
1 & 0 & -1 & 0 & 0 & -1 & 0 & 0 & 0  \\
-1 & 0 & 1 & 0 & 0 & 1 & -1 & 0 & 0 \\
0 & 0 & 0 & -1 & -1 & 0 & 0 & 0 & 0 \\
0 & 0 & 0 & 2 p & 1 & -1 & 0  & 0 & 1 \\
0 & 0 & 0 & 0  & 0 & 0 & 0 & -1 & -1 \\
0 & 0 & 0 & 0  & -1 & 1 & -1 & 2 & -1 \\
1 & -1 & 0 & 0  & 0 & 0 & 0 & 0 & 0 \\
0 & 0 & 0 & 0  & 0 & 0 & 1 & 0 & 0 \\
0 & 0 & 0 & 1  & 1 & 0 & 0 & 0 & 0 \\
0 & 0 & 0 & 1-p & 0 & 0 & 0 & 0 & 0 \end{smallmatrix}\right). \nonumber
\ee
As we have not merged any cycle reactions, no cycles are lost $\wedge_{\blackwhiteffor} = 0$. This is readily seen by making the cycle vector $\pmb{c}$ one reaction shorter 
\bea
\pmb{c}_{\blackffor} &=& (1,1,1,0,0,0,0,0,0)^T, \\
\mathbb{S}_{\blackffor} \pmb{c}_{\blackffor} &=& \pmb{0}.
\eea
Thus, co-production $\text{\footnotesize{$\Upsilon$}} = 1$ then yields $\rotatebox[origin=c]{180}{e}_{\bullet \!  \! \circ}=1$, as evidenced by an additional left nullvector
\be
\pmb{\ell}_{\blackffor}^{(3)} = \left(6,-5,1,3,9,6,3,6,4,-3,\frac{6 - 18 p}{1-p} \right)^T
\ee
To see if this is the elusive hidden conservation law\footnote{$\text{CQ}_3$ was taken from the Supp. Matt. of Ref\cite{Liu_PRE_2024} where it is derived. It differs from the expression for $\text{CQ}_3$ due to a small typo for the coefficient of [$\ce{OH}^\bullet$]} $\text{CQ}_3$ reported in the literature\cite{Liu_PRE_2024}, we substitute ($p=0.40541.. $):
\bea
\pmb{\ell}_{\blackffor}^{(3)} &=& \left(6,-5,1,3,9,6,3,6,4,-3, 2.18.. \right)^T \nonumber \\
CQ_3 &\approx& \left(6,-5,1,3,9,6,3,6,4,-3, 2.21 \right)^T \nonumber
\eea
The numerical estimate of the measured quantity $\text{CQ}_3$ thus closely approximates the genuine conservation law. We can thus confirm that the observation of $\text{CQ}_3$ was due to a genuine conserved quantity in the (chemostatted) model. The existence and exact nature - co-production conservation - of this conservation law has thereby been clarified. \\

In general, the SID algorithm proposed in that work\cite{Liu_PRE_2024} detects conservation laws of first order differential equations $d_t \pmb{x} = \pmb{f}(\pmb{x})$ by creating a list of $K$ independent phase-space functions  $\pmb{b} = b^{(1)}(\pmb{x}),...,b^{(K)}(\pmb{x})$. $H(\pmb{x})= \pmb{\theta}^T b(\pmb{x})$  is then a conserved quantity, if, for all
$\pmb{x}$:\footnote{As SID is a numerical algorithm, this is done in practice by checking $P$ randomly chosen points. As long as the phase space functions and the flow equations are polynomial, as is the case here, sufficiently large $P$ will ensure that $H(x)$ is conserved for all $\pmb{x}$} 
\bea
g(\pmb{x})^T \pmb{\theta} &=& 0, \\
g(\pmb{x}) &=& \nabla \pmb{b} \pmb{f}(\pmb{x}). 
\eea
Applied to CRNs where $\pmb{x} = [\pmb{\ce{X}}]$, $\pmb{f}(\pmb{x}) = \mathbb{S} \pmb{J} $ and with a list of candidate functions $\pmb{b} = [\pmb{\ce{X}}]$, the above equation reduces to (the transpose of) the condition $\pmb{\ell}^T \mathbb{S} \pmb{J} = 0$ mentioned before. We conclude that the analysis provided herein is sufficient to explain all true linear conservation laws the SID algorithm might find in a CRN. There can be further conservation laws due to special choices of parameters (Sec. \ref{appendix:parametric_collinearity}).   \\


\section{Example: 
\texorpdfstring{$\wedge_{\blackwhiteffor}$}{L} vs \texorpdfstring{$\rotatebox[origin=c]{180}{e}_{\blackwhiteffor}$}{lt}}
\label{appendix:wedge}

Coproduction can result in an emergent conservation law or a broken cycle. Here we illustrate a minimal example of a broken cycle.

\begin{figure}[tbhp!]
\centering
\includegraphics[width=1.0\linewidth]{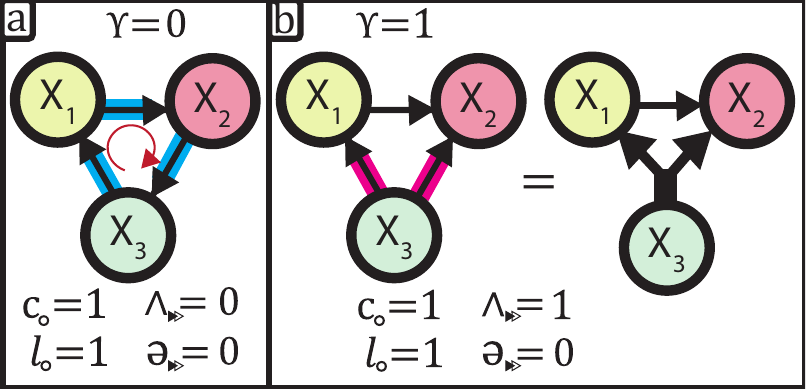}
\caption{a) a cyclic CRN composed of irreversible reactions. Cycle reactions are highlighted in blue. Reversing any of the reaction arrows in this cyclic CRN affords CRN b) with co-production index 1. After merging collinear reactions, the original reaction cycle can no longer be performed. The networks depicted are equal '=' with respect to their description of a dynamical system on a macroscopic level: they represent the exact same ODEs.}
\label{fig:a broken cycleb}
\end{figure}

For this, we first start with a  CRN displaying the cycle to be broken
\be
\ce{X_1} \overset{1}{\rightarrow} \ce{X_2} \overset{2}{\rightarrow} \ce{X_3} \overset{3}{\rightarrow} \ce{X_1} ,
\ee
for which 
\be
\mathbb{S}_{\blackffor} = \begin{pmatrix} -1 & 0 & 1  \\ 
1 & -1 & 0  \\
0 & 1 & -1 \end{pmatrix}.
\ee
Here, there are no reactions to be merged, and so $\text{\footnotesize{$\Upsilon$}} = \rotatebox[origin=c]{180}{e}_{\blackwhiteffor}  = \wedge_{\blackwhiteffor}
 = 0$. Furthermore, there is $c=1$ cycle and $\ell_\circ=1$ conservation law
\bea
\pmb{c} &=& (1,1,1)^T \\
L^{(1)} &=&  [\ce{X_1}] + [\ce{X_2}] + [\ce{X_3}]
\eea

Reversing any of the reactions will yield 
\bea
\ce{X_1} \overset{1}{\rightarrow} \ce{X_2} \overset{2}{\leftarrow} \ce{X_3} \overset{3}{\rightarrow} \ce{X_1} \label{equation:crnbroken},\\
\mathbb{S}_\triangleright = \begin{pmatrix} -1 & 0 & 1  \\ 
1 & 1 & 0  \\
0 & -1 & -1 \end{pmatrix}.
\eea
For which a cycle (right nullvector) is $\pmb{c} = (1,-1,1)^T$. 
We find that $\text{\footnotesize{$\Upsilon$}}=1$, because reactions $R_2$ and $R_3$ are now collinear. Upon merging these we obtain
\bea
\ce{X_1} \overset{1}{\rightarrow} \ce{X_2},&& \ce{X_3} \overset{2}{\rightarrow} p \ce{X_1} + \left(1-p \right) \ce{X_2}, \nonumber \ \ \\
\mathbb{S}_{\blackffor} &=& \begin{pmatrix} 
-1 & p  \\ 
1 & -1   \\
0 & 1-p \end{pmatrix}.
\eea
and by this merger $\wedge_{\bullet \!  \! \circ} = 1$ cycle is thus lost.


\section{Applying laws for isospectral and concealed species}
\label{appendix:isospectral_concealed}

\begin{figure}[tbhp!]
\centering
\includegraphics[width=1.0\linewidth]{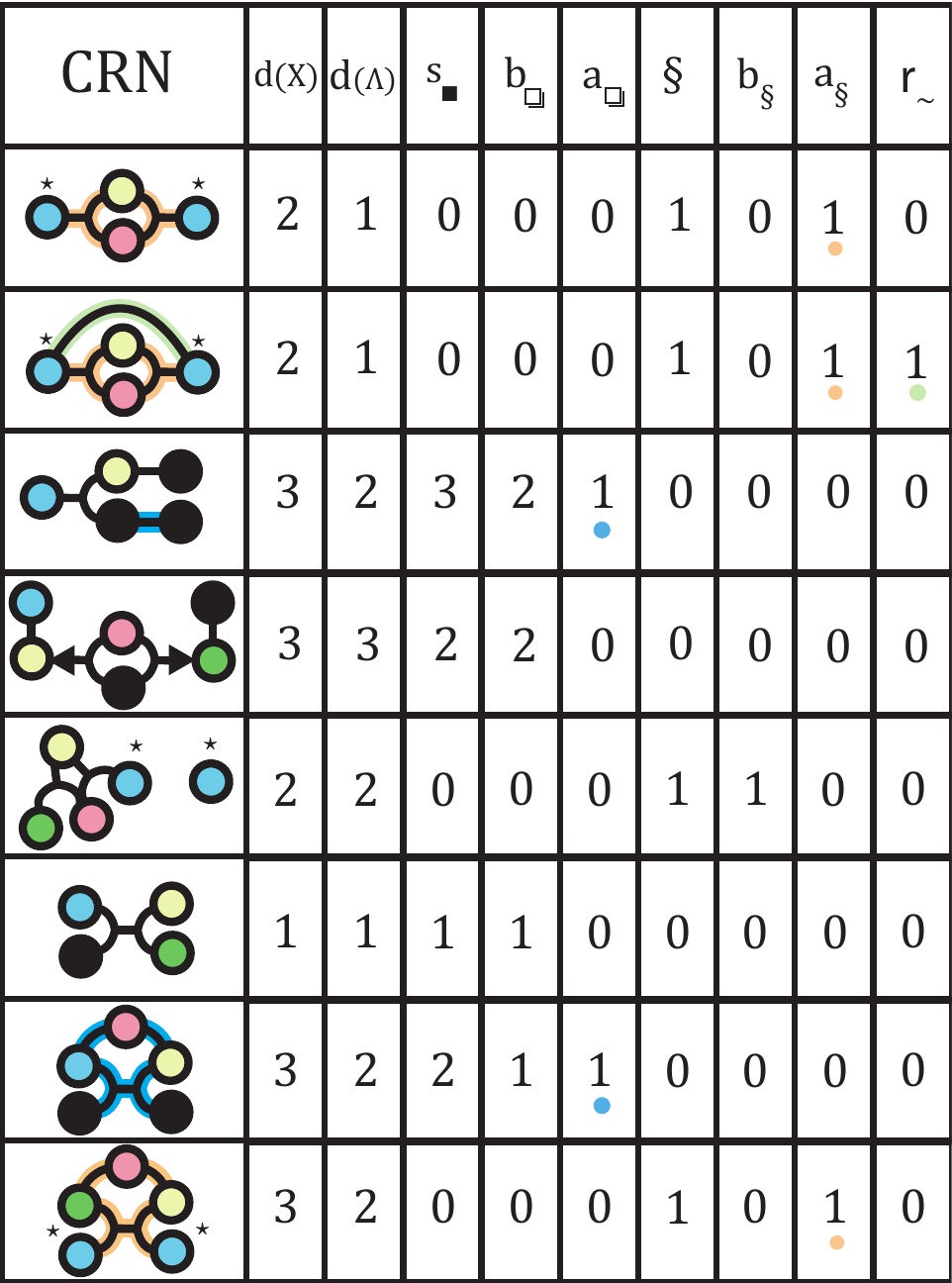}
\caption{Illustration of laws for isospectral (Eq. \eqref{equation:isospectralspecieslaw}) and concealed (Eq. \eqref{equation:concealedspecieslaw}) species for several CRNs. Dark nodes correspond to spectroscopically inactive species. Species that are indistinguishable are marked with a star. New cycles - leading to loss of data dimension - are highlighted in orange and blue. Data dimensions: $d(\ce{X}), d(\Lambda)$ (spectral), $d(\ce{X})$ (species). $d(\Lambda)$ spectral data dimension, $s_{\blacksquare}(\Lambda): \#$ concealed species, $b_{\square} (\Lambda): \#$ concealed conservation laws, $a_{\square} (\Lambda): \#$ apparent cycles, $\text{\textsection}(\Lambda):$ isospectral index, $r_\sim (\Lambda): \#$ isospectral isomerization reactions, $b_{\text{\textsection}}(\Lambda): \#$ isospectral broken conservation laws, $a_{\text{\textsection}}(\Lambda) \#$, isospectral emergent cycles.}
\label{fig:effectiveCRNextended}
\end{figure}

Fig. \ref{fig:effectiveCRNextended} furnishes further examples of the application of the isospectral species law and the concealed species law.

\section{Parametric collinearity}
\label{appendix:parametric_collinearity}
We have considered collinearity that is due to structure alone. For certain combinations of parameters and initial conditions, a CRN may become symmetric and behave like a lower-dimensional CRN. As a concrete example, we 
consider
\be
\ce{C} \overset{1}{\leftrightarrows} \ce{A} + \ce{B} \overset{2}{\rightleftarrows} \ce{D} .
\ee
For which we previously considered that we have $r_\circ=2$ linearly independent currents
\bea
J_{1,\circ} &=& \kappa_1 [\ce{A}] [\ce{B}] - \kappa_3 [\ce{C}] = J_{1,\ffor} - J_{3,\ffor} \nonumber \\ 
J_{2,\circ} &=& \kappa_2 [\ce{A}] [\ce{B}] - \kappa_4 [\ce{D}] = J_{2,\ffor} - J_{4,\ffor} \nonumber
\eea

and, hence, \textit{a priori}, no collinearity should exist. If we now fix $[\ce{C}]_0 = [\ce{D}]_0$, $\kappa_1=\kappa_3$, $\kappa_2=\kappa_4$, then the rates become collinear $J_{1,\circ} = J_{2,\circ}$ and dynamics for $[\ce{C}]$ and $[\ce{D}]$ become equivalent, hence we have a conserved quantity: 
\be
L = [\ce{C}]-[\ce{D}] = 0.
\ee 
This relates to a symmetry in the CRN $\ce{C} \leftrightarrow \ce{D}$, which vanishes when we let $[\ce{C}]_0 \neq [\ce{D}]_0$. 

This requirement of 3 parametric constraints is readily attained in chemical practice: if $\ce{C}$ and $\ce{D}$ are enantiomers in an achiral environment, then the 2 constraints on rate constants will naturally be verified. The constraint $[\ce{C}]_0 = [\ce{D}]_0$ would require that we either add $\ce{C}$ and $\ce{D}$ as a racemic mixture, or do not add them at all. We refer to Ref. \cite{laurent_emergence_2021,Laurent2022} for pertinent further decompositions of CRNs with symmetries due to chiral species. 

Since parametric collinearity occurs for specific choices of rate constants and initial conditions, it will not generically be detected by algorithmic procedures that seek out conserved quantities. 

\section{Change in fixed-point stability}
\label{appendix:fixedpointstab}
Special instances of parametric co-production can occur for a fixed point that changes in stability upon variation of a parameter $p$. If the Jacobian $\mathbb{J}(\pmb{x},p)$ acquires a single positive eigenvalue upon variation of $p$ by the passage of a negative eigenvalue through the origin at a point $p^*$, then at $p^*$ the Jacobian has an additional eigenvalue $0$. This implies $J(\pmb{x}^*,p^*)$ has an additional left nullvector $\pmb{\ell}^*$, such that locally
\bea
\pmb{\ell}^* \mathbb{J}(\pmb{x}^*,p^*) = 0 , \\
\pmb{\ell}^* \Delta \pmb{x} = \pmb{\ell}^* \mathbb{J}(\pmb{x}^*,p^*) \Delta \pmb{x} = 0 .
\eea
I.e. a change in fixed-stability that goes through the origin yields a parametric conservation law. This law applies locally, but may also have global validity. For linear systems, $\mathbb{J}$ does not depend on $\pmb{x}^*$ and the law is always global. Consider 
\bea
\ce{A} \overset{1}{\rightarrow} \ce{B} \overset{2}{\rightarrow} \ce{C} \overset{3}{\rightarrow} 2 \ce{A}, 
\ce{C} \overset{\emptyset}{\rightarrow} \emptyset
\eea
which exhibits one cycle ($c=1$). 
Recognizing $\copro = 1$, we merge reactions and obtain 
\bea
&\ce{A} \rightarrow \ce{B} \rightarrow \ce{C} \rightarrow 2 p \  \ce{A}, \ \ (0<p<1) \\
&p = \frac{k_3}{k_3 + k_\emptyset}
\eea
for $p<1/2$ and $p>1/2$, we now have a broken cycle $\wedge_{\blackwhiteffor} = 1$. For $p=1/2$, we do have a cycle, and from the coproduction law \eqref{Delta_SL1b} it then follows that we have a co-production emanant $\rotatebox[origin=c]{180}{e}_{\blackwhiteffor}=1$ instead, which for this simple example becomes 
\bea 
\rotatebox[origin=c]{180}{\pmb{e}} &=& (1,1,1) , \\
L^* &=& [\ce{A}] + [\ce{B}] + [\ce{C}] .
\eea
The Jacobian $\mathbb{J}$ is
\bea
\mathbb{J}(\pmb{x}^*) &=& \begin{pmatrix} -k_1 & 0 & 2 k_3 \\
k_1 & -k_2 & 0 \\
0 & k_2 & - k_3 - k_\emptyset
\end{pmatrix} \\
&=& \begin{pmatrix} -k_1 & 0 & 2 k_3 \nonumber \\
k_1 & -k_2 & 0 \\
0 & k_2 & - \frac{k_3}{p}
\end{pmatrix}
\eea
We can readily check that at the critical point $p^*=1/2$, $\rotatebox[origin=c]{180}{\pmb{e}}$ is a left nullvector of $\mathbb{J}$
\bea
\rotatebox[origin=c]{180}{\pmb{e}} \ \mathbb{J}(\pmb{x}^*,p^*) &=& (1,1,1) \begin{pmatrix} -k_1 & 0 & 2 k_3 \\
k_1 & -k_2 & 0 \\
0 & k_2 & - 2 k_3
\end{pmatrix} \nonumber \\ 
&=& (0,0,0) 
\eea
Another possibility to pass from stability to instability is for a pair of eigenvalues to traverse the imaginary axis (i.e. a Hopf bifurcation) \cite{Vassena2025a}, for which no conservation law emerges. In this sense, the presence or absence of parametric conservation laws can be a diagnostic structural feature for system behavior.

\section{Co-production beyond mass-action}
\label{appendix:co-production beyond mass action}

A key prerequisite for coproduction is that currents can become linearly dependent, which under mass-action naturally allows to relate co-production to reaction stoichiometry. This prerequisite is not satisfied in every setting, for instance parameter-rich kinetics\cite{Vassena2023symbolic} axiomatically excludes such collinearities through additional parameters. 

Nevertheless, coproduction is not restricted to mass action. Moreover, in chemical practice, many non-mass action rate expressions are approximations derived from mass-action expressions. When we perform those same derivations starting from a CRN with co-production, we find that co-production is readily conserved. As such, co-production is not limited to mass action and can be encountered in many further settings. We provide a few derivations below.  

\subsection{Michaelis-Menten mechanism}

Consider the following enzymaticatically catalyzed co-production scheme, involving two reactions described as irreversible
\bea
\! \! \! \! \! \! \! \! &\mathbb{S}_{\circ}: \ce{X} + \ce{E} \overset{1}{\leftarrow} \ce{EX} \overset{2}{\leftrightarrows} \ce{E} + \ce{S} \overset{3}{\leftrightarrows} \ce{EY} \overset{4}{\rightarrow} \ce{Y} + \ce{E} . \nonumber \\
\! \! \! \! \! \! \! \! &\mathbb{\partial}_{\circ}: \Gamma_1 \overset{1}{\leftarrow} \Gamma_2 \overset{2}{\leftrightarrows} \Gamma_3 \overset{3}{\leftrightarrows} \Gamma_4 \overset{4}{\rightarrow} \Gamma_5 . \nonumber \\
\! \! \! \! \! \! \! \! &\mathbb{\partial}_{\bowtie}: \Gamma_1 , \ \ \ \Gamma_2 \overset{2}{\leftrightarrows} \Gamma_3 \overset{3}{\leftrightarrows} \Gamma_4, \  \  \Gamma_5 . \nonumber 
\eea
From the CLL (Eq. \eqref{equation:copro-linkage-law}) we find $r^+=2, \lambda_{\bowtie} = 3, \lambda_{\bowtie}^{0} = 2$, and thus $\text{\footnotesize{$\Upsilon$}}=1$.  Since $r=4, c=0, \text{\footnotesize{$\Upsilon$}}=1$, we have $d=3$.

Now consider we have as our approximation the usual Michealis-Menten setup
\bea
\! \! \! \! \! \! \! \! d_t [\ce{EX}] &=& -\left(\kappa_1 + \kappa_2^-\right) [\ce{EX}] + \kappa_2^+ [\ce{E}][\ce{S}] = 0 \nonumber \\
\! \! \! \! \! \! \! \! d_t [\ce{EY}] &=& -\left(\kappa_4 + \kappa_3^-\right) [\ce{EY}] + \kappa_3^+ [\ce{E}][\ce{S}] = 0 \nonumber \\
\ \! \! \! \! \! \! \! \! [\ce{E}]^0 &=& [\ce{E}] \left(1 + \frac{\kappa_2^+ [\ce{S}]}{\kappa_1 + \kappa_2^-} + \frac{\kappa_3^+ [\ce{S}]}{\kappa_4 + \kappa_3^-} \right)
\eea
And we now have as an approximate description
\bea
\ce{X} &+& \ce{E} \overset{1}{\leftarrow} \ce{E} + \ce{S}  \overset{4}{\rightarrow} \ce{Y} + \ce{E} .  \\
J_1 &=& \kappa_1 [\ce{EX}] = \frac{\kappa_1 [\ce{E}]^0 [\ce{S}]}{\left(1 + \frac{\kappa_2^+ [\ce{S}]}{\kappa_1 + \kappa_2^-} + \frac{\kappa_3^+ [\ce{S}]}{\kappa_4 + \kappa_3^-} \right)} \nonumber \\
J_4 &=& \kappa_4 [\ce{EY}] = \frac{\kappa_4 [\ce{E}]^0 [\ce{S}]}{\left(1 + \frac{\kappa_2^+ [\ce{S}]}{\kappa_1 + \kappa_2^-} + \frac{\kappa_3^+ [\ce{S}]}{\kappa_4 + \kappa_3^-} \right)}  \nonumber 
\eea
Again, we have $\text{\footnotesize{$\Upsilon$}}=1$, and thus now $d=1$. The dimension reduction by coproduction is here preserved upon taking the Michaelis-Menten approximation. 

\subsection{Lindemann-Haldane mechanism}

The Haldane-Lindemann scheme describes the activation and deactivation of a species by a molecular collision (usually in the gas phase) with a species $\ce{M}$, with the activated species undergoing a further unimolecular reaction. A coproduction analogue is then 
\bea
&\mathbb{S}_{\circ}: \ \ce{A} + \ce{M} \overset{1}{\leftrightarrows} \ce{A}^* + \ce{M}, \\
&\ce{X} \overset{2}{\leftarrow} \ce{A}^* \overset{3}{\rightarrow} \ce{Y}. \\
&\partial_{\circ}: \ \ \Gamma_1 \overset{1}{\leftrightarrows} \Gamma_2, \ \ \ \Gamma_4 \overset{2}{\leftarrow} \Gamma_3 \overset{3}{\rightarrow} \Gamma_5 \\
&\partial_{\bowtie}: \ \ \Gamma_1 \overset{1}{\leftrightarrows} \Gamma_2, \ \ \ \Gamma_4 , \ \ \ \Gamma_3 , \ \ \ \Gamma_5
\eea
From the CLL (Eq. \eqref{equation:copro-linkage-law}) we find $r^+=2, \lambda_{\bowtie} = 4, \lambda_{\bowtie}^{0} = 3$, hence $\text{\footnotesize{$\Upsilon$}}=1$. Since $r=3, c=0, \text{\footnotesize{$\Upsilon$}}=1$ we have $d=2$.

We may now write
\bea
d_t [\ce{A}^*] &=& \kappa_1^+ [\ce{A}] [\ce{M}] - \kappa_1^- [\ce{A}^*] [\ce{M}] \nonumber \\  &-&  \kappa_2 [\ce{A}^*] - \kappa_3 [\ce{A}^*] = 0 \nonumber \\
\ [\ce{A}^*] &=& \frac{\kappa_1^+ [\ce{A}][\ce{M}] }{\kappa_1^- [\ce{M}] + \kappa_2  + \kappa_3 }
\eea
and this affords effective expressions for the currents
\bea
J_2 &=& \kappa_2 \frac{\kappa_1^+ [\ce{A}][\ce{M}] }{\kappa_1^- [\ce{M}] + \kappa_2  + \kappa_3 } \\
J_3 &=& \kappa_3 \frac{\kappa_1^+ [\ce{A}][\ce{M}] }{\kappa_1^- [\ce{M}] + \kappa_2  + \kappa_3 } 
\eea
which retains $\text{\footnotesize{$\Upsilon$}}=1$.

It may be noted more generally that these derivations involve making approximations to yield an expression for some quasi-stationary concentration, which is then substituted. Clearly, this substitution cannot lift the collinearity. By the same token it is easy to verify that analogous arguments hold to preserve coproduction in other commonly encountered rate laws, for instance in the Eigen-Wilkins mechanism (associative substitution) and in dissociative substitution.

\section{Markov Processes and co-production}
\label{appendix:markov-processes}

A Markov process (with countable state space) can always be mapped to a CRN by mapping states to species. Our results, such as the co-production law - thus naturally apply to Markov processes. We will illustrate some applications of co-production below.

\subsection{1d Biased diffusion}
Let us consider a biased random walk on a 1d lattice with n sites, with site 0 and site $n-1$ being absorbing states. 
When jumping away from a site $[k]$, we jump to $[k-1]$ with probability $p$, and to site $[k+1]$ with probability $(1-p)$.
We can represent the two possible jumps to neighboring sites as irreversible reactions, which we can merge
\bea
\! \! \! \! \! [k-1] &\leftarrow& [k] \rightarrow [k+1], \ \ \ (1 \leq k \leq n-1) \nonumber \\ 
\! \! \! \! \! [k] &\rightarrow& p  [k-1] + (1-p) [k+1], \ \ 
\eea
There are $n-2$ sites where we can jump away from, and we can thus perform $n-2$ such mergers.
A system with $n$ sites ('species'), thus has $n-2$ independent reactions ($c=0$) and thus there are $\ell = 2$ conservation laws. A stoichiometric matrix for $n=7$ looks as follows
\hspace*{-0.5cm}\vbox{\bea
\! \! \! \! \mathbb{S}_{\blackffor} = \begin{pmatrix} 
p   & 0    & 0   & 0   & 0 \\
-1  & p    & 0   & 0   & 0 \\
1-p & -1   & p   & 0   & 0 \\
0   & 1-p  & -1  & p   & 0 \\
0   & 0    & 1-p & -1  & p \\
0   & 0    & 0   & 1-p & -1 \\
0   & 0    & 0   & 0   & 1-p 
\end{pmatrix} 
\eea}
Whose successive rows correspond to $[0],[1],...,[6]$. Conservation of probability directly yields $\pmb{\ell}^{(1)} = (1,1,1,1,1,1,1)$ as left nullvector and conserved quantity
\be
\pmb{\ell}^{(1)} \cdot \pmb{p} = \sum_k p_k (t) = 1
\ee
For the second conserved quantity $\pmb{\ell}^{(2)} = (b_0, b_1, .., b_{n-1})$, we can freely set $b_0 = 0, b_1=1$ by scaling and subtraction of $\pmb{\ell}^{(1)}$. Then from inspection of $\pmb{\ell}^{(2)} \mathbb{S} = \pmb{0}$
\bea
 - 1 + (1-p) b_2 = 0, \\
p b_{k-1} - b_k + (1-p) b_{k+1} = 0, \\
b_k = \frac{1-p}{p} b_{k-1} + 1
\eea
from which we obtain 
\bea
q &=& \frac{1-p}{p}, \\
b_k &=& q^{k - 1} + \frac{q^k - 1}{q - 1}
\eea
At long times, only site $0$ and site $k-1$ will be populated. Let us denote with an overbar $\bar{p}_{k}$
\be
\bar{p}_{k} = \lim_{t \rightarrow \infty} p_k (t)
\ee
The conservation law $\pmb{\ell}^{(2)}$ informs of a constraint on the dynamics of the random walk, and in the long-time limit provides the stationary occupations
\bea
L^{(2)} &=& \pmb{\ell}^{(2)} \cdot \pmb{p}(0) = \pmb{\ell}^{(2)} \cdot \pmb{p}(t) \\
&=& \pmb{\ell}^{(2)} \cdot \bar{\pmb{p}} = b_{n-1} \bar{p}_{n-1}. \nonumber
\eea
For illustration, let $p=0.2$. We then have
\be
\pmb{\ell}^{(2)} = \left(0, 1, \frac{5}{4}, \frac{21}{16}, \frac{85}{64}, \frac{341}{256}, \frac{1365}{1024} \right)
\ee
We initially occupy site $[2]$ and $[3]$ with equal probability $p_2(0)=p_3(0)=1/2$, so that
\be
L^{(2)} = \frac{5}{4} p_2(0) + \frac{21}{16} p_3(0) = \frac{41}{32}
\ee
For $n=7$ sites, our stationary probability for reaching absorbing site $[6]$ is now
\bea
\bar{p}_6 = L^{(2)} / b_6 &=& \frac{1024}{1365} \frac{41}{32} \\
&=& \frac{1312}{1365} \approx 0.961 \nonumber
\eea

\subsection{2d diffusion}
This logic applies more universally. Consider some connected lattice on which diffusion is taking place, with $n$ distinct sites, among which $k \geq 1$ are absorbing states. There are then $n-k$ nonabsorbing sites, whose outgoing reactions can for each be merged to a single reaction, so that there are $n-k$ independent ($c=0$) reactions. From SL1, there are then $\ell = k$ conservation laws, i.e. there are as many conservation laws as there are sinks. Furthermore, we can always choose for our conservation laws a basis such that each has a single nonzero coefficient corresponding to a sink.

Let us now consider a 2d example, namely a 3-by-3 lattic with periodic boundary conditions. We thus have reactions of the form
\bea
\ [x,y] &\rightarrow& [x \pm 1,y], \ \  [x,y] \rightarrow [x, y \pm 1] \nonumber \\
\ [x,y] &\rightarrow& \frac{1}{4} [x + 1,y] + \frac{1}{4} [x - 1,y] \nonumber \\
&+& \frac{1}{4} [x,y + 1] + \frac{1}{4} [x,y - 1] 
\eea
where coordinates are evaluated modulo 3. Let us now make $[1,0]$ and $[2,2]$ absorbing sites. We then have
\be
\! \! \! \! \! \! \! \! \mathbb{S}_{\blackffor} = \begin{pmatrix}
          -1  & \frac{1}{4} & \frac{1}{4} & 
           0 & 0 & \frac{1}{4} & 0  \\  
           
          \frac{1}{4} & \frac{1}{4} & 0 & 
           \frac{1}{4} & 0 & 0 & \frac{1}{4} \\
           
          \frac{1}{4} & -1 & 0 & 
           0 & \frac{1}{4} & 0 & 0 \\  
           
          \frac{1}{4} & 0 & -1 & \frac{1}{4} & 
           \frac{1}{4} & \frac{1}{4} & 0 \\ 
           
          0 & 0 & \frac{1}{4} &  -1 & 
           \frac{1}{4} & 0 & \frac{1}{4} \\ 
           
          0 & \frac{1}{4} & \frac{1}{4} & \frac{1}{4} & 
           -1 & 0 & 0 \\ 
           
          \frac{1}{4} & 0 & \frac{1}{4} & 0 & 0 & -1 & \frac{1}{4} \\

          0 & 0 & 0 & \frac{1}{4} & 0 & \frac{1}{4} & -1 \\
          
          0 & \frac{1}{4} & 0 & 0 & \frac{1}{4} & \frac{1}{4} & \frac{1}{4} 
          
      \end{pmatrix}
\ee   
Successive rows correspond to $[0,0],     [1,0], [2,0]$, $[0,1],    [1,1], [2,2]$, $[0,2], [1,2],   [2,2]$.  One convenient basis for the left nullspace is
\bea
\pmb{\ell}^{(1)} = (6,10,5,5,6,4,4,5,0)\\ 
\pmb{\ell}^{(2)} = (4,0,5,5,4,6,6,5,10) 
\eea
from $\pmb{\ell}^{(1)}$ it can immediately be seen that the stationary probability to reach absorbing site $[1,0]$ $(\bar{p}_{1,0})$ from a single starting site can be $0.4$, $0.5$ or $0.6$. From $\pmb{\ell}^{(2)}$ we draw the same conclusion for $\bar{p}_{2,2}$, and we can furthermore see that $\bar{p}_{2,2} = 1 - \bar{p}_{1,0}$.

\subsection{n-mer adsorption}
\label{appendix:nmer_matrix}
Evaluating $\mathbb{F}_{p,n}$ for $n=3$ $p=1,2,3$, we obtain
\bea
\mathbb{F}_{1,3} &=& \begin{pmatrix} 2 & 0 & 0 \\
1 & 1 & 0 \\
\frac{2}{3} & \frac{2}{3} & \frac{2}{3} \\
\end{pmatrix}, \ \mathbb{F}_{2,3} = \begin{pmatrix} \frac{1}{2} & 0 & \frac{3}{4} \\
\frac{2}{5} & \frac{2}{5} & \frac{3}{5} \\
\frac{1}{3} & \frac{1}{3} & \frac{5}{6} 
\end{pmatrix}, \nonumber \\
\mathbb{F}_{3,3} &=& \begin{pmatrix} \frac{2}{7} & 0 & \frac{6}{7} \\
\frac{1}{4} & \frac{1}{4} & \frac{3}{4} \\
\frac{2}{9} & \frac{2}{9} & \frac{8}{9} 
\end{pmatrix} 
\eea
Which upon substitution in Eq.  ~\eqref{equation:gensol_nmer} yield the coefficients of the conservation laws $\ell^{(1)},\ell^{(2)},\ell^{(3)}$ in Eq.~\eqref{equation:coeffcons_law_nmer} in the main text.

\section{Co-production Curtin-Hammett extensions for elaborate reaction mechanisms}
\label{appendix:chextension}

We can extend the law obtained for the simple CRN $\ce{A} \leftarrow \ce{B} \leftrightarrows \ce{C} \rightarrow \ce{D}$ in various manners. For our illustration, we will retain i) 2 irreversibly produced products $\ce{X}, \ce{Y}$, ii) co-production 1 ($\copro = 1$), in such a manner that for $t \rightarrow \infty$ only products remain. Under these conditions, we can always obtain a pair of conservation laws.

Furthermore, we can always evaluate our pair of conservation laws at $t=0$ and $t=\infty$ and thereby extract expressions $[\ce{X}]_{\infty}, [\ce{Y}]_{\infty}$ that are linear in all initial conditions.

To illustrate this, let us add one more step, to obtain the CRN
\be
\ce{X}_1 \overset{1}{\leftarrow} \ce{X}_2 \overset{2}{\leftrightarrows} \ce{X}_3 \overset{3}{\leftrightarrows} \ce{X}_4 \overset{4}{\rightarrow} \ce{X}_5 \label{equation:CRNext}
\ee
after merging reactions pairwise, we can write the CRN with one less reaction
\bea
\ce{X}_2 \rightarrow \frac{k_1}{k_1 + k_2^+} \ce{X}_1 + \frac{k_2^+}{k_1 + k_2^+} \ce{X}_3 , \\
\ce{X}_3 \rightarrow \frac{k_2^-}{k_3^+ + k_2^-} \ce{X}_2 + \frac{k_3^+}{k_3^+ + k_2^-} \ce{X}_4, \\ 
\ce{X}_4 \rightarrow \frac{k_3^-}{k_3^- + k_4} \ce{X}_3 + \frac{k_4}{k_4 + k_3^-} \ce{X}_5 .
\eea
for which one can check that the following are conserved
\bea
L_1 &=& [\text{X}_1] + [\text{X}_2] + [\text{X}_3] + [\text{X}_4] + [\text{X}_5] \nonumber \\ 
L_2 &=& \frac{k_2^+ + k_1}{k_1} [\text{X}_1] + [\ce{X}_2] - \frac{k_2^-}{k_3^+} [\text{X}_4] \nonumber \\  &-& \frac{k_2^-}{k_3^+} \frac{k_4 + k_3^-}{k_4} [\text{X}_5] 
\eea
To derive $L_2$, one can proceed by setting $\ell^{(2)}_3 = 0, \ell^{(2)}_2 = 1$, and subsequently substituting for each reaction a remaining coefficient that makes the reaction conserve $L_2$. Equivalently, one can find left nullvectors of the stoichiometric matrix after merging reactions, $\mathbb{S}_{\blackffor}$, which here becomes
\be
\mathbb{S}_{\blackffor} = \begin{pmatrix} 
\frac{k_1}{k_1 + k_2^+} & 0 & 0 \\
-1 & \frac{k_2^-}{k_2^- + k_3^+} & 0 \\
\frac{k_2^+}{k_1 + k_2^+} & -1 & \frac{k_3^-}{k_4 + k_3^-} \\
0 & \frac{k_3^+}{k_2^- + k_3^+} & -1 \\
0 & 0 & \frac{k_4}{k_4 + k_3^-}
\end{pmatrix}
\ee
Upon rearranging $L_1,L_2$, we can eliminate $[\ce{X}_1]$
\bea
\! \! \! \! \! \! \! \! L_3 &=& L_1 - \frac{k_2^+ + k_1}{k_1} \\
\! \! \! \! \! \! \! \! &=& \frac{1}{k_1 k_3^+ k_4} ( k_2^+ k_3^+ k_4 [X_2] \nonumber \\ 
&+& (k_1 + k_2^+) k_3^+ k_4 [X_3] \nonumber \\
\! \! \! \! \! \! \! \! &+& (k_1 k_2^- +  k_2^+ k_3^+ + k_1 k_3^+) k_4 [X_4] \nonumber \\ 
\! \! \! \! \! \! \! \! &+&  Z [X_5] ) \nonumber \\
Z &=& k_1 k_2^- \left(k_4 + k_3^-\right) +   k_3^+ k_4 \left(k_2^+ + k_1 \right) \nonumber
\eea
Where wlog we let $[\ce{X}_5]^0 = 0$. Since $[\ce{X}_2]^{\infty}=[\ce{X}_3]^{\infty}=[\ce{X}_4]^{\infty}=0$, we can then write
\bea
\ \! \! \! \! \! \! \! \! [\ce{X}_5]^{\infty} &=& W_5 \left( k_2^+ k_3^+ \right) [\ce{X}_2]^{0}  \\ 
\! \! \! \! \! \! \! \! &+&  \left( k_2^+ k_3^+ + k_1 k_3^+ \right) [\ce{X}_3]^{0} \nonumber \\
 \! \! \! \! \! \! \! \! &+& \left(k_2^+ k_3^+ + k_1 k_3^+  + k_1 k_2^- \right) [\ce{X}_4]^{0} ) \nonumber \\
 \! \! \! \! \! \! \! \! W_5 &=& \frac{k_1 k_3^+}{Z} 
\eea
we can repeat this derivation for $[\ce{X}_1]^{\infty}$, or alternatively leverage the symmetry of the CRN, to obtain
\bea
\ \! \! \! \! \! \! \! \! [\ce{X}_1]^{\infty} &=& W_1 ( k_2^- k_3^- [\ce{X}_4]^{0}  \\ 
\! \! \! \! \! \! \! \! &+&  \left( k_2^- k_3^- + k_2^- k_4 \right) [\ce{X}_3]^{0} \nonumber \\
 \! \! \! \! \! \! \! \! &+& \left(k_2^- k_3^- + k_2^- k_4 + k_3^+ k_4  \right) [\ce{X}_2]^{0} ) \nonumber \\
  \! \! \! \! \! \! \! \! W_1 &=& \frac{k_4 k_3^+}{Z} 
 \eea
for the quotient, we then obtain a similar functional form as Eq. \eqref{equation:Curtin-hammett-extend}
\bea
\! \! \! \! \! \! \! \! \frac{[\ce{X}_1]^{\infty}}{[\ce{X}_5]^{\infty}} &=& \frac{k_4 k_3^- k_2^-}{k_1 k_2^+ k_3^+} Q  \\ 
\! \! \! \! \! \! \! \! &=& e^{\delta \Delta G^{\ddagger}} Q \nonumber \\
\! \! \! \! \! \! \! \! Q &=& \frac{Q_1}{Q_5} 
\eea
where now
\bea
\! \! \! \! \! \! \! \! Q_1 &=& [\ce{X}_4]^{0} + \left(1+\frac{k_4}{k_3^-}\right)[\ce{X}_3]^{0}  \\
\! \! \! \! \! \! \! \! &+& \left(1 + \frac{k_4}{k_3^-} + \frac{k_4}{k_2^-} K_3^{-1} \right)[\ce{X}_2]^{0} , \nonumber \\
\! \! \! \! \! \! \! \! Q_5 &=& [\ce{X}_2]^{0} + \left(1+\frac{k_1}{k_2^+}\right)[\ce{X}_3]^{0} \\ 
\! \! \! \! \! \! \! \! &+& \left(1 + \frac{k_1}{k_2^+} + \frac{k_1}{k_3^+} K_2 \right)[\ce{X}_4]^{0} , \nonumber \\
K_2 &=& \frac{k_2^-}{k_2^+}, \ \ \ K_3 = \frac{k_3^-}{k_3^+} .
\eea
As before, in the limit where isomerization is fast compared to conversion to product $k_2^-,k_2^+,k_3^-,k_3^+ \gg k_1, k_4$, we recover the Curtin-Hammett limit $Q \rightarrow 1$.

It may furthermore be remarked that in mechanisms where reactants are initially trapped in a precursor that irreversibly converts into reactant, that this precursor can be aborbed into the initial concentration for any evaluation of terminal composition in a linear system. 
I.e. suppose we add to our CRN \eqref{equation:CRNext} the reactions
\bea
\ce{P}_2 \rightarrow \ce{X}_2 , \ \ \ 
\ce{P}_3 \rightarrow \ce{X}_3 , \ \ \ 
\ce{P}_4 \rightarrow \ce{X}_4 .
\eea
then our expressions for conserved quantities $L_1, L_2$ retain validity if we perform the substitutions
\be
\ [\ce{X}_k] \rightarrow [\ce{X}_k] + [\ce{P}_k]. 
\ee
$[\ce{X}_1]^{\infty}, [\ce{X}_5]^{\infty}$ can be updated by performing the same substitutions. Precursors themselves may be subject to coproduction
\be
\ce{P} \rightarrow p_2 \ce{X}_2 + p_3 \ce{X}_3 
+ \ce{X}_4 
\ee
which can be accommodated by substituting $[\ce{X}_k]$ with $[\ce{X}_k] + p_k [\ce{P}]$ allowing facile further extension of the results.

\section{Coproduction linkage law from FTLA}
\label{appendix:coproductionFTLA}

Let us consider an ensemble of $\lambda_{\bowtie}$ static linkage classes $\mathcal{L}_1, ..., \mathcal{L}_{\lambda_{\bowtie}}$ and let us consider all $r^+$ reactions between them, and reactions leaving from them (note that any reversible reactions are already absorbed within linkage classes), and for each of $R_1^+, ..., R_{r^+}^+$ we mark the corresponding (single) linkage class from which they leave with $-1$ in their corresponding columns in a matrix $\Lmir_{\bowtie}^+$, analogous to a stoichiometric matrix, e.g.
\be
\Lmir_{\bowtie}^+ = \begin{pmatrix}
-1 & -1 & 0 \\
0 & 0 & -1 \\
0 & 0  & 0
\end{pmatrix},
\ee
Which acts on 
$\mathcal{L}_1,\mathcal{L}_2,\mathcal{L}_3$. We can now span the left nullspace of $\Lmir_{\bowtie}^+$ by $\lambda^{0}_{\bowtie}$ unit vectors, one for each static linkage class with no outgoing reactions. Among $r^+$ outgoing irreversible reactions, any  in excess of one attached to the same linkage class yields a right nullvector of the form $e_k - e_{k+1}$. The number of such nullvectors by definition equals the coproduction. FTLA then yields the equivalent of the coproduction-linkage law Eq. \eqref{equation:copro-linkage-law} 
\be
\text{rk}(\Lmir_{\bowtie}^+) = \lambda_{\bowtie} - \lambda^{0}_{\bowtie} = r^+ - \text{\footnotesize{$\Upsilon$}} \label{equation:ftlalmir}
\ee
Where \\
$r^+: \#$ irreversible reactions with reactant in description (outgoing irreversible reactions) \\
$\text{\footnotesize{$\Upsilon$}}: $ Co-production ($\text{dim(ker}(\Lmir^+_{\bowtie})$)) \\
$\lambda_{\bowtie}: \#$ static linkage classes \\
$\lambda_{\bowtie}^0: \#$ static linkage classes with no outgoing irreversible reactions. ($\text{dim(coker}(\Lmir^+_{\bowtie})$))

\bibliography{BibFile}

\end{document}